\documentclass{IEEE-Con-Sys-mag_rev}

\newcommand{\myparagraph}[1]{\paragraph{#1.}}

\newcommand{\LDdiffdotA}{\dot{S}}
\newcommand{\LDdiffdotB}{\dot{I}}
\newcommand{\LDdiffdotC}{\dot{R}}
\newcommand{\LDdiffdotD}{\dot{\x}}
\newcommand{\LDdiffdotE}{\dot{x}_1}
\newcommand{\LDdiffdotF}{\dot{x}_n}
\newcommand{\LDdiffdotG}{\dot{x}_i}
\newcommand{\LDdiffdotH}{\dot{y}}
\newcommand{\LDdiffdotI}{\dot{x}}
\newcommand{\LDdiffdotJ}{\dot{T}}
\newcommand{\LDdiffdotK}{\dot{L}}
\newcommand{\LDdiffdotL}{\dot{u}}
\newcommand{\LDdiffdotM}{\dot{m}}

\newcommand{\LDdiffdotO}{\dot{b}}

\newcommand{\LDdiffdotS}{\dot{x}_2}
\newcommand{\LDdiffdotT}{\dot{p}}
\newcommand{\LDdiffdotU}{\dot{z}}
\newcommand{\LDdiffdotV}{\dot{w}}
\newcommand{\LDdiffdotW}{\dot{\xx}}
\newcommand{\LDdiffdotX}{\dot{\vecx}}

\newcommand{\LDdiffdotZ}{\dot{z}_1}
\newcommand{\LDdiffdotAA}{\dot{z}_2}

\usepackage{amsmath,amssymb,amsthm}

\usepackage{soul} 

\usepackage{sidecap}

\usepackage{tikz}
\usepackage{wrapfig}

\newcommand{\TB}{\vspace{-0.05ex}}
\newcommand{\TiE}{\setlength{\itemsep}{-0.5ex}}
\newenvironment{TightItem}{\TB\TB\begin{itemize}\TiE}{\end{itemize}\TB}

\newcommand{\bit}{\begin{TightItem}}
\newcommand{\eit}{\end{TightItem}}

\newcommand{\comment}[1]{}

\newcommand{\ns}{\null\vspace{-15pt}}
\newcommand{\red}[1]{{\textcolor{red}{#1}}}

\newcommand{\donotinclude}[1]{}

\newcommand{\collaboration}[1]{}

\usetikzlibrary{arrows}

\newcommand{\R}{{\mathbb R}}  

\newcommand{\sign}{\mbox{sign}\,}
\newcommand{\be}[1]{\begin{equation}\label{#1}}
\newcommand{\ee}{\end{equation}}
\newcommand{\beq}{\begin{eqnarray}}
\newcommand{\eeq}{\end{eqnarray}}
\newcommand{\beqn}{\begin{eqnarray*}}
\newcommand{\eeqn}{\end{eqnarray*}}
\newcommand{\bi}{\begin{itemize}}
\newcommand{\ei}{\end{itemize}}
\newcommand{\ben}{\begin{enumerate}}
\newcommand{\een}{\end{enumerate}}

\newcommand{\x}{x}  
\newcommand{\f}{f}  
\newcommand{\sarr}{\rightarrow } 
\newcommand{\U}{{\mathbb U}}
\newcommand{\Y}{{\mathbb Y}}
\newcommand{\X}{{\mathbb X}}
\newcommand{\barx}{{\bar \x}}

\newcommand{\seq}{=}

\newcommand{\ff}{f} 
\newcommand{\xx}{x} 
\newcommand{\il}{\ell}
\newcommand{\baru}{{\bar u}}

\usepackage[numbers,comma,sort&compress]{natbib}

\newcommand{\FIGDIR}{FIGURES_transients_paper}
\newcommand{\picc}[2]{\begin{center}\pic{#1}{#2}\end{center}}
\newcommand{\pic}[2]{\includegraphics[scale=#1]{\FIGDIR/#2}}

\newcommand{\minp}[2]{\begin{minipage}{#1\textwidth}#2\end{minipage}}

\usepackage{fancybox}

\usepackage{amsfonts}
\usepackage{amsmath}
\usepackage{amssymb}
\usepackage[english]{babel}
\usepackage{booktabs}
\usepackage{enumerate}
\usepackage{graphicx,epsfig}
\usepackage{mathtools}
\usepackage{multicol}
\usepackage{setspace}
\usepackage{subfig}
\usepackage{tabularx}
\usepackage{xcolor}

\newcommand{\GEF}{GEF}
\newcommand{\GAP}{GAP}

\newcommand{\fractxt}[2]{#1/#2}
\newcommand{\fractext}[2]{(#1)/(#2)}
\newcommand{\fractextnp}[2]{#1/#2}

\newcommand{\vecx}{x}

\newcommand{\vecu}{u}
\newcommand{\vecf}{f}

\newcommand{\abs}[1]{\left\vert #1 \right\vert}

\title{Dynamic response phenotypes and model discrimination in systems and synthetic biology}

\author{Eduardo D. Sontag\\
  Northeastern University}

\def\twodigits#1{\ifnum#1<10 0\fi\the#1}
\makeatletter
  \newcommand*\short[1]{\expandafter\@gobbletwo\number\numexpr#1\relax}
\makeatother
\usepackage{datetime2}

\newcommand{\ds}{\displaystyle}
\newcommand{\s}{s}
\newcommand{\pp}{c}

\newcommand{\truered}[1]{{\textcolor{red}{#1}}}
\newcommand{\green}[1]{{\textcolor{green}{#1}}}

\newcommand{\blue}[1]{{\textcolor{blue}{#1}}}
\newcommand{\dgreen}[1]{{\textcolor{darkgreen}{#1}}}
\definecolor{darkgreen}{RGB}{51,102,0}           

\newcommand{\chemotaxisvideoslink}[1]{}

\newcommand{\bfig}{\begin{figure}[ht]}
\newcommand{\efig}{\end{figure}}

\newcommand{\bfigstar}{\begin{figure*}[ht]}
\newcommand{\efigstar}{\end{figure*}}

\jvol{XX}
\jnum{XX}
\paper{8}
\jmonth{June}
\jname{IEEE CONTROL SYSTEMS}
\pubyear{2020}

\begin{document}

\maketitle

\dois{}{}

\begin{sidebar}{What Can We Learn from How Biological Systems Respond Over Time?}
\label{sidebar:whatcanwelearn}

Many biological systems are defined not only by where they eventually settle, but by how they respond over time. A brief surge, a delayed response, a return to baseline, or sensitivity to relative rather than absolute changes can determine outcomes in settings ranging from cell signaling and immune responses to cancer treatment and population dynamics.

Control and systems theory provide a natural framework for addressing these questions because they shift attention from biological components in isolation to the relationships among inputs, internal states, network structure, and measured outputs. In systems and synthetic biology, this perspective is especially valuable when detailed parameter identification is impractical or when biological function depends on transient rather than steady-state behavior.

This article shows how such time-dependent response patterns can be used as practical clues to the hidden organization of a biological system. Instead of requiring a detailed model with precisely known parameters, researchers can often rule out broad classes of possible mechanisms simply by observing which response patterns do, or do not, occur. The examples considered here illustrate how control-theoretic tools can help identify which biological network architectures are consistent or inconsistent with experimentally observed dynamics.  Carefully designed experiments using changing, repeated, or gradually varying stimuli can provide even more information, helping distinguish mechanisms that would look identical in conventional steady-state measurements.

The broader message is that biological dynamics are not merely complications on the way to equilibrium. They are a source of information. Paying attention to the shape and timing of responses can improve model selection, guide experiments, and reveal organizing principles shared across very different biological systems.

\end{sidebar}

\begin{summary}
  \summaryinitial{M}any biological systems encode function not primarily in steady states,
but in the structure of transient responses elicited by time-varying
stimuli.  Overshoots, biphasic dynamics, adaptation kinetics, fold-change
detection, entrainment, and cumulative exposure effects often determine
phenotypic outcomes, yet are poorly captured by classical steady-state
or dose--response analyses.  This article develops an
input--output perspective on such \emph{dynamic phenotypes}, emphasizing
how qualitative features of transient behavior constrain underlying
network architectures independently of detailed parameter values.

A central theme is the role of sign structure and interconnection logic,
particularly the contrast between monotone systems and architectures
containing antagonistic pathways.  We show how incoherent feedforward
(IFF) motifs, operating across distinct time scales, provide a simple
and recurrent mechanism for generating non-monotonic and adaptive
responses across multiple levels of biological organization, from
molecular signaling to immune regulation and population dynamics.
Conversely, monotonicity imposes sharp impossibility results that can be
used to falsify entire classes of models from transient data alone.

Beyond step inputs, we highlight how periodic forcing, ramps, and
integral-type readouts such as cumulative dose responses offer powerful
experimental probes that reveal otherwise hidden structure, separate
competing motifs, and expose invariances such as fold-change detection.
Throughout, we illustrate how control-theoretic concepts, including
monotonicity, equivariance, and input--output analysis, can be used not
as engineering metaphors, but as precise mathematical tools for
biological model discrimination.

Taken together, the examples surveyed here argue for a shift in emphasis
from asymptotic behavior to transient and input-driven dynamics as a
primary lens for understanding, testing, and reverse-engineering
biological networks.

\end{summary}

\section{Introduction}

Understanding how network structure shapes dynamics is a central
challenge in systems and synthetic biology. Signal-transduction
pathways and feedback loops govern how biological systems, from
molecular circuits involving ligands, transcription factors, and
genes, to interacting cell populations in immunology or oncology, to
large-scale social and epidemiological networks, process information
and respond to stimuli. A key question is how the architecture of
these networks constrains their possible dynamical behaviors,
particularly the transient responses that arise before steady-state
behavior manifests. This question complements a central concern of control engineering: determining what can be inferred about an underlying system from its responses to carefully chosen inputs. In biological systems, where parameters are often uncertain and the network structure is only partially known, qualitative response characteristics can be particularly informative. In this work, we investigate what qualitative
features of a network's structure, meaning its interconnection pattern,
graph-theoretic properties, and feedback organization, can be inferred
from ``dynamic phenotypes,'' understood here as the time-dependent
characteristics of responses to rich families of probing inputs beyond
simple step changes.

This article offers a highly selective perspective, focusing largely on
the author's own work and experiences. It should not be taken as a
comprehensive account of the field, to which many researchers have
made foundational and influential contributions. Our aim is to
illustrate, through a few representative examples, how network
structure can impose sharp constraints on nonlinear system behavior
and give rise to qualitative dynamical phenomena not present in linear
systems. We highlight three classes of behavior: adaptation,
fold-change detection (or scale invariance), and non-monotonic or
subharmonic responses, sketching both the underlying mathematical
principles and their relevance across biological scales. Detailed
proofs and full mathematical developments are not presented here;
instead, we point readers to the cited literature for complete
treatments.

\myparagraph{A remark on mathematics as a universal language}

A recurring question in the quantitative life sciences is how
mathematical methods can be applied across diverse biological
contexts, and why systems that appear unrelated at first glance can
nevertheless be described using similar mathematical structures.
One illustrative example of this universality arises when comparing
immune-tumor interactions with ecological predator-prey
systems, as illustrated in Figure~\ref{fig:immune_and_predators}.
Mathematics provides a common language for understanding
interacting populations, whether in the tumor microenvironment or in
ecosystems such as lynx and hares. In both settings, one population
expands in response to the other: predators proliferate when prey are
abundant, and immune effector cells activate and grow in number when
tumor cells present recognizable antigens. Conversely, both systems
exhibit suppressive dynamics in which predators reduce prey levels and
immune responses constrain tumor growth. These reciprocal interactions
generate rich transient and oscillatory behaviors shaped by nonlinear
feedbacks, time delays, and resource limitations. At the same time,
the analogy clarifies important differences: tumors can evolve escape
mechanisms not found in ecological prey, and the immune system
includes regulatory pathways with no direct ecological
counterpart. Nevertheless, viewing immune-tumor and predator-prey
interactions through a shared mathematical framework reveals how
similar structural motifs can govern dynamics across biological
scales.
\bfigstar
\minp{0.5}{\picc{0.1}{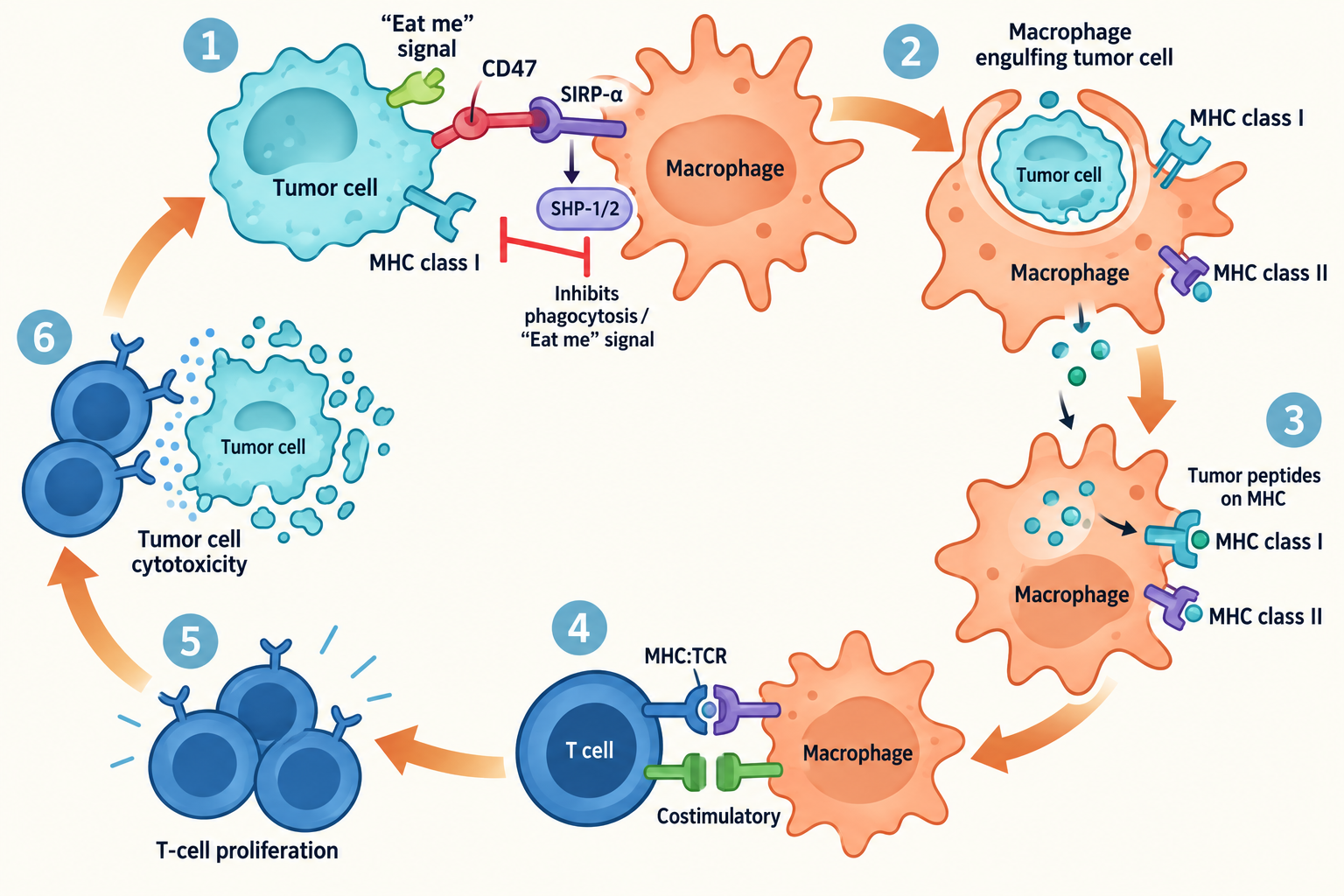}}%
\minp{0.5}{\picc{0.1}{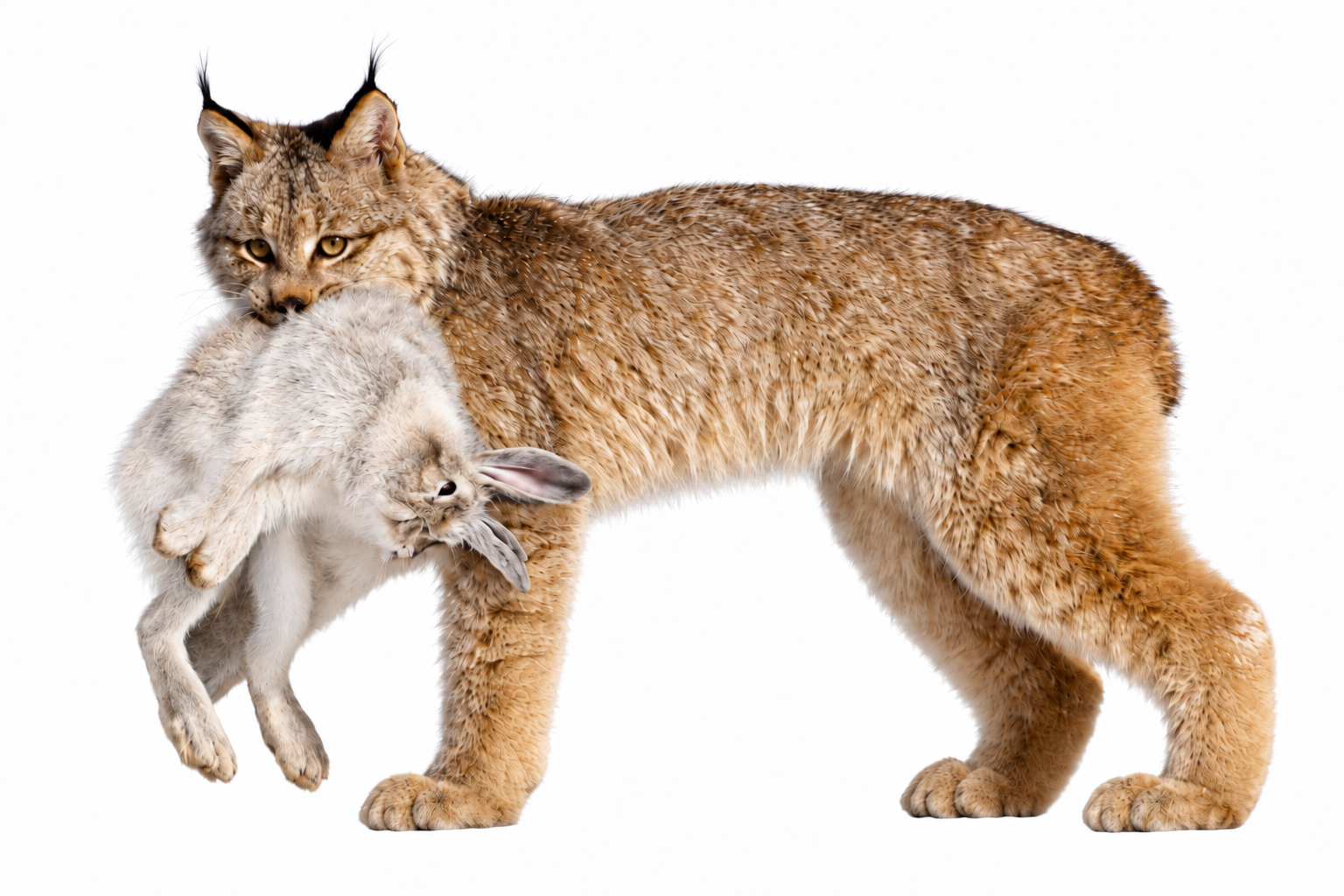}}
\caption{Predator-prey dynamics can model biological phenomena as
  varied as immune-tumor interactions and ecological predator-prey dynamics.
Left illustration shows a typical interaction among immune cells and
tumors: 
tumor-cell CD47 engaging macrophage SIRP-$\alpha$/SHP-1/2 to oppose the ``eat me'' signal; macrophage engulfment of tumor material; MHC class I/II and tumor-peptide presentation with costimulation to a T cell; then T-cell proliferation and tumor-cell cytotoxicity.
The right illustration depicts a Canada lynx (Lynx canadensis) carrying its
primary prey, a snowshoe hare (Lepus americanus). 
}
\label{fig:immune_and_predators}
\efigstar

\myparagraph{Organization of the article}

In the remainder of this introduction, we highlight the importance of
transient behaviors, drawing illustrative examples from multiple
levels of biological organization, and introduce the mathematical
framework used throughout the paper.
The main body of the article is organized as a sequence of thematic
``vignettes,'' each anchored in concrete biological case studies. They
address: (1) the role of monotonicity in shaping transient,
as well as asymptotic, system behavior, with particular emphasis on
motifs such as incoherent feedforward structures and nonlinear
feedback mechanisms; (2) the use of these concepts for the reverse
engineering of signaling pathways from measured responses; (3)
mechanisms of adaptation; (4) the use of periodic inputs to achieve
finer discrimination among adapting motifs; (5) Weber's law and scale
invariance, together with their implications for bacterial chemotaxis;
and (6) cumulative dose responses as an additional experimentally
accessible ``dynamic phenotype.'' 
Appendices collect several mathematical details and technical arguments that
are not readily available in the cited literature.

\subsection{It's not all in the asymptotics: transients matter!}

This article emphasizes \textit{transient responses} and dynamical behaviors
induced by \textit{time-varying inputs}, including periodic, pulsed, and
otherwise nonstationary signals, rather than focusing exclusively on
asymptotic stability properties. In this respect, our perspective
departs from the traditional emphasis in control theory on equilibrium
stabilization and long-term convergence.

That said, the systematic study of transient behavior is well
established in classical control. Canonical examples include
overshoot, rise time, settling time, and transient amplification in
linear systems; input-output gain bounds and induced norms in robust
control; and the use of lead-lag compensation to shape transient
responses without altering steady-state behavior. More recently,
transient performance has been a central concern in safety-critical
control: control-barrier functions are routinely combined with
control-Lyapunov functions to enforce state and output constraints
while still guaranteeing stabilization, effectively providing
dynamical ``guardrails'' that regulate trajectories during transients
rather than only at equilibrium. Other examples include model
predictive control, where finite-horizon performance criteria
explicitly prioritize transient behavior, and adaptive or
gain-scheduled controllers, where responses are shaped differently
across operating regimes.

What distinguishes the biological setting is not the absence of such
questions, but rather their centrality. In many biological systems,
function is encoded not in steady states but in the temporal profiles
of responses: pulse timing, pulse amplitude, adaptation kinetics, fold
changes, cumulative exposure, and entrainment to rhythmic
inputs. Moreover, biological systems are intrinsically nonlinear,
stochastic, and subject to significant parameter variability,
and their transient behaviors
often carry semantic meaning, for example, distinguishing cell fates,
triggering developmental decisions, or coordinating collective
motion. These phenomena cannot be adequately understood by
linearization around equilibria alone, nor by appealing solely to
familiar nonlinear behaviors such as hysteresis, limit cycles, or
chaos.

The focus on transient and input-driven dynamics therefore highlights
a broader role for nonlinearity: not merely as a source of complex
asymptotic behavior, but as a structural mechanism for robustness,
invariance, and information processing under time-varying
stimulation. In this sense, the biological questions addressed here
align naturally with, but also extend beyond, classical
control-theoretic paradigms. For systems and synthetic biology, the resulting challenge is therefore not to replace classical control concepts, but to extend and deploy them in settings where transient response itself may encode biological function. 

\myparagraph{Example: cell fate upon stress}
In mammalian cells, one striking example of the role of transient
dynamics is provided by their responses to certain types of stresses.
The tumor suppressor gene \emph{p53} is often referred to as the
``guardian of the genome'' and is mutated or functionally inactivated
in more than 50\% of human cancers. p53 functions primarily as a
stress-responsive transcription factor that is rapidly activated in
response to diverse cellular insults, including DNA damage, oncogene
activation, hypoxia, and metabolic stress. Once activated, p53
regulates the expression of hundreds of target genes involved in DNA
repair, cell-cycle checkpoints, apoptosis, senescence, and cellular
metabolism, thereby acting as a central integrator of stress signals
that determine cell fate~\cite{riley07,riley07hmm}.

The review~\cite{p53_phenotype} surveys experimental and theoretical
work demonstrating that distinct \emph{dynamic phenotypes} of p53
activation, rather than simply its absolute abundance, are closely
linked to different cellular outcomes following stress, such as DNA
damage. In particular, (i) sustained trains of p53 pulses are commonly
associated with the activation of apoptotic programs leading to
programmed cell death; (ii) isolated or transient p53 pulses tend to
trigger reversible cell-cycle arrest, during which DNA repair pathways
are engaged and, if repair is successful, cells may re-enter the cell
cycle and resume proliferation; and (iii) convergence of p53 activity
to a persistently elevated steady state correlates with the induction
of cellular senescence, a stable and largely irreversible fate
characterized by permanent cell-cycle arrest while maintaining
metabolic and secretory activity. These distinct temporal patterns are
illustrated schematically in Figure~\ref{fig:p53}.
\bfig
\picc{0.1}{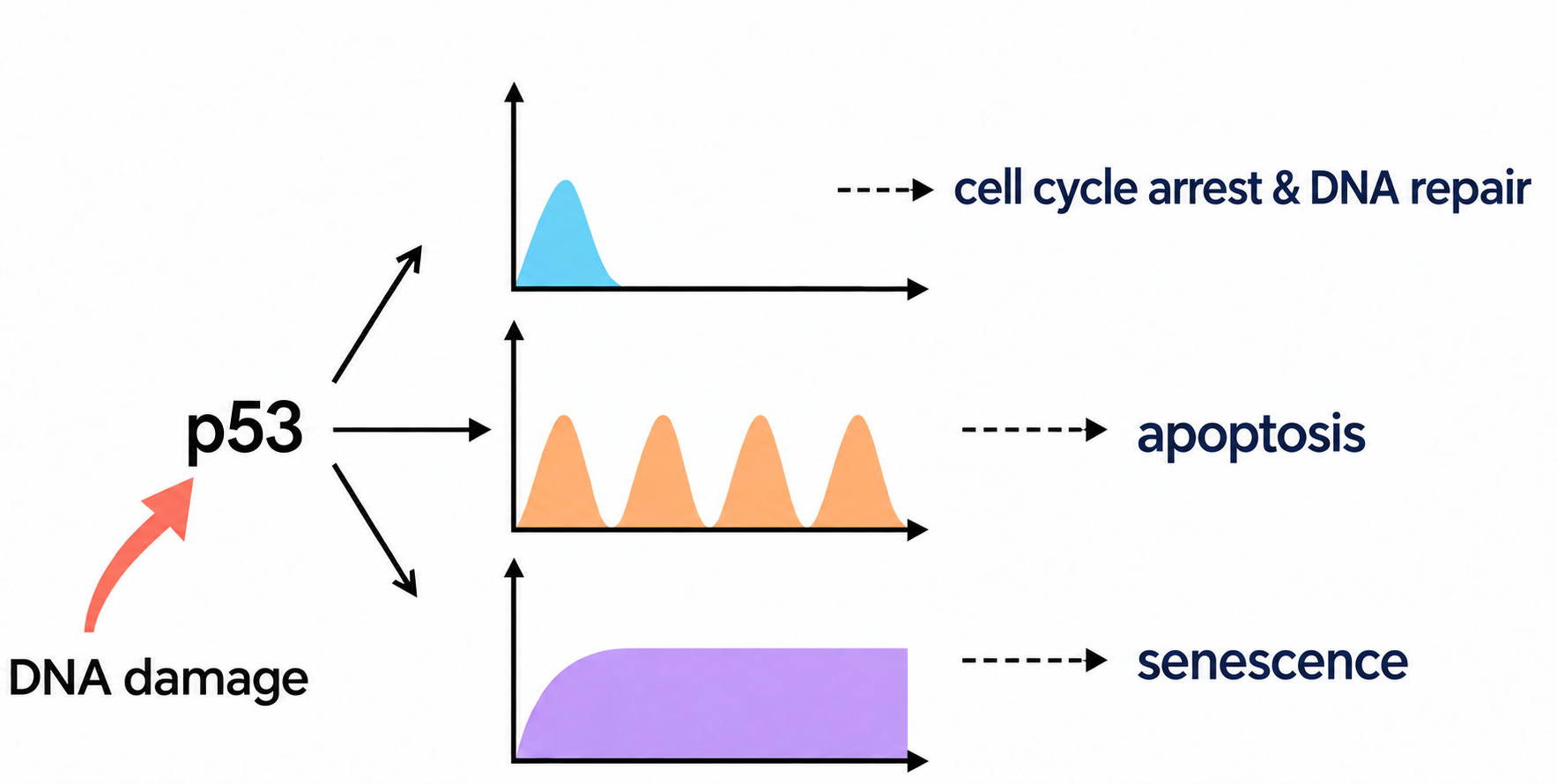}
\caption{Phenotypic consequences of distinct p53 activation dynamics.
Figure generated using  ChatGPT, inspired by the paper~\cite{p53_phenotype}.}
\label{fig:p53}
\efig

Taken together, these observations underscore the critical role of p53
\emph{dynamics}, including pulse frequency, duration, and temporal
structure, in encoding information that directs cell fate decisions.
They provide a paradigmatic example of how biological systems use
transient and time-varying signaling behaviors, rather than static
signal levels alone, to implement robust and context-dependent
decision-making processes.

We next highlight the central importance of transient dynamics through
a small set of additional motivating examples drawn from diverse areas
of biology. A recurring theme across these examples is the appearance
of non-monotonic responses. 
These often arise from the presence of \emph{incoherent feedforward (IFF)
motifs}. Such motifs consist of two distinct pathways linking an input
to an output: a typically fast, activating pathway and a slower
pathway that acts in opposition. The interplay between these competing
influences naturally produces transient overshoots, delayed
suppression, and other non-steady behaviors that cannot be inferred
from asymptotic analysis alone. Despite their structural simplicity,
IFF motifs provide a powerful and widely recurring mechanism for
shaping time-dependent responses in biological systems. As we
illustrate throughout the article, they, together with more classical negative
feedback loops, offer a systems-level
explanation for many seemingly disparate transient phenomena
encountered in cellular signaling, immunology, and population
dynamics.

\myparagraph{Transient dynamics of an infectious population during an epidemic}

Epidemic dynamics provide a particularly clear illustration of why
\emph{transient behavior}, rather than asymptotic stability, is often
the dominant object of interest. A classical starting point is the SIR
model, in which a population is divided into \emph{susceptible} ($S$),
\emph{infectious} ($I$), and \emph{removed} ($R$) individuals. Infectious
individuals transmit the disease to susceptibles, who then become
infectious themselves, while infectious individuals eventually recover
or are removed (through immunity or death). On the time scales of
interest here, we assume no return flow from the removed class back to
the susceptible class. See Figure~\ref{fig:sir} (left).
\bfigstar
\minp{0.4}{\picc{0.45}{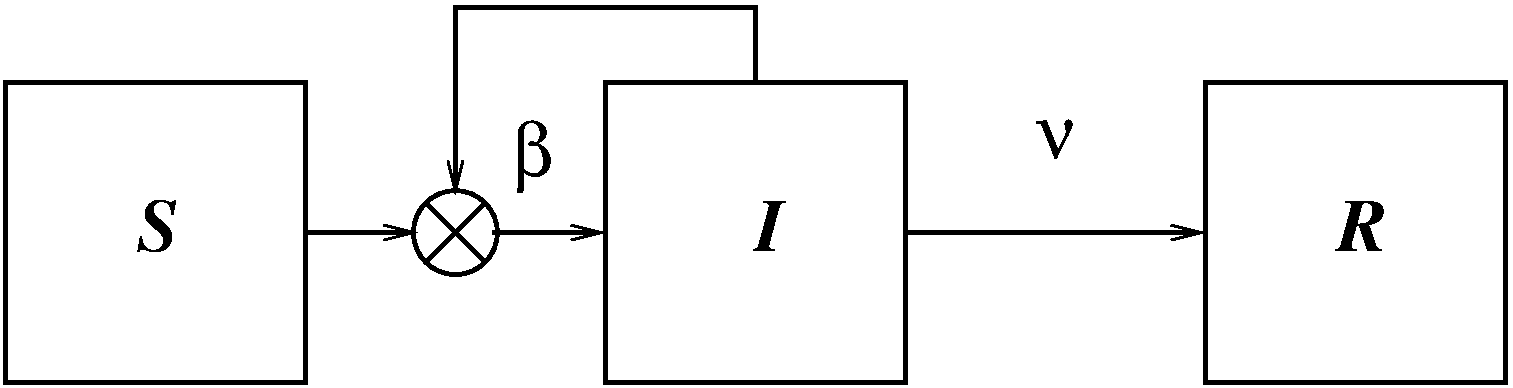}}%
\minp{0.15}{%
  \setlength{\unitlength}{4144sp}%
\centering
\begin{picture}(777,705)(3946,-4171)
\put(4051,-4111){\line( 1, 1){450}}
\multiput(4411,-3571)(5.62500,-5.62500){33}{\makebox(1.5875,11.1125){\tiny.}}
\put(4681,-3691){\vector( 0,-1){310}}
\put(4051,-4111){\vector( 1, 0){540}}
\put(3801,-4162){$\beta$}
\put(4636,-3616){$S$}
\put(4631,-4156){$I$}
\end{picture}
}%
\minp{0.4}{\picc{0.1}{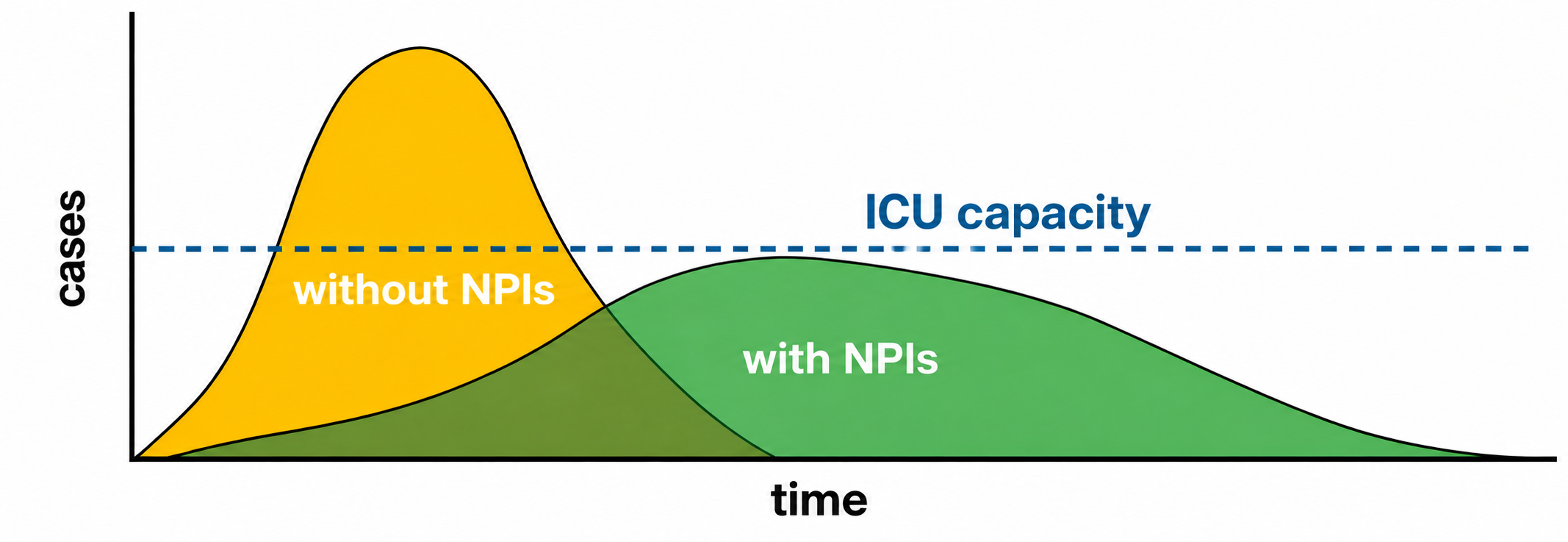}}
\caption{SIR model of epidemics.
  Left: Block diagram representation of the SIR model.
  Center: Incoherent feedforward structure implicit in the SIR model.
  Right: Flattening the curve in epidemics:~the objective is to use
  non-pharmaceutical interventions (NPIs) in order to reduce
the maximum number of infected individuals in such a way that demand
for intensive care units can be met.}
\label{fig:sir}
\efigstar

Under the standard mass--action assumption in a well-mixed population,
the dynamics are
\begin{eqnarray*}
\LDdiffdotA &=& -\beta S I, \label{eq:SIR_S}\\
\LDdiffdotB &=& \beta S I - \nu I, \label{eq:SIR_I}\\
\LDdiffdotC &=& \nu I, \label{eq:SIR_R}
\end{eqnarray*}
where $\beta>0$ denotes the transmission (or contact) rate and
$\nu>0$ the recovery rate. Since $R(t)$ can be obtained by integrating
$I(t)$, the essential dynamics are captured by the $(S,I)$ subsystem.

From the standpoint of asymptotic behavior, the SIR model is
mathematically unremarkable: regardless of parameter values, one always
has
\[
I(t)\rightarrow 0,
\qquad
S(t)\searrow S_\infty>0
\quad \text{as } t\rightarrow\infty .
\]
Thus, on sufficiently long time scales, months or years after an
outbreak, the infection inevitably dies out. From a purely
stability-oriented viewpoint, this would suggest that there is little
of interest to analyze.

In practice, however, \emph{the transient dynamics are everything}.
The primary epidemiological concern is not the eventual disappearance
of the disease, but rather the magnitude and timing of the outbreak
peak. When the basic reproduction number
\[
\mathcal R_0=\frac{\beta S_0}{\nu}
\]
exceeds one, the number of infectious individuals initially grows,
reaches a maximum, and then declines. The height of this peak,
$\|I\|_\infty=\max_t I(t)$, determines whether healthcare systems are
overwhelmed.

This motivates the widely adopted public-health objective of
``flattening the curve,'' illustrated schematically in
Figure~\ref{fig:sir}, where the goal is to reduce the peak
number of infections so that it remains below treatment capacity.

From a control-theoretic perspective, epidemic mitigation acts primarily
through modifications of the effective transmission rate $\beta$, using
non-pharmaceutical interventions (NPIs) such as social distancing,
masking, or mobility restrictions. Importantly, changes in $\beta$ do
not influence the system through a single pathway. An increase in
$\beta$ has an immediate positive effect on the growth rate of $I$, but
it also accelerates the depletion of susceptible individuals $S$, which
in turn reduces future infection rates. Conversely, a larger susceptible
pool amplifies infection growth.

This interaction gives rise to an \emph{incoherent feedforward (IFF) motif}.
Seen in this light, $\beta$ acts as an input that simultaneously
promotes infection directly and, indirectly through its effect on $S$,
contributes to suppressing future infection growth. A larger $S$
increases $I$, while increased $I$ decreases $S$.
(As discussed later, we use IFF to refer to the signed
feedforward structure of the interaction graph, rather than to the presence or
absence of feedback loops.)
This structure is
summarized schematically as shown in Figure~\ref{fig:sir}.
The resulting non-monotonic behavior of $I(t)$, initial growth followed
by decline, is therefore a direct consequence of this incoherent
feedforward interaction. Crucially, this behavior is not an asymptotic
property but a transient one. It is precisely this transient overshoot
that public health interventions seek to shape.

This example reinforces a broader theme: in many biological and social
systems, including epidemics, the most relevant questions concern peak
responses, timing, and cumulative burden rather than long-term
equilibria. The presence of incoherent feedforward interactions provides
a mechanistic explanation for why non-monotonic transients arise
naturally and why interventions that modestly alter parameters such as
$\beta$ can have disproportionately large effects on peak outcomes.
Incoherent feedforward motifs will arise repeatedly in this article and serve as a central organizing theme.

We do not pursue epidemic modeling further in this article; instead, we
refer the readers to the special issues
\cite{special_issue_annual_reviews_epidemics1,special_issue_annual_reviews_epidemics2,special_issue_annual_reviews_epidemics3},
which survey a broad range of applications of control-theoretic
methods to epidemic mitigation. In particular, the role of
intervention timing, such as the scheduling of non-pharmaceutical
interventions or ``lockdowns'' to minimize the peak number of
infectious individuals $\max_t I(t)$, is studied, for example,
in~\cite{explicit_lockdowns_medrxiv_2021}. Adaptive and feedback-based
control strategies aimed at reducing this peak are explored, for
example, in~\cite{alradhawi_sadeghi_sontag_cdc2021_covid}.

\myparagraph{Transients in response to chemotherapy}

Although chemotherapy can induce rapid reductions in tumor burden,
these benefits are often short-lived; tumor regrowth frequently
follows as resistance develops.
The resulting response is therefore non-monotonic and strongly time
dependent, making transient dynamics central to therapeutic outcomes
(see Figure~\ref{fig:waxmab_iffl}, right).
One plausible contributor to this lack of durability is that cytotoxic
chemotherapy is not tumor-specific and may damage components of the
host immune system, thereby weakening immune surveillance and
long-term tumor control.
For example, administration of cyclophosphamide (CPA) has been shown
to reduce immune activity: experimental data reveal significant
decreases in marker gene expression associated with natural killer
(NK) cells, dendritic cells (DCs), and macrophages within the tumor
microenvironment following CPA treatment~\cite{wu2014metronomic}.
Thus, a therapeutic input can have opposing effects: it may
directly reduce tumor burden while simultaneously impairing immune
mechanisms that would otherwise contribute to sustained suppression,
as illustrated schematically in Figure~\ref{fig:waxmab_iffl} (left).
From a systems-theoretic perspective, this antagonistic interaction is
naturally represented by an \emph{incoherent feedforward (IFF) motif}:
increasing drug dose enhances direct tumor cell killing while
concurrently diminishing immune-mediated clearance
(Figure~\ref{fig:waxmab_iffl}, center).
\bfigstar
\minp{0.4}{\picc{0.1}{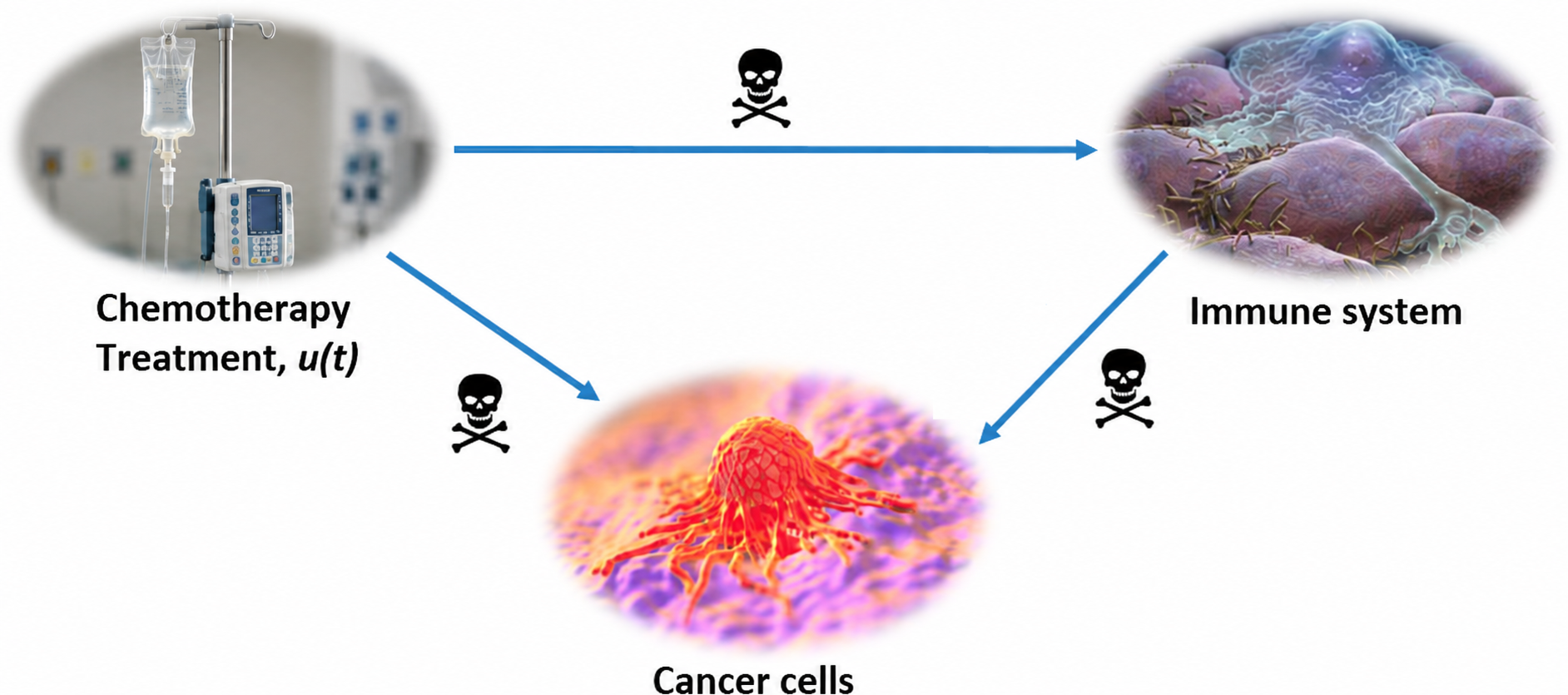}}%
\minp{0.15}{%
  \ \ \
\setlength{\unitlength}{4144sp}%
\begin{picture}(777,705)(3946,-4171)
\put(4051,-4111){\line( 1, 1){445}}
\put(4681,-3691){\line( 0,-1){310}}
\put(4051,-4111){\line( 1, 0){500}}
\put(3801,-4162){$u$}
\put(4416,-3586){\line(1,-1){160}}
\put(4636,-3616){$I$}
\put(4621,-4046){$-$}
\put(4526,-4156){$|$}
\put(4611,-4176){$C$}
\end{picture}%
}%
\minp{0.45}{\picc{0.35}{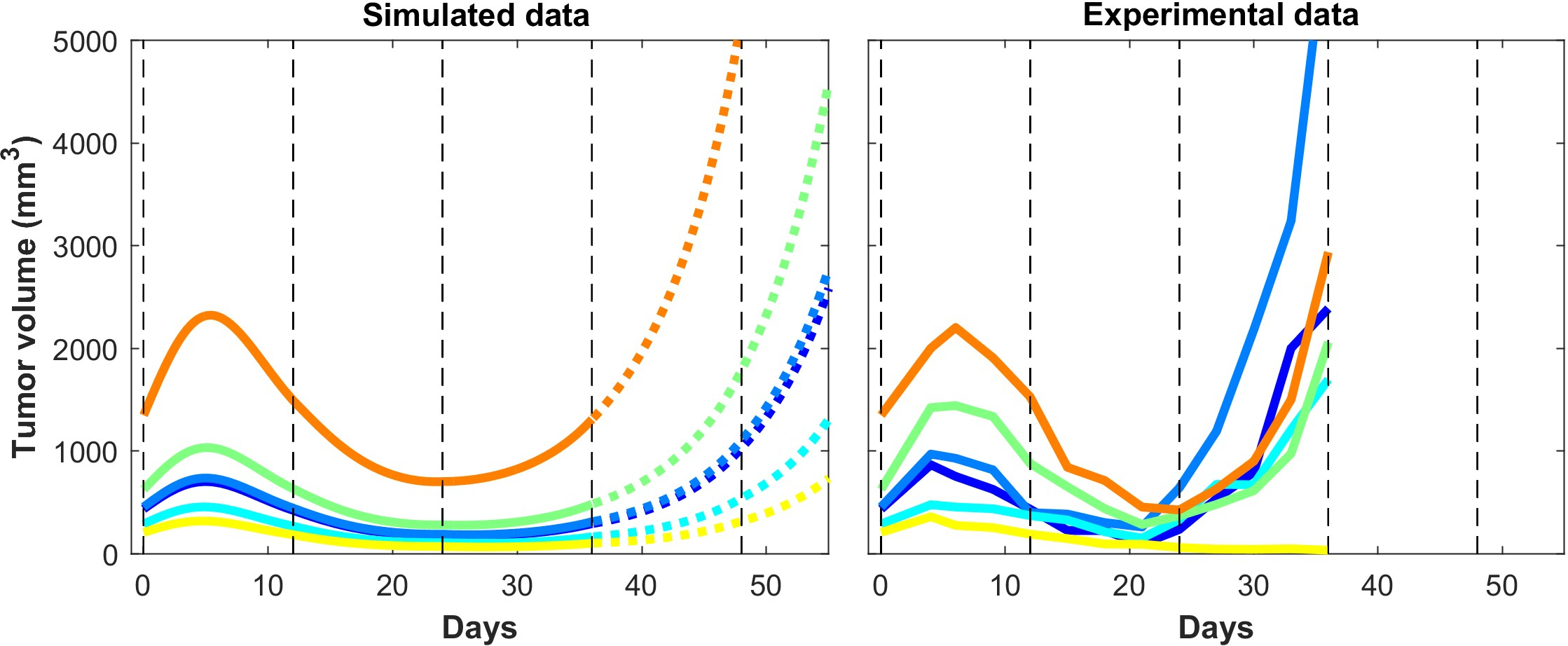}}
\caption{%
  Left: Chemotherapy directly kills cancer cells, but collateral
  damage (``friendly fire'') also affects certain immune cell
  populations, thereby weakening immune-mediated regulation of tumor
  growth. 
  Center: A simplified incoherent feedforward (IFF) representation of
  these interactions, where $u$ denotes drug concentration, 
  $I$ immune components, and $C$ cancer cells.
  This schematic is intentionally coarse-grained: chemotherapy can
  also exert immunostimulatory effects, and tumors actively recruit
  and modulate immune populations. 
  Right: Representative treatment responses for a metronomic regimen
  with doses administered every 12 days. Solid curves show model
  fits (using parameters in~\cite{2019_phong_tran1}) to experimental tumor-volume data from
  six mice reported in~\cite{wu2014metronomic}. Following an initial
  reduction, tumor regrowth occurs after a delay, likely reflecting
  pharmacokinetic and pharmacodynamic effects as well as diminished
  immunogenic cell death. Independent measurements of immune activity
  in~\cite{wu2014metronomic} (not shown) confirm suppression of immune markers
  during treatment.}
\label{fig:waxmab_iffl}
\efigstar

A longstanding paradigm in cancer chemotherapy is the \emph{maximum
tolerated dose} (MTD) strategy, in which cytotoxic drugs are
administered at the highest doses that can be safely tolerated, with
the aim of maximizing immediate tumor cell kill.
However, high-dose chemotherapy often damages not only tumor cells but
also key components of the immune system. Such immune impairment can
substantially weaken immune-mediated tumor control, thereby promoting
immunosuppression and increasing the likelihood of tumor relapse and
the emergence of drug resistance~\cite{2019_phong_tran1}. These
limitations help explain why the benefits of MTD regimens are
frequently transient and accompanied by significant side effects.
This perspective motivates \emph{metronomic chemotherapy}, an
alternative strategy that emphasizes sustained, low-dose,
high-frequency drug administration. Rather than maximizing acute
cytotoxicity, metronomic regimens aim to balance direct tumor cell
killing with preservation, and in some cases enhancement, of
anti-tumor immune responses, including immunogenic cell death. Because
immune effects are mediated through complex processes within the tumor
microenvironment, such as the recruitment of immunosuppressive
populations and cytokine signaling, chemotherapy can exert both pro-
and anti-immune influences, giving rise to strongly time-dependent and
potentially non-monotonic treatment responses.
Motivated by these considerations, in~\cite{2019_phong_tran1}
we developed a mathematical framework for analyzing metronomic
chemotherapy strategies. The resulting phenomenological model
aggregates immunostimulatory and immunosuppressive effects into a
small number of effective variables, including drug concentration and
tumor volume, enabling systematic investigation of dosing schedules
and their transient consequences. The model was calibrated using
experimental data from~\cite{wu2014metronomic} on metronomic
cyclophosphamide treatment in an implanted GL261 glioma model at a
dose of 140~mg/kg. Despite its simplicity, a single parameter set was
sufficient to reproduce tumor responses across a wide range of
treatment schedules, including non-monotonic transients consistent
with an incoherent feedforward structure. More broadly, this work
illustrates how a focus on transient dynamics and network motifs can
reveal mechanistic insights that are not apparent from steady-state or
conventional dose–response analyses alone.

\myparagraph{Transient responses and pseudoprogression in immune checkpoint blockade}

The clinical development of immune checkpoint inhibitors, beginning
with ipilimumab (anti--CTLA-4), highlighted the importance of
transient response dynamics in cancer therapy.  In a subset of
patients, early radiographic assessments revealed an apparent
worsening of disease, characterized by enlargement of existing lesions
or the appearance of new lesions, followed later by tumor
stabilization or regression.
Such patterns were initially unexpected
under conventional cytotoxic response paradigms, but are now widely
recognized as manifestations of \emph{pseudoprogression}.
Mechanistically, this early increase in measured tumor burden is often
attributed not to net malignant growth, but to immune-cell
infiltration, inflammatory edema, and tissue remodeling triggered by
immune activation.  As a result, standard response criteria based
solely on early changes in lesion size may misclassify effective
immunotherapy as treatment failure.  This issue is not merely
academic: premature discontinuation of therapy or inappropriate
removal of patients from clinical trials can occur if transient
responses are interpreted using classical criteria alone.
These observations motivated the development of immune-adapted
response frameworks.  The immune-related response criteria
(irRC)~\cite{Wolchok2009irRC} were among the first to explicitly
accommodate atypical patterns of response by incorporating new lesions
into an aggregate tumor burden rather than defining progression solely
by their appearance.  Subsequently, the iRECIST
guidelines~\cite{Seymour2017iRECIST} formalized a two-stage notion of
progression, distinguishing between \emph{unconfirmed} immune
progression (iUPD) and \emph{confirmed} immune progression (iCPD),
thereby allowing continued treatment in clinically stable patients
until progression is verified on follow-up imaging.

From a dynamical systems perspective, immune checkpoint blockade
illustrates how \emph{transient behavior}, rather than asymptotic or
steady-state response, can be the dominant determinant of therapeutic
interpretation.  Early inflammatory expansion followed by contraction
is not an anomaly, but a signature of immune-mediated tumor control.
The ipilimumab experience thus provides a paradigmatic example of why
time-resolved, input-driven dynamics must be incorporated into both
clinical decision-making and mathematical models of cancer therapy.

At present, it remains difficult to identify which specific mechanisms
are primarily responsible for the observed non-monotonic response
patterns in immune checkpoint therapies. The immune–tumor interface
involves a dense web of interacting pathways, many of which combine
activating and inhibitory effects operating on different time
scales. In particular, numerous incoherent feedforward (IFF) motifs, as
well as classical negative feedback motifs, arise naturally in the
regulation of immune activation, immune suppression, and tumor evasion
in the context of immune checkpoint blockade. These motifs provide a
plausible systems-level explanation for transient worsening, delayed
responses, and other non-monotonic behaviors observed clinically. A
schematic overview of some of these interactions is shown in
Figure~\ref{fig:checkpoints}. 

\bfig
\picc{0.20}{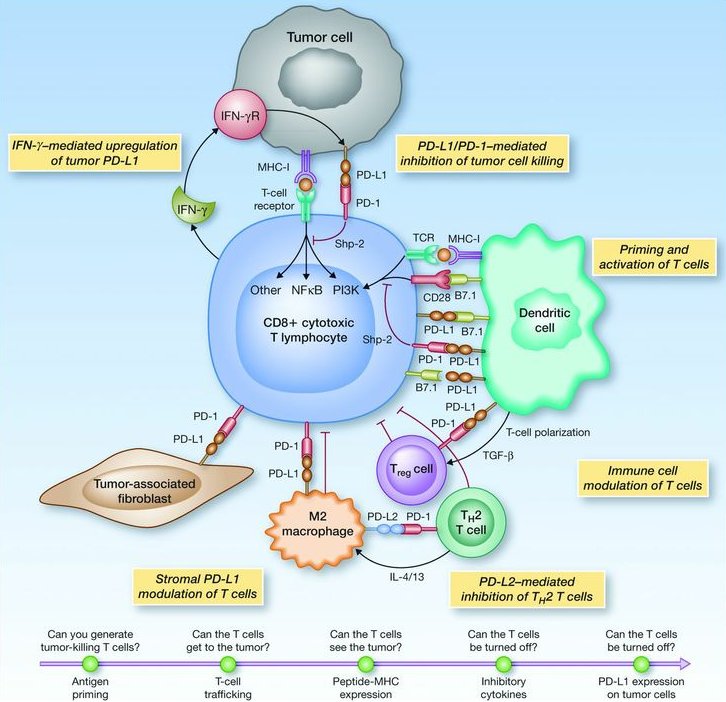}

\setlength{\unitlength}{6000sp}%
\begin{picture}(777,705)(2946,-4171)
\put(4051,-4111){\vector( 1, 1){450}}
\put(4781,-3691){\line( 0,-1){310}}
\put(4681,-3996){\line( 1,0){210}}
\put(4051,-4111){\vector( 1, 0){540}}
\put(3451,-4132){cancer cell}
\put(4616,-3616){checkpoint}
\put(4671,-4156){CD8}
\end{picture}
\caption{Top: Schematic representation of interactions among immune
  components and between immune cells and tumor cells in the context
  of immune checkpoint regulation. Figure reproduced
  from~\cite{2023_frontiers_immune_interactions}, freely available through
  Creative Commons Attribution License (CC BY). 
  Shown are both activating and
  inhibitory pathways, including the PD-L1/PD-1
  inhibitory axis originating from tumor cells, which counterbalances
  T cell activation mediated by the MHC-I/TCR pathway. 
  Bottom: The coexistence
  of such antagonistic pathways naturally gives rise to incoherent
  feedforward and feedback motifs that can shape transient and
  non-monotonic response dynamics. }
\label{fig:checkpoints}
\efig

\myparagraph{Immune detection of velocity of antigen presentation}

In vertebrates, immunity emerges from the coordinated action of innate
and adaptive components, protecting against pathogens and contributing
to tumor immune surveillance. A basic requirement is discrimination:
mounting strong responses to dangerous ``nonself'' cues while
maintaining tolerance to self, a theme that traces back to classic
self/nonself ideas (e.g., thymic negative selection of T cells) and
the foundational proposals of Burnet and Talmage
\cite{Kuby_immunology_book,AbbasLichtmanPillai_immunology_book,burnet1957b,talmage1957,pradeu-book}.

A purely static self/nonself picture is, however, difficult to
reconcile with several well-known observations that point to an
essential role for \emph{dynamics}. The immune system stably tolerates
commensal microbiota despite continual exposure to foreign molecular
patterns; slow-growing tumors can evade detection even when expressing
immunogenic antigens; chronic stimulation can drive functional
hyporesponsiveness (e.g., reduced NK activation); and repeated
exposure to inflammatory triggers can produce tolerance or
``reprogramming,'' as in endotoxin tolerance in macrophages
\cite{pradeu-book,grossmanberke1980,pradeu2013,west_heagy2002}. Likewise,
lymphocyte anergy and the clinical phenomenology of allergy
desensitization suggest that not only antigen identity, but also
\emph{how stimulation changes in time}, can determine whether
responses persist or shut down
\cite{grossman1992,burks2012,pradeu-book}.
In a different context, Bocharov and colleagues~\cite{bocharov2004} 
studied the effect of varying exponential rates of growth
of Hepatitis B and Hepatitis C viral infections and found
non-monotonic immune responses.

Motivated by such phenomena, several authors have argued that immune
recognition should incorporate \emph{temporal features}—in effect,
sensitivity to rates of change as well as levels. In this view,
sustained effector responses are more likely to follow sufficiently
rapid increases in stimulation, whereas slowly varying or chronic
inputs can induce adaptation and tolerance. Representative
formulations include the ``tunable activation threshold'' proposal of
Grossman and Paul \cite{grossman1992}, along with later related
perspectives such as the ``discontinuity theory'' \cite{pradeu2013} and
growth-threshold ideas \cite{arias2015}.

In this spirit, the phenomenological model introduced in
\cite{two-zone-journal} uses a simple circuit architecture to support
dynamic discrimination of immune challenges. A central ingredient is
an incoherent feedforward (IFF) motif.
The IFF motif can render the response
sensitive to exponential growth-like increases in
stimulation—consistent with experimental evidence that rapidly
increasing antigenic drive promotes reactivity.
This article suggested a plausible immunological implementation
via interactions between effector and regulatory T-cell
populations, Figure~\ref{fig:two-zone}.
\bfigstar
\minp{0.3}{\picc{1.25}{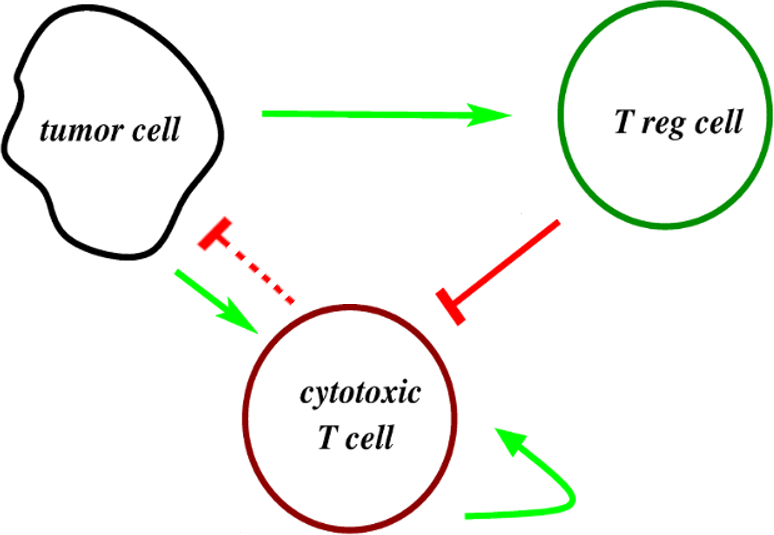}}%
\minp{0.2}{%
\setlength{\unitlength}{6000sp}%
\begin{picture}(777,705)(3946,-4171)
\put(4051,-4111){\vector( 1, 1){450}}
\put(4781,-3691){\line( 0,-1){310}}
\put(4681,-3996){\line( 1,0){210}}
\put(4051,-4111){\vector( 1, 0){540}}
\put(3601,-4132){antigen}
\put(4636,-3616){T reg}
\put(4671,-4156){CD8}
\end{picture}
}%
\minp{0.25}{\picc{0.1}{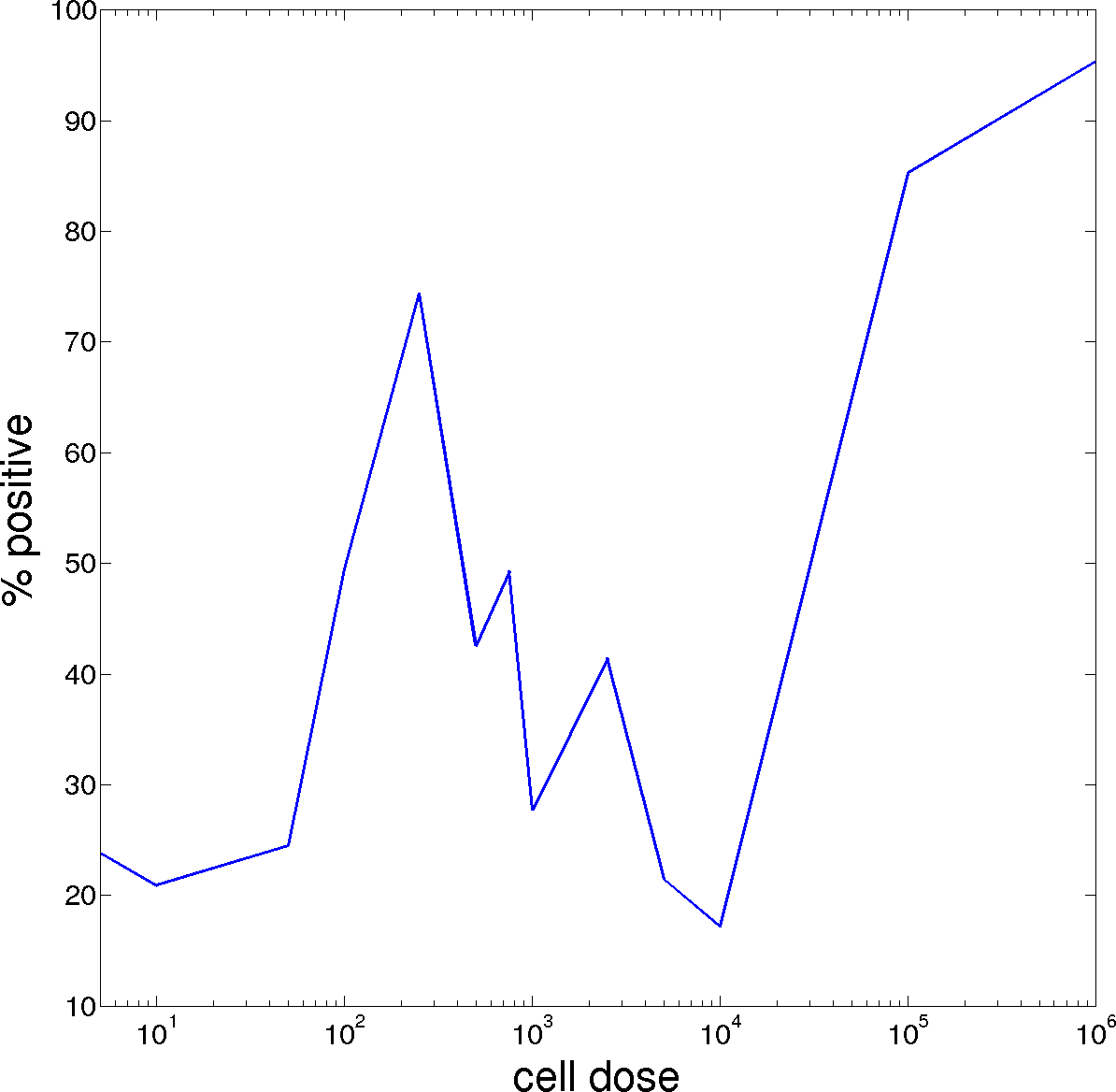}}%
\minp{0.25}{\picc{0.1}{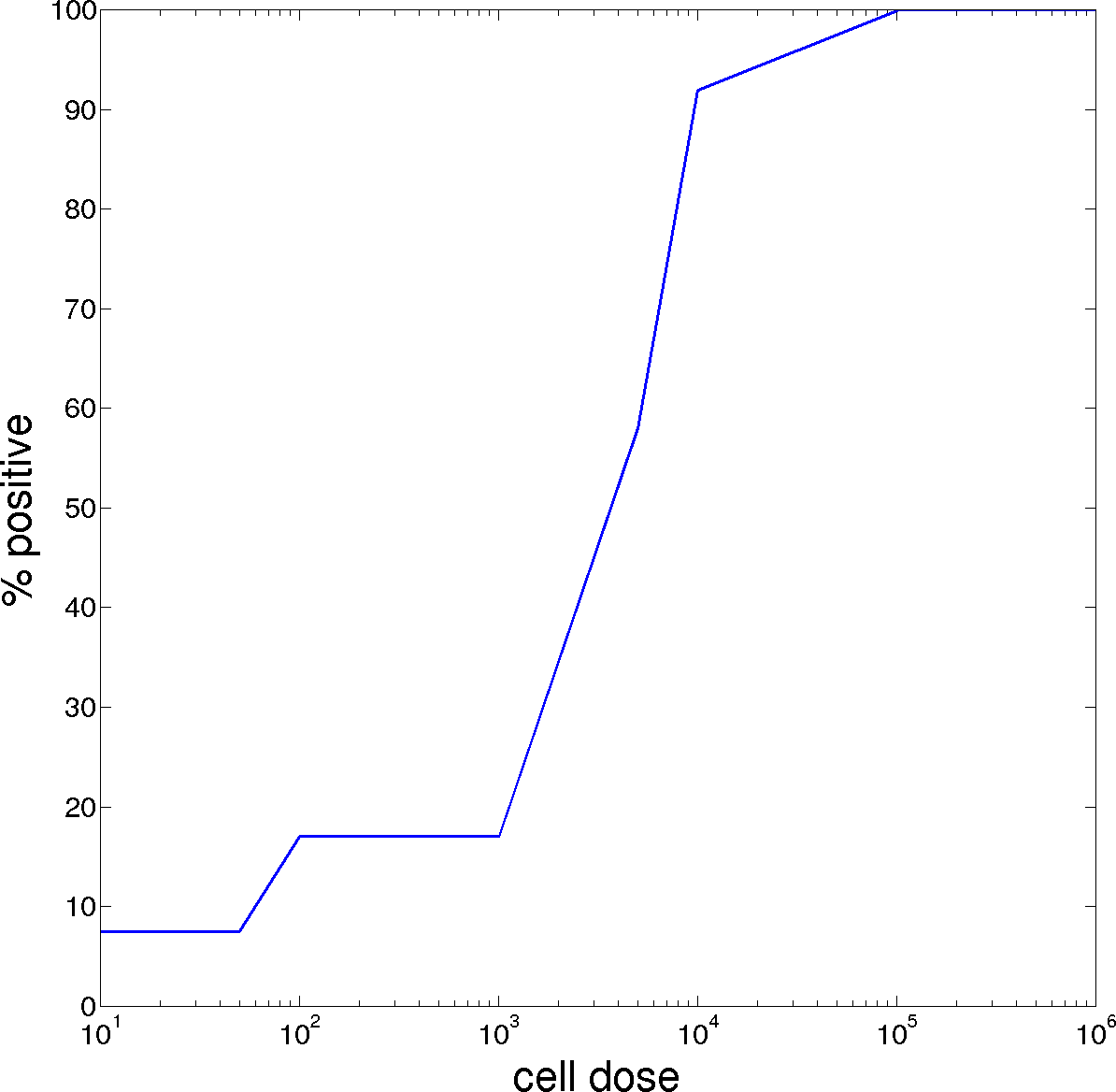}}%
\caption{Left:~conceptual diagrams showing T-cell activation by
  antigens in tumor cells combined with recruitment of regulatory T-cells which in turn repress activation.
  Self-activation of T cell populations, and a
  slower negative feedback from immune cells into tumor cells are also
  included in the mathematical model in~\cite{two-zone-journal}.
  Right: plots showing non-monotonic response in the presence of an
  inhibitory node, and monotone response when the inhibitory node is
  removed, plotted using data from~\cite{gatenby1981}.}
\label{fig:two-zone}
\efigstar

The diagrams in Figure~\ref{fig:two-zone} are deliberately simplified
schematics of a complex interaction network, intended only to clarify the
dynamic behavior. The key point is that
increasing antigen exposure can promote rapid activation and
proliferation of cytotoxic CD8$^+$ T cells while also, through slower
and indirect pathways, enhancing the expansion and suppressive
function of regulatory T cells (Tregs), which in turn inhibit CD8$^+$
effector activity. An important component of this antagonism arises
from competition for shared cytokine resources, particularly
interleukin-2 (IL-2), which is produced by activated effector T cells
but preferentially consumed by Tregs due to their high constitutive
expression of the IL-2 receptor $\alpha$ chain (CD25).  This
coarse-grained diagram therefore abstracts multiple interacting
mechanisms, including antigen presentation by dendritic cells,
cytokine-mediated feedback, metabolic and receptor-level competition,
and threshold modulation, into an effective feedforward inhibitory
structure. Such motifs provide a conceptual framework for
understanding non-monotonic and transient immune responses,
emphasizing that immune outcomes depend not only on antigen magnitude
but also on timing, kinetics, and the balance of shared regulatory
resources.

When combined with a positive feedback on T cell activation and
inhibitory regulation, the model reproduces a nontrivial dependence of
outcome on tumor growth rate, including regimes of elimination and
tolerance reminiscent of the experimentally reported ``two-zone tumor
tolerance'' and ``sneaking through'' phenomena.
The seminal study by Gatenby, Basten, and Creswick~\cite{gatenby1981}
reported a striking non-monotonic relationship between tumor
inoculum size and 
tumor take, a phenomenon they termed sneaking through. Using a
syngeneic Meth-A fibrosarcoma model in BALB/c mice, the authors showed that very
small numbers of tumor cells were able to establish progressive tumors almost as
effectively as very large inocula, whereas intermediate inocula were more
frequently rejected. This effect was shown to be strongly T-cell dependent: it
disappeared in nude mice and in animals depleted of T cells, reappeared upon
T-cell reconstitution, and was abolished by cyclophosphamide treatment, which at
the time was known to preferentially eliminate suppressor T-cell
populations.
See Figure~\ref{fig:two-zone} and discussion in~\cite{two-zone-journal}.
Furthermore, repeated exposure to low doses of irradiated tumor cells induced a
state of tumor-specific susceptibility, closely resembling classical
low-zone tolerance. Collectively, these observations support the
interpretation that early tumor growth can actively induce an immunosuppressive
state, allowing tumors to evade immune surveillance during their initial
expansion.

A historical remark about the meaning
of the regulatory node is warranted because these older articles can be difficult to interpret in light of the contemporary literature.
In the immunological framework of the 1970s and 1980s, the immunosuppressive
population implicated in sneaking through was referred to as \emph{T suppressor
(T\textsubscript{s}) cells}. Although the T\textsubscript{s} concept later fell
out of favor due to difficulties in phenotypic identification and experimental
reproducibility, it is now widely accepted that many of the functions attributed
to these cells correspond to those of modern \emph{regulatory T cells}
(T\textsubscript{reg}), in particular CD4\textsuperscript{+}CD25\textsuperscript{+}
FOXP3\textsuperscript{+} T cells. Regulatory T cells mediate peripheral tolerance
by actively suppressing effector T-cell responses through mechanisms including
the secretion of inhibitory cytokines such as IL-10 and TGF-$\beta$, metabolic
competition, and inhibitory receptor signaling. These mechanisms are fully
consistent with the tumor-specific, cyclophosphamide-sensitive suppression
observed in the sneaking-through experiments. From a modern perspective, the
results of Gatenby \emph{et al.\ } can thus be reinterpreted as an early
demonstration of tumor-induced regulatory immunity, in which small or slowly
growing tumor burdens preferentially recruit or induce regulatory T cells,
leading to immune tolerance rather than rejection. This reinterpretation places
the sneaking-through phenomenon within contemporary theories of cancer
immunoediting and highlights its relevance to current views of dynamic immune
regulation in tumor--host interactions.

Are these dynamic effects relevant to therapy?
K\"undig and collaborators further sharpened the case for
a dynamic view of immune recognition by demonstrating that \emph{the
temporal profile of antigen exposure alone} can decisively shape
immune reactivity \cite{johansen2008}. Using both dendritic cell
vaccination protocols and controlled \emph{in vitro} stimulation of T
cells, they showed that antigen delivery schedules with identical
cumulative dose but different kinetics elicited markedly different
responses. In particular, antigenic stimulation that increased
exponentially over several days induced substantially stronger CD8$^+$
T-cell activation and antiviral immunity than either a single bolus
dose or repeated equal daily doses. At the cellular level, they found
that IL-2 production was minimal under constant stimulation, increased
under linearly rising stimulation, and was maximal under exponential
stimulation, indicating that T cells are capable of decoding
higher-order temporal features of antigen exposure. These results led
the authors to conclude that antigen kinetics constitute an
independent informational dimension in immune signaling, distinct from
antigen identity or total dose. On the basis of this principle, K\"undig
and coauthors later obtained a patent proposing vaccination strategies
based on exponentially increasing antigen presentation to enhance
CD8$^+$ T-cell responses, explicitly emphasizing that immunogenicity
can be amplified in a manner largely independent of absolute antigen
dose \cite{kundig2008patent}. Taken together, these findings provide
experimental support for models in which immune circuits—often
containing incoherent feedforward or autocatalytic structures—preferentially amplify rapidly accelerating stimuli,
consistent with the broader theme that transient 
(growth-rate–dependent) signals play a central role in immune decision
making.

\begin{sidebar}{Formalism from control theory: I/O systems}
  \label{sidebar:formalism}
  
Although our discussion will remain informal, referring to the
literature for rigorous details and most proofs, we will, for
concreteness, frame all results in the setting of finite-dimensional
deterministic continuous-time systems with inputs and outputs in the
standard sense of control theory,
$\LDdiffdotD = \f(\x,u)$, $y = h(\x,u)$ or in coordinates:
\beqn
  \LDdiffdotE\;=\;\frac{d{ x_1}(t)}{dt} &=&
             f_1({ x_1}(t),\ldots ,{ x_n}(t),
             \truered{u_1(t),\ldots ,u_m(t)})\\
&&\vdots
 \\
 \LDdiffdotF\;=\;\frac{d{ x_n}(t)}{dt} &=&
             f_n(
        \underbrace{{ x_1}(t),\ldots ,{ x_n}(t)}_{\mbox{\bf states}} ,
       \truered{\underbrace{{u_1(t)},\ldots ,{u_m(t)}}_{\mbox{\bf inputs}}})
 \\
\blue{y_j(t)}&=&
         h_j(x(t),u(t))  \quad\quad \leftarrow  \mbox{\blue{\bf output variables}.}
\eeqn
The functions $\f=(\ff_1,\ldots ,\ff_n)^T$ and $h$ respectively describe
the dynamics and the readout map.
Here, the forcing function $u=u(t)$ is a generally time-dependent
external input (in various biological contexts, one might refer to  $u$
as a ``stimulus'' or an ``excitation''),
$\x(t)=(\xx_1(t),\ldots ,\xx_n(t))$ is an 
$n$-dimensional vector of state variables, and $y(t)$ is the output
(``response,'' ``measurement,'' or ``reporter'') variable
such as, for example, $y(t) = x_n(t)$, which would simply be a readout
of the value of $x_n$.
In molecular biology, the components $x_i$ of $x$ might represent 
concentrations of chemical species (proteins, mRNA, metabolites); in
epidemiology, they may represent different populations
(young/mature/old, immune/susceptible/infected, etc.).

In order to impose constraints such as positivity of variables, we
introduce the following additional notation.
States, inputs, and outputs are constrained to lie in
particular subsets $\X$, $\U$, and $\Y$ respectively, of
Euclidean spaces $\R^n,\R^m,\R^q$. 
In order to avoid keeping track of domains of existence of maximal
solutions, we will assume that for each piecewise-continuous input
$u:[0,\infty )\sarr \U$, and each initial state $\xi \in \X$, there is a
unique solution $\x:[0,\infty )\sarr \X$ with initial
condition $\x(0)\seq\xi $, which we write as 
$
\varphi(t,\xi ,u)
$,
and we denote the corresponding output $y:[0,\infty )\sarr \Y$,
given by $h(\varphi(t,\xi ,u),u(t))$, as
$
\psi (t,\xi ,u)
$.
See any textbook, e.g.~\cite{mct} for precise definitions and elementary properties.

\end{sidebar}
\section{Monotonicity}

Systems whose dynamics preserve a partial order are called
\emph{monotone systems}. Such systems inherit monotone response
properties when initialized at steady state: for example, a
nondecreasing input can never produce a biphasic (U-shaped or inverted
U-shaped) output
response. Monotone systems have been studied since at least the 1980s
by Hirsch, Matano, Smith, Smale, and many others; see the
expositions~\cite{smith,Hirsch-Smith} and the general
discussion in~\cite{almostmonotone_journal}.
They are dynamically
``well-behaved'' in several important senses. For instance, if the
system has a unique equilibrium, then every bounded trajectory
converges to it~\cite{dancer98}. When multiple equilibria are present,
and under a mild technical assumption of \emph{strong monotonicity},
solutions generically converge to equilibria, as stated in Hirsch's
Generic Convergence Theorem~\cite{Hirsch2,Hirsch}. In particular, such
systems cannot exhibit chaotic dynamics nor even stable limit cycles, a
fact also proved for the special case of cooperative systems (defined
below) in~\cite{hadeler83}. Many of these results extend to
delay-differential systems and to systems governed by partial
differential equations.
We employ ODE models in our discussion, see the Sidebar ``Formalism from control theory: I/O systems.''

Traditionally, monotone systems were defined only for autonomous
systems (i.e., without inputs or outputs). The paper~\cite{monotoneTAC}
introduced a natural extension of the concept to systems with inputs
and outputs.
This extension enables an interconnection-based framework
for verifying monotonicity in large networks and facilitates the
analysis of non-monotone feedback loops, as illustrated in an example
below.
Suppose given partial orders on the sets $\X$, $\U$, and $\Y$ of
states, inputs, and outputs. We denote all these orders by the same
symbol $\preceq$, but the meaning will be clear from the context.
A monotone system is one for which, for all inputs and initial states,
\[
u(t) \preceq v(t) \; \forall\;t,
\;x \preceq z
\;\;\Rightarrow \;\;\varphi(t,x,u) \preceq \varphi(t,z,v)\;\forall\; t
\]
and also the output map $h$ is monotone:
\[
x_1 \preceq x_2\;\Rightarrow \; h(x_1) \preceq h(x_2)\,.
\]
Typically, the partial order is specified by a convex pointed cone $K$ through
the following rule:
\[
b \, \succeq a \; \Leftrightarrow b-a\in K\,.
\]
A special type of cone is provided by the possible orthants
$K_{\sigma}\subset\R^n$, with $\sigma\in\{1,-1\}^n$, where
$x\in K_\sigma$ means that, for each $i$, $\sign(x_i)=\sigma_i$ or $x_i=0$.
A system that is monotone with respect to cones of the form
$K_\sigma$ is said to be \textit{orthant-monotone}.
For example, when $K = K_{(1,1,\cdots1)}$, the main orthant in $\R^n$, we obtain the
\textit{NorthEast (NE) order} for which ``$x\preceq z$'' means that 
$x_i\leq z_i$ for each $i=1,\ldots,n$.
A system that is monotone with respect to the NE order in the spaces
of input, states, and outputs is said to be a \textit{cooperative}
system.
Under a coordinate change $x_i\mapsto \sigma_i x_i$ (and similar
transformations on inputs and output spaces), any orthant
monotone system becomes a cooperative system.
Cooperative systems are characterized by the property that, for every $i$,
$f_i$ is nondecreasing with respect to each $x_j$, $j\not=i$ and with
respect to all input coordinates, and (if there are outputs) each
coordinate $h_j(x)$ is nondecreasing with respect to each coordinate
of $x$.
(Figure~\ref{fig:main_order}).
\bfigstar
\minp{0.5}{\picc{0.4}{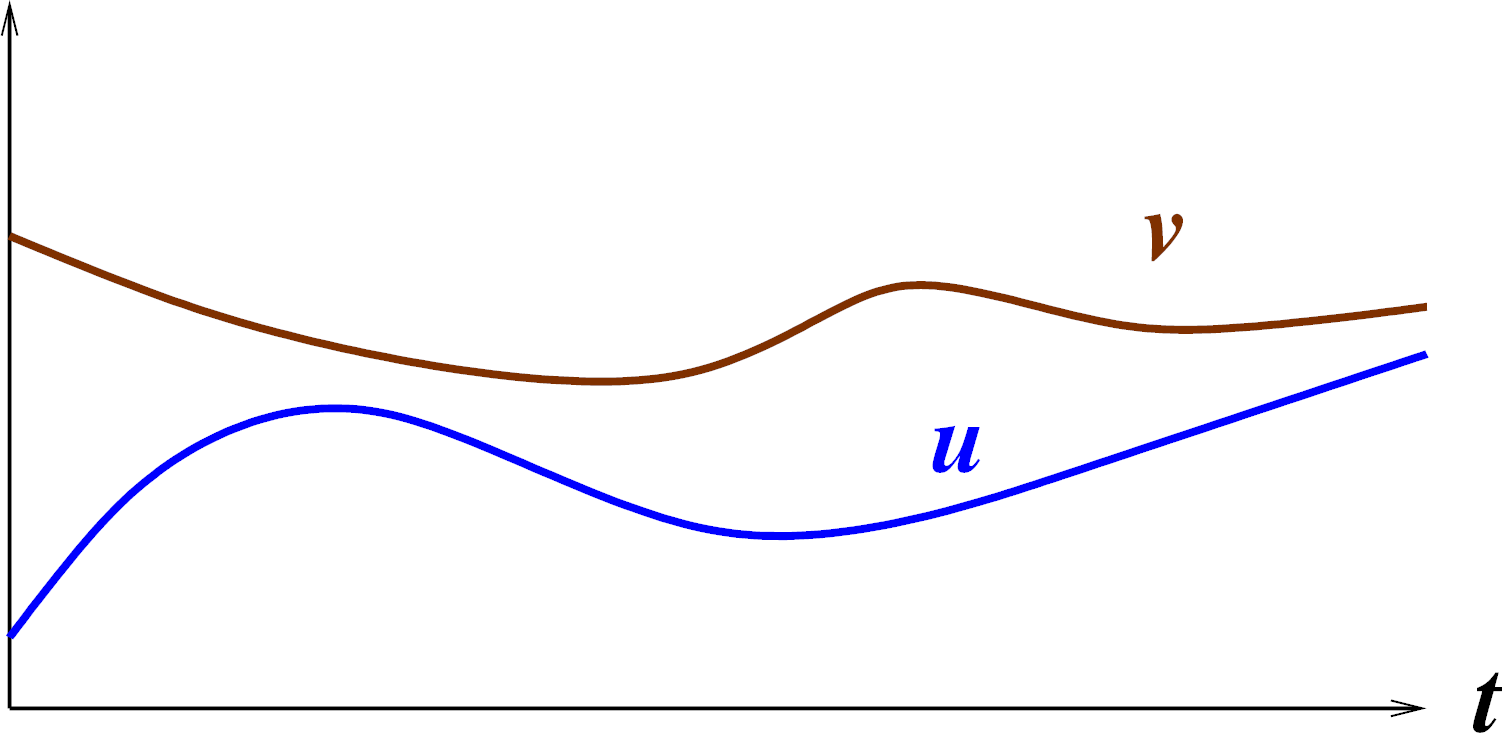}}%
\minp{0.5}{
\begin{center}
\setlength{\unitlength}{1500sp}%
\begin{picture}(4854,4854)(1759,-7603)
\thicklines
\put(3376,-7100){$x(0)\preceq z(0)$}
{\color{red}%
\put(2978,-7154){\circle*{200}}
\put(3601,-6561){\circle*{200}}
\put(2978,-7154){\vector( 1, 3){1200}}
\put(3601,-6561){\vector( 2, 3){2400}}
\color{black}}%
\put(3600,-3336){$x(t) \preceq z(t)$}%
\end{picture}
\end{center}
}%
\caption{Cooperativity. Suppose that $x(0)\preceq z(0)$, meaning (NE
  order) that $x_i(0)\leq z_i(0)$ for each coordinate, so that $z(0)$
  is to the ``NorthEast'' of $x(0)$, and that two inputs $u,v$ are
  given such that $u_j(t)\leq v_j(t)$ for all $t\geq0$ (illustrated
  here with scalar inputs). Then $x(t)\preceq z(t)$ for all $t\geq0$.}
\label{fig:main_order}
\efigstar

Intuitively, monotone systems ``behave like one-dimensional systems,'' at least locally in state space. This intuition can be made precise using Birkhoff's theory of contractions in Hilbert's projective metric on positive cones; see~\cite{sontag2026kcooperativeintuition}.
Let us illustrate the simplified dynamics of monotonicity through a simple example
in the very special case of two-dimensional cooperative systems.
We claim that in such a system there cannot exist any periodic orbits.
Figure~\ref{fig:2d_cooperative} sketches the proof.
(In dimensions greater than three, periodic orbits can exist in
monotone systems, but they must be unstable.)

\bfig
\picc{0.3}{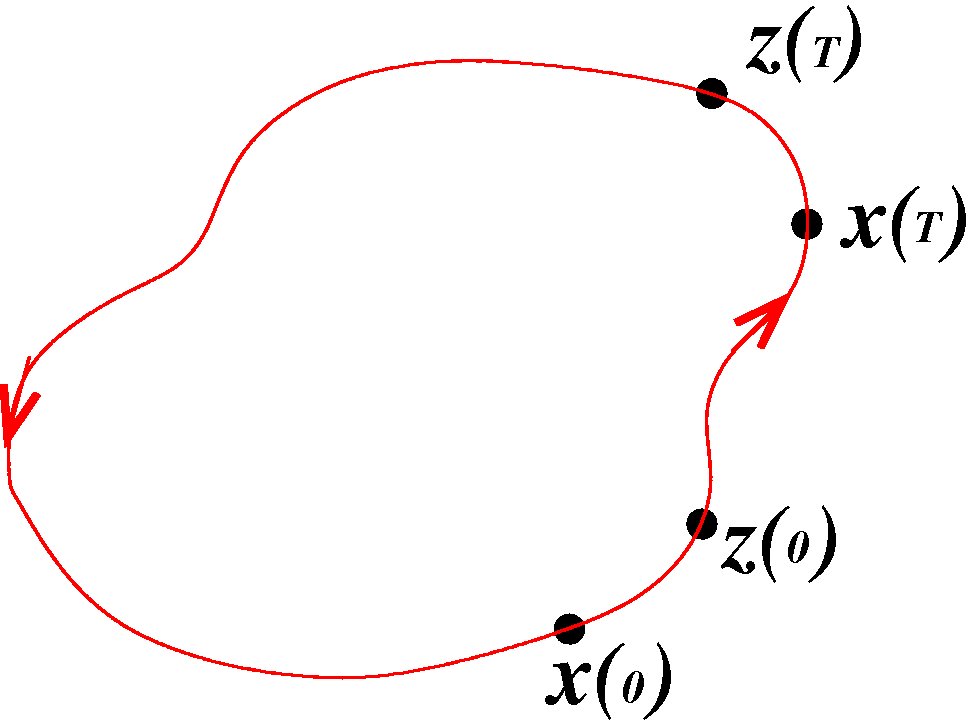}
\caption{In a cooperative planar system, there cannot exist any
  periodic solution. A sketch of proof is as follows. Suppose that
  there is a clockwise-oriented solution. (The counter-clockwise case is
  analogous.) Pick two initial states as shown: $x(0)\preceq z(0)$. As time
  evolves, the solution from $x(0)$ eventually will be at a point
  $x(T)$ as shown, where $T$ is picked so that the $x_1(T)$-coordinate is
  maximized and among such points the $x_2(T)$ coordinate is
  maximized. Cooperativity would imply that $x(T)\preceq z(T)$, which
  cannot happen since solutions cannot cross.}
\label{fig:2d_cooperative}
\efig

\subsection{Signed graphs and orthant-monotone systems}

We assume here that $\X$ and $\U$ are open subsets of $\R^n$ and
$\R^m$
and that the partial derivatives 
\[
\frac{\partial \ff_j}{\partial \xx_i}(\xx,u)
\quad\mbox{and}\quad
\frac{\partial \ff_j}{\partial u}(\xx,u)
\]
have a constant sign (either $\geq 0$ or $\leq 0$) for all $(\xx,u)\in \X\times \U$.
(For cooperative systems, these are all nonnegative.)
For those derivatives that are not identically zero, we
write $\varphi_{ij}$ and $\gamma _{i}$ for their signs ($\pm1$):
\[
\varphi_{ij} := \mbox{sign} \frac{\partial \ff_j}{\partial \xx_i}(\xx,u)
\quad\mbox{and}\quad
\gamma _{i}:= \mbox{sign} \frac{\partial \ff_i}{\partial u}(\xx,u)
\]
and let $\varphi_{ij}=0$ or $\gamma _{i}=0$ if the corresponding derivative is
identically zero. 
The effect of any given variable $x_j$ on the rate of change of
a different variable $x_i$ is ``activating'' or ``inhibiting''
depending on Jacobian signs:
\bi
\item
$x_j$ \emph{activates} $x_i$ \ (``$x_j\rightarrow x_i$'') \ if
$\displaystyle\frac{\partial \LDdiffdotG}{\partial x_j} = \displaystyle\frac{\partial f_i}{\partial x_j}(x,u)>0$
\item
$x_j$ \ \emph{inhibits} $x_i\;$ \ \ (``$x_j\,\dashv \,x_i$'') 
\ if
$\displaystyle\frac{\partial \LDdiffdotG}{\partial x_j} = \displaystyle\frac{\partial f_i}{\partial x_j}(x,u)<0$
\ei
and similarly for the effects of inputs $u_j$'s on the $x_i$'s.
(If outputs are of interest, similar terminology and notation are used.)
We summarize these effects using a directed signed (edges are labeled by
activation or repression arrows (or by ``$+$'' or ``$-$'' signs) graph;
see, for example, Figure~\ref{fig:signedgraph}.
\bfig
\picc{0.3}{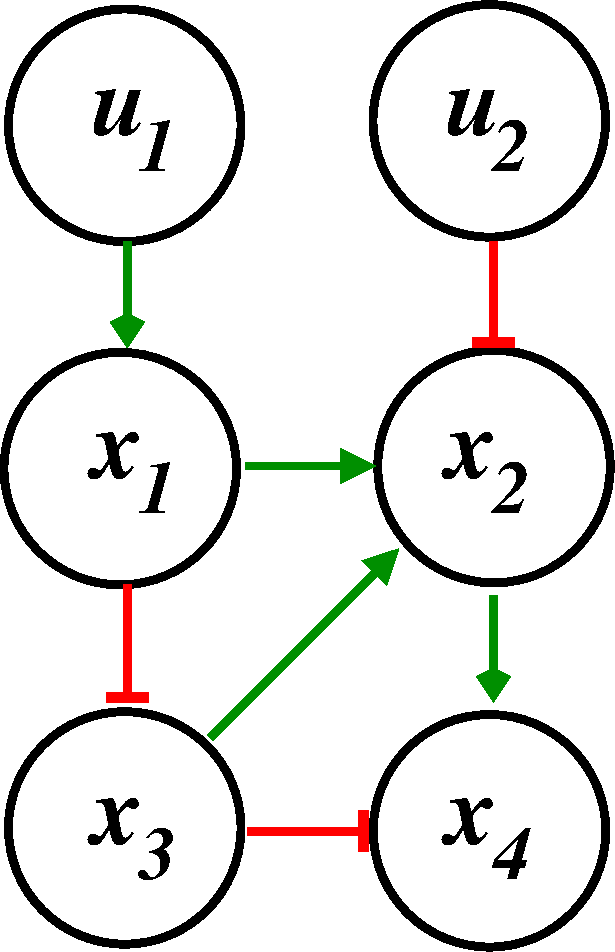}
\caption{An example of a signed graph associated with a system. Here
$\frac{\partial \ff_1}{\partial u_2}\equiv0$,
$\frac{\partial \ff_4}{\partial \xx_2}>0$,
$\frac{\partial \ff_3}{\partial \xx_1}<0$, and so forth.}
\label{fig:signedgraph}
\efig

A \emph{(graph) path} $\pi $ from the input $u$ to a node $\xx_j$ means, by
definition, a 
sequence of $k$ indices 
$
\il _1,\il _2,\ldots ,\il _k=j
$
such that $\gamma _{\il _1}\not= 0$
and 
$
\varphi_{\il _i,\il _{i+1}}\not= 0
$
 for all $i=1,\ldots ,k-1$.  We denote by
$s(\pi )$ the sign of the path, defined as the product
$
s(\pi ) := \gamma _{\il _1}\,\varphi_{\il _1\il _2}\,\varphi_{\il _2\il _3}\,\ldots \,\varphi_{\il _{k-1}\il _k} \,.
$
Similarly, a path from a node $\xx_i$ to a node $\xx_j$ means, by definition, a
sequence of $k$ indices 
$
\il _1,\il _2,\ldots, \il _k=j
$
 such that 
$
\varphi_{i,\il _1}\not= 0
$
and 
$
\varphi_{\il _i,\il _{i+1}}\not= 0
$
 for all $i=1,\ldots ,k-1$.  We denote by
$s(\pi )$ the sign of the path, defined as the product
\[
s(\pi ) := \varphi_{i\il _1}\,\varphi_{\il _1\il _2}\,\varphi_{\il _2\il _3}\,\ldots \,\varphi_{\il _{k-1}\il _k} \,.
\]
If there is a path from the input $u$ to a node
$\xx_j$, we say that $\xx_j$ is \emph{(graph) reachable}.
If there is a path from a node $\xx_i$ to the output node $\xx_n$, we say that
the node $\xx_i$ is \emph{(graph) observable}.

\subsection{Connections to spin models (``coherence'' or ``non-frustration'')}

Given a directed graph with edges labeled positive or negative,
a \textit{spin assignment} or \textit{switching function} is an assignment
of a label to each node $v_i$, written as a sign [or, in physics, an \(\uparrow\)
or \(\downarrow\)]:
\[
\sigma _i\in \{\pm 1\} \quad \mbox{[or } \sigma _i\in \{\uparrow,\downarrow\} \mbox{]}
\]
and such an assignment is said to be \emph{consistent} if every edge
with sign $J_{ij}\in \{\pm\}$ is consistent, meaning that
$\sigma _j = J_{ij}\sigma _i$, see Figure~\ref{fig:consistent}.
The graph is said to be \textit{balanced} if there exists at least one consistent
assignment among all possible $2^n$ spin assignments.
\bfigstar
\begin{center}
\setlength{\unitlength}{1600sp}%
\begin{picture}(3377,746)(1758,-2761)
\put(2101,-2386){\circle{670}}
\put(4792,-2358){\circle{670}}
\put(2101,-2611){\dgreen{\vector( 0, 1){450}}}
\put(2701,-2386){\vector( 1, 0){1500}}
\put(4801,-2611){\dgreen{\vector( 0, 1){450}}}
\put(2626,-2800){\dgreen{\small consistent}}%
\end{picture}%
\hspace{20pt}\setlength{\unitlength}{1600sp}%
\begin{picture}(3377,746)(1758,-2761)
\put(4792,-2358){\circle{670}}
\put(2101,-2386){\circle{670}}
\put(2701,-2386){\vector( 1, 0){1500}}
\put(4801,-2611){\dgreen{\vector( 0, 1){450}}}
\put(2101,-2161){\red{\vector( 0,-1){450}}}
\put(2500,-2800){\red{\small inconsistent}}%
\end{picture}%
\hspace{20pt}\setlength{\unitlength}{1600sp}%
\begin{picture}(3377,746)(1758,-2761)
\put(2101,-2386){\circle{670}}
\put(4792,-2358){\circle{670}}
\put(2701,-2386){\line( 1, 0){1500}}
\put(4201,-2236){\line( 0,-1){300}}
\put(2101,-2611){\dgreen{\vector( 0, 1){450}}}
\put(4801,-2161){\red{\vector( 0,-1){450}}}
\put(2626,-2800){\dgreen{\small consistent}}%
\end{picture}%
\end{center}
\caption{Three signed edges and consistent or inconsistent assignments}
\label{fig:consistent}
\efigstar

It is an easy exercise to show that \textit{a graph is balanced if and only
if every undirected loop (ignore directions but keep signs)
has a positive parity}, that is, contains an even number of negative
edges. Equivalently, any two (undirected) paths between two nodes have
same parity. 
(We did not define ``undirected'' paths. These are paths in the graph
$G\cup G^\top$ where we include an edge from node $j$ to node $i$ if
there is an edge from $i$ to $j$, and assign the same sign to it. If
there was already an edge from $j$ to node $i$, and it had the
opposite sign, the graph is not balanced.)

Yet another equivalent property is also easy to prove, and has often been
used in the theory of social interactions, dating at least to
a 1954 paper by Harary (see the discussion
in~\cite{Roberts1977}).
A graph is \textit{balanced if and only if the vertices can be partitioned
into two subsets so that every edge joining vertices within a class is
positive and every edge joining vertices in different classes is
negative}, see Figure~\ref{fig:balanced_two_sets}.
In this form, the concept of balancing plays an important role in the
study of collective phenomena such as the emergence of consensus or
polarization in human societies or social decisions in animal groups,
as discussed in~\cite{2023_naomi_signed_networks}.
\bfig
\picc{0.3}{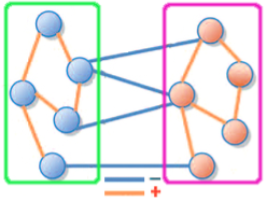}
\caption{A balanced graph (directions of arrows ignored) partitioned
  into two sets of consistent nodes. Orange edges are positive and
  blue edges are negative. Think of nodes as representing
  individuals: individuals like others in their own
  group, but dislike individuals in the other group.}
\label{fig:balanced_two_sets}
\efig

One can prove the following key equivalence:
\textit{A system is orthant-monotone if and only if its interaction graph is balanced.}

\subsection{Predictability of behavior in balanced networks}

With regard to transient dynamics, balanced graphs exhibit highly
robust and unambiguous responses to perturbations at specific nodes.
Consider the two graphs shown in Figure~\ref{fig:predictability}. In
the left-hand graph, every path connecting nodes~1 and~4 has a net
positive sign, so the graph is balanced. If node~1 is instantaneously
perturbed upward, both the path \(1 \rightarrow 3 \rightarrow 4\) and
the alternative path \(1 \dashv 2 \dashv 4\) transmit a net positive
effect to node~4: along the second path, the increase in node~1
decreases node~2, thereby reducing repression of node~4 and ultimately
causing node~4 to increase. Thus, all paths convey the same qualitative
information, and the effect on node~4 is unambiguous.

The situation differs for the graph on the right. The path
\(1 \rightarrow 3 \rightarrow 4\) remains net positive, but the
alternative path \(1 \rightarrow 2 \dashv 4\) has a net negative sign.
This graph is therefore not balanced. A perturbation at node~1 now
yields conflicting qualitative predictions: one path indicates that
node~4 should increase, whereas the other predicts a decrease. This
``incoherent feedforward'' structure makes the effect of the
perturbation indeterminate from graph topology alone; it will depend on
which path dominates, a feature determined by the specific algebraic
form and parameter values of the interactions.
\bfig
\picc{0.55}{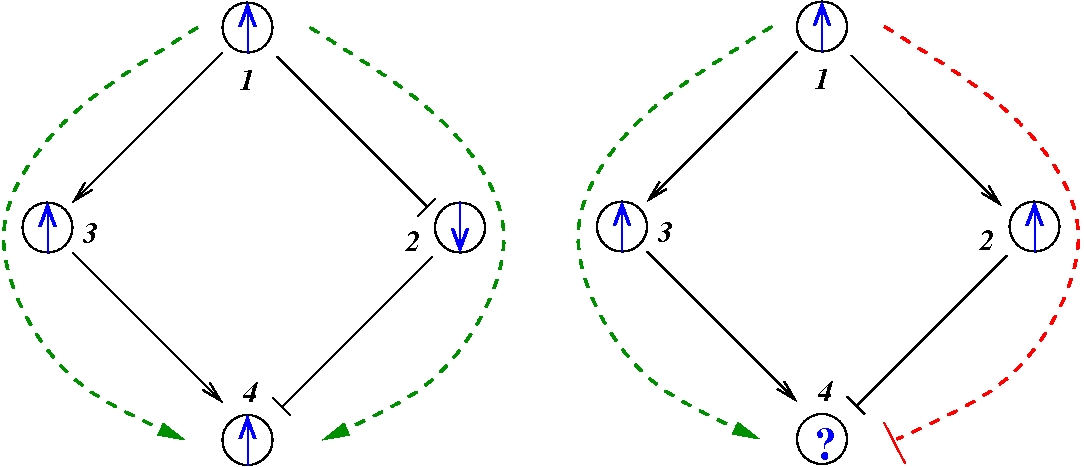}
\caption{Balanced and unbalanced graphs (dashed lines show net parity along each path). Left graph is balanced; right graph is not.}
\label{fig:predictability}
\efig

Intuitively, balancing ensures a predictable, unambiguous, global
effect of node perturbations.
As a concrete example of the second graph, suppose that the equations
for the system are as follows:
\[
\frac{dx_1}{dt}=0,\;
\frac{dx_2}{dt}=x_1,\;
\frac{dx_3}{dt}=x_1,\;
\frac{dx_4}{dt}=x_4(k_3x_3-k_2x_2),
\]
where the reaction constants $k_2$ and $k_3$ are two positive numbers.
The initial conditions are taken to be $x_1(0)=x_4(0)=1$,
and $x_2(0)=x_3(0)=0$, and we ask how the solution $x_4(t)$ will
change when the initial value $x_1(0)$ is perturbed.
With $x_1(0)=1$, the solution is $x_4(t)=\exp{\alpha t^2/2}$, where $\alpha =k_3-k_2$.
On the other hand, if $x_1(0)$ is perturbed to a larger value, let us say
$x_1(0)=2$,
then $x_4(t)=\exp{\alpha t^2}$.  This new value of $x_4(t)$ is larger than
the original unperturbed value $\exp{\alpha t^2/2}$ provided that $\alpha >0$,
but it is smaller than it if, instead, $\alpha <0$.  In other words, the sign of
the sensitivity of $x_4$ to a perturbation in $x_1$ cannot be predicted from
knowledge of the graph alone, but it depends on whether $k_2<k_3$ or $k_2>k_3$.
Compare this with the balanced case in the left graph of Figure~\ref{fig:predictability}.
A concrete example is obtained if we modify the $x_2$ equation to
$\fractextnp{dx_2}{dt}=\fractextnp{1}{(1+x_1)}$.
Now the solutions are $x_4(t)=\exp{\beta _1t^2}$
and $x_4(t)=\exp{\beta _2t^2}$ respectively, with $\beta _1=k_3/2-k_2/4$
and $\beta _2=k_3-k_2/6$, so we are guaranteed
that $x_4$ is larger in the perturbed case, a conclusion that holds true no
matter what the numerical values of the (positive) constants $k_i$ are.

\subsection{Biological networks, balancing, and monotonicity}

Systems molecular biology seeks to understand the behavior of
biochemical networks composed of proteins, RNA, DNA, metabolites, and
other molecular species. These networks mediate control and signaling
in development, regulation, and metabolism by processing environmental
cues, coordinating internal events such as gene expression, and
generating appropriate cellular responses. Unlike many areas of
applied mathematics and engineering, the study of dynamics in biology,
and especially in cell biology, must confront the substantial
uncertainty inherent in models of intracellular biochemical networks.
This uncertainty arises from environmental fluctuations as well as
variability among cells of the same type. For instance, concentrations
of enzymes, transcription factors, metabolites, mRNAs, and many other
molecular components can vary widely both across cells within a single
organism and across individuals. Physical factors such as pH,
temperature, and other environmental conditions further influence
biological processes, while ecological interactions with other
individuals introduce additional sources of variability.

Yet, despite these uncertainties, biological behaviors are often
remarkably predictable and robust. From a mathematical standpoint,
uncertainty manifests as difficulty in measuring key model parameters,
such as kinetic constants or cooperativity indices, and therefore
makes it impossible to obtain fully specified models. This motivates
the development of analytical tools that are robust in the sense that
they yield meaningful conclusions based solely on the qualitative
features of a network, ideally without relying on precise parameter
values or even exact reaction forms. Achieving such robustness is
challenging, since dynamical behavior may undergo bifurcations or
other phase transitions that depend sensitively on parameter values.
Balancing, and more generally monotonicity, provides a degree of
predictability and robustness to network responses, as discussed
above.

As surveyed in \cite{almostmonotone_journal}, it has long been
understood that system behavior depends critically on the network's
topology and on the signs (activating or inhibiting) of its
feedforward and feedback interconnections.
See, for example,
\cite{novick57,
monod61,
lewis77,
segel84,
deangelis86,
thomas90,
goldbeter96,
keener98,
murray2002,
alon02,
keshet05}%
.
For example, Figures~\ref{3figs}(a-c) illustrate the three possible
types of feedback loops that involve two interacting chemicals.
\bfig
\minp{0.15}{\picc{0.3}{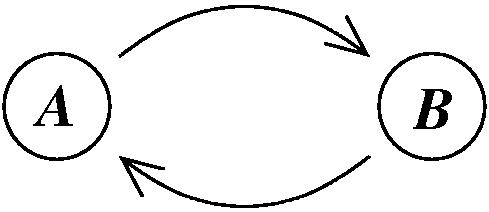}}%
\minp{0.15}{\picc{0.3}{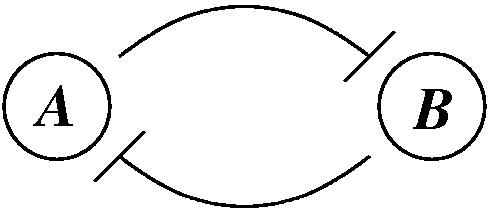}}%
\minp{0.15}{\picc{0.3}{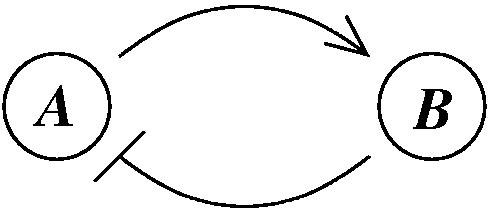}}
 \caption{Left to right: mutual activation,
 mutual inhibition,
activation-inhibition.}
\label{3figs}
\efig
A mutual activation configuration is shown in Figure~\ref{3figs}(a):
a positive change in $A$ results in a positive change in $B$, and
vice versa. Such configurations are associated with signal
amplification and the production of switch-like biochemical
responses. A mutual inhibition configuration is shown in
Figure~\ref{3figs}(b): a positive change in $A$ represses $B$, and
repression of $B$ in turn enhances $A$. These configurations allow
systems to exhibit multiple discrete alternative stable steady states
and thereby provide a mechanism for biochemical memory. Both
Figures~\ref{3figs}(a) and~\ref{3figs}(b) represent positive-feedback
systems %
\cite{ptashne92,
plathe95,
cinquin02,
gouze98,
thomas01,
remi03,
monotoneMulti,
pnasangeliferrellsontag04}.

In contrast, activation-inhibition configurations such as that in
Figure~\ref{3figs}(c) are essential for generating periodic behaviors,
including circadian rhythms and cell-cycle oscillations. They may
occur alone or in combination with multistable positive-feedback
subsystems. These configurations also underlie adaptation, disturbance
rejection, and the tight regulation (homeostasis) of physiological
variables %
\cite{rapp75,
hastings-tyson-webester-periodic-goodwin-jde77,
tyson-othmer,
thomas81,
keshet05,
goldbeter96,
mct,
kholodenko00,
murray2002,
sha03,
pomerening-ferrell,
monotoneLSU}.
Unlike positive-feedback systems, negative-feedback systems are not
balanced. In case (c), a positive change in $A$ is counteracted by the
feedback loop, resisting the perturbation.

Balancing, or the lack of it, also influences the behavior of graphs
that do not contain feedback loops. For example,
\cite{alon02,alon03,alon3} analyze the distinct signal-processing
characteristics of balanced (``coherent'') compared to nonbalanced
(``incoherent'') feedforward motifs. Thus, balancing can play a
central role in understanding the qualitative dynamics of biochemical
networks.

Of course, there is no a priori reason to expect a system to have a balanced
interaction graph. Yet we speculate that (1) systems that are
``nearly balanced'' may be, statistically, more biologically
advantageous than those that are far from monotone, in the sense that
they tend to exhibit more regular dynamical behavior; and (2) real
biological networks may lie much closer to being balanced than random
networks with the same numbers of vertices and the same distribution
of positive and negative edges.  To the best of our knowledge, there
is \textit{no precise mathematical formulation}, let alone a proof,
of these conjectures. Nonetheless, several intriguing observations
point in this direction, which we briefly review
in the Sidebar ``Near-monotonicity in biological networks''
in the hope of
stimulating further research.

\begin{sidebar}{Near-monotonicity in biological networks}
  \label{sidebar:nearmonotone}

Since balanced structure in biological networks may be advantageous,
one might conjecture that natural biological networks tend to be more
balanced than expected by chance. To explore this hypothesis, we
introduced a measure of frustration or imbalance, the consistency
deficit (CD) of a graph $G$, defined as the smallest number of edges
that must be removed from $G$ to render the remaining graph
balanced. An algorithm for computing this quantity was presented
in~\cite{biosystems06}.

We applied this method to the gene regulatory network of the yeast
Saccharomyces cerevisiae, as compiled in~\cite{alon02} and obtained
from~\cite{yeastnet} (the data in~\cite{alon02} originate from the YPD
database~\cite{Costanzo2001}). In this network, nodes correspond to
genes, and edges are directed from transcription factors, or
transcription-factor complexes, to the genes they regulate. The network
contains 690 nodes and 1082 edges, of which 221 are negative and 861
are positive. (We treated the single ``neutral’'' edge as positive;
the conclusions remain essentially unchanged if it is treated as
negative or simply removed.)  Using the approximation algorithm
of~\cite{biosystems06}, we estimated the CD to be 43. The exact
algorithm of~\cite{huffnerWEA07} subsequently refined this estimate,
yielding the precise value CD$=$41. Thus, \textit{removing only about
4\% of the edges suffices to make the network balanced}.
Equally striking is the structural effect of these deletions. The
original graph has 11 connected components: one large component of
size 664, one of size 5, three of size 3, and six of size 2. After
deleting the 41 edges, all components remain connected. The deleted
edges lie entirely within the largest component and involve a total of
65 nodes from that component.

To assess whether a small CD could arise merely by chance, we also
applied the algorithm to random graphs with 690 nodes and 1082
uniformly chosen edges, of which 221 edges were designated as negative
(also uniformly). For these random graphs, approximately 12.6\% of
edges (136.6$\pm$5) had to be removed to achieve balance.
To examine how this scales with network size, we generated random
graphs with $N$ nodes and $1.57N$ edges, of which $0.32N$ were
negative. For $N > 10$, roughly $N/5$ edges needed to be removed to
obtain a balanced graph, consistent with the value observed for $N =690$.
Thus, the CD of the yeast network lies about \textit{15 standard
  deviations} below the mean CD for random graphs, an extremely unlikely event by chance alone.
Both the network’s \textit{topology} (underlying graph structure) and
its \textit{edge signs} contribute to this unusually high degree of
balance. To disentangle these effects, we conducted the following
numerical experiment. We randomly flipped the signs of 50 positive and
50 negative edges, producing a network with the same topology and the
same total number of positive and negative edges as the original, but
with 100 edges assigned new, random signs. In this modified network,
about 8.2\% of edges (88.3$\pm$7.1) must be deleted to achieve
balance, intermediate between the original yeast network and random
graphs.
Flipping more signs (100 positives and 100 negatives) yields a still less balanced network, requiring 115.4$\pm$4.0 deletions (about 10.7\% of edges), although this remains noticeably lower than the value for a fully random network.

Analogous analyses were carried out for an \textit{E. coli} gene
regulatory network, a \textit{S. cerevisiae} gene regulatory network,
and a signaling network in a CA1 hippocampal neuron. These studies
likewise found that all three intracellular regulatory networks are,
in a meaningful sense, far more balanced than would be expected by
chance~\cite{maayan07}.
The three networks examined share several structural features: similar
node-to-link ratios, comparable proportions of positive and negative
edges, and ``small-world'' characteristics, including high clustering
coefficients and characteristic path lengths close to those of random
networks. They also exhibit connectivity distributions that are well
described by power laws, indicating an enrichment of highly connected nodes
(hubs). In such networks, certain genes or proteins directly
regulate, or are regulated by, many others.
By comparing the numbers of positive and negative feedback and
feedforward loops in the real networks versus sign-shuffled versions,
the study showed that the biological networks contain significantly
more positive loops than would be expected in randomized counterparts.
\end{sidebar}

In summary, biological networks appear to be far more balanced than
one would expect by chance alone. What are the dynamical implications
of this observation? Formulating a precise theorem is challenging,
since even systems with only a few feedback loops can exhibit highly
complex behavior. Nevertheless, it would be extremely valuable to make
this intuition rigorous, for example, by showing that in a randomly
selected family of systems that are ``close'' to balanced, one
consistently observes some form of regularity in their dynamics. A
preliminary computational investigation was carried out
in~\cite{sontag_laubenbacher_jarrah07}, where numerical experiments on
discrete-time, discrete-state (Boolean) systems revealed a correlation
between a measure of imbalance and the lengths of their periodic
orbits.

In~\cite{maayan07}, we argue that cellular regulatory networks exhibit
remarkably stable behavior despite noisy and rapidly changing
environments and that a growing body of evidence suggests that this
robustness may arise from the fact that intracellular networks are
often ``close to monotone'' in their structure. In particular, real
biological networks appear enriched in positive feedback and
feedforward loops and depleted in negative ``inconsistent'' loops,
which are known to promote oscillations or other unstable
dynamics. Although negative feedback is traditionally viewed as
essential for homeostasis, empirical network topologies reveal that
such loops are relatively uncommon. Instead, stability in many
cellular systems is achieved through constitutive degradation or
deactivation processes, for example, by broad-specificity phosphatases
in signaling pathways, that dampen signals without requiring explicit
negative feedback. This structural bias toward sign-consistent
interactions provides a plausible mechanism for the stable dynamical
behavior observed in cells.
Monotone or near-monotone architectures also confer advantages beyond
stability, including ordered responses, predictable input-output
behavior, and greater modularity under evolutionary pressures. Networks
composed primarily of positive feedforward and feedback connections
tend to maintain consistent qualitative relationships between distant
components: perturbing a node or modifying reaction rates leads to
graded changes in downstream activity that may eventually decay due to
housekeeping negative regulators. In contrast, networks containing
inconsistent loops can exhibit oscillations or even chaotic responses to
small parameter changes. Observed biochemical network architectures
align with these theoretical expectations. For example, in mammalian
signaling pathways, the number of protein kinases greatly exceeds that of
corresponding phosphatases, and kinases function predominantly as
activators while phosphatases act as broadly acting, relatively
unregulated negative hubs. This asymmetry naturally favors
sign-consistent circuitry and may provide a genetic basis for the
prevalence of monotone-like motifs in cellular signaling networks.

\subsection{Decompositions or embeddings into monotone systems}

Sometimes non-monotone systems can be interpreted as negative feedback
interconnections of monotone subsystems. For example, a system with a
non-balanced interaction graph, as shown in Figure~\ref{fig:decompose_monotone},
can be decomposed into two subsystems, each with a balanced graph,
although their interconnection destroys the overall balance.
\bfigstar
\setlength{\unitlength}{1800sp}%
\begin{center}
\begin{picture}(9757,4619)(1816,-5108)
\thicklines
\put(2101,-1561){\circle{540}}
\put(2101,-3286){\circle{540}}
\put(4501,-1561){\circle{540}}
\put(4501,-3286){\circle{540}}
\put(1951,-2911){\line( 1, 0){300}}
\put(2101,-1936){\line( 0,-1){975}}
\put(2476,-3286){\line( 1, 0){1650}}
\put(4126,-3136){\line( 0,-1){300}}
\put(3001,-2236){\blue{\line(-5, 3){606.618}}}
\put(3301,-2461){\blue{\circle{540}}}
\put(2230,-2070){\blue{\line(3,4){330}}}
\put(4201,-3061){\blue{\vector(-4, 3){516}}}
\put(4501,-2911){\vector( 0, 1){975}}
\put(4126,-1561){\vector(-1, 0){1650}}
\put(2401,-4411){\line( 0,-1){600}}
\put(2401,-5011){\line( 1, 0){600}}
\put(3001,-5011){\line( 0, 1){600}}
\put(3001,-4411){\vector(-1, 0){525}}
\red{\put(3676,-4861){-}
\put(3601,-4411){\line( 0,-1){600}}
\put(3601,-5011){\line( 1, 0){600}}
\put(4201,-5011){\vector(-1, 1){525}}}
\put(2551,-4786){+}%
\green{\put(6751,-4861){\framebox(1725,375){}}}
\green{\put(9676,-4861){\framebox(1725,375){}}}
\green{\put(6376,-3961){\framebox(3450,3150){}}}
\green{\put(10626,-2911){\framebox(975,1275){}}}
\put(6556,-4786){\small monotone}%
\put(9496,-4786){\small monotone}
\put(7126,-4411){\line( 1, 0){300}}
\put(7276,-4411){\line( 0, 1){150}}
\put(7276,-4261){\line( 1, 0){3000}}
\put(10276,-4261){\line( 0,-1){225}}
\put(7276,-4861){\line( 0,-1){225}}
\put(7276,-5086){\line( 1, 0){3000}}
\put(10276,-5086){\line( 0, 1){225}}
\multiput(10126,-5011)(6.00000,6.00000){16}{.}
\multiput(10250,-4950)(6.00000,-6.00000){10}{.}
\put(6601,-1561){\circle{540}}
\put(6601,-3286){\circle{540}}
\put(9001,-1561){\circle{540}}
\put(9001,-3286){\circle{540}}
\put(6601,-1936){\line( 0,-1){975}}
\put(9001,-2911){\vector( 0, 1){975}}
\put(6976,-3286){\line( 1, 0){1650}}
\put(8626,-3136){\line( 0,-1){300}}
\put(8626,-1561){\vector(-1, 0){1650}}
\put(6451,-2911){\line( 1, 0){300}}
\put(11090,-2191){\blue{\circle{540}}}
\put(11101,-1786){\vector( 0, 1){525}}
\put(9901,-3286){\line( 1, 0){1200}}
\put(11101,-3286){\line( 0, 1){225}}
\put(11101,-1261){\line( 0, 1){750}}
\put(11101,-511){\line(-1, 0){4500}}
\put(6601,-511){\line( 0,-1){150}}
\put(11101,-3061){\vector( 0, 1){525}}
\put(9376,-3286){\vector( 1, 0){525}}
\put(6601,-661){\vector( 0,-1){525}}
\put(6376,-661){\line( 1, 0){450}}
\end{picture}%
\end{center}
\caption{The graph on the left is not balanced (compare black-square
  positive path and red-triangle negative path) but can be decomposed
  into two balanced systems, as a negative feedback interconnection of
  two balanced graphs corresponding to two monotone systems.}
\label{fig:decompose_monotone}
\efigstar

Such decompositions do not always exist. However, it is often possible
to embed a non-monotone system into a higher-dimensional monotone
system, sometimes of roughly twice the original dimension, an idea
that can be traced at least to the work
of~\cite{gouze_hadeler94}. This idea underlies the introduction of the
formalism of monotone systems with inputs and
outputs~\cite{monotoneTAC}, which made it possible to develop a
general interconnection theory. The article ~\cite{monotoneTAC}
established global stability results for monotone systems under
negative feedback. That work was undertaken to address a specific
question concerning the robustness of stability and dynamical
responses in a signaling context involving mitogen-activated protein
(MAPK) cascades.
MAPK proteins form a family of enzymes that relay signals from the
cell surface to the nucleus and regulate essential cellular processes
such as proliferation, differentiation, and survival. The MAPK
signaling pathway plays critical roles in normal physiology and is
also a major contributor to disease states, including cancer, where
its dysregulation motivates several therapeutic interventions.
The approach was further developed
in~\cite{enciso_smith_sontagJDE06,enciso_angeli_sontag2014}.
The embedding procedure also forms
the basis of a widely used method for estimating reachability regions
and providing safety
guarantees~\cite{2015_coogan_arcak_mixed_monotone} in the framework of
mixed monotone systems, as surveyed
in~\cite{2020_coogan_mixed_monotone_survey}.

Let us illustrate a typical result in this setting. Suppose we have
two systems with scalar inputs and outputs, with
$\U_1=\Y_2\subseteq\R$ and $\U_2=\Y_1\subseteq\R$. 
Each system is assumed to be monotone and to possess a well-defined
\textit{characteristic}, denoted $k_i(u)$, which
is an analog of a ``DC gain'' but understood in the nonlinear
monotone-systems sense of a characteristic map.
By this we mean that, for every constant input $u$, there exists a
unique equilibrium state $x_u\in\X$ and this equilibrium is
globally asymptotically stable for the system $\LDdiffdotI = f(x,u)$ with that
fixed input. The characteristic is defined as the function $k(u):= h(x_u)$.
Depending on the biological context, such functions 
$k$ may be referred to as ``signal versus input concentration curves,''
``dose-response curves,'' ``receptor activity plots,'' or other similar
terms.
See Figure~\ref{fig:dcgain}.
\bfigstar
\minp{0.6}{\picc{0.45}{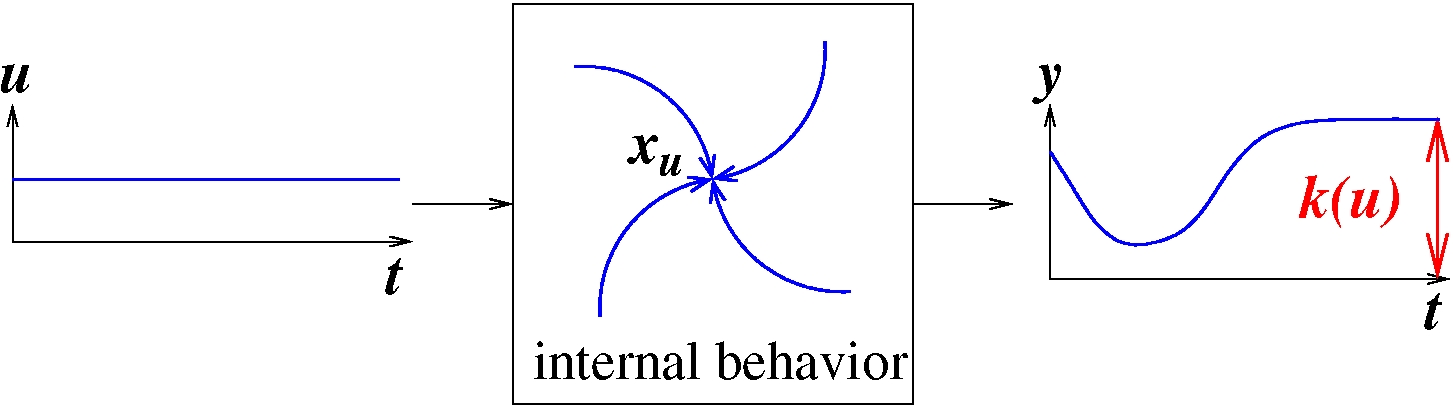}}%
\minp{0.4}{\picc{0.4}{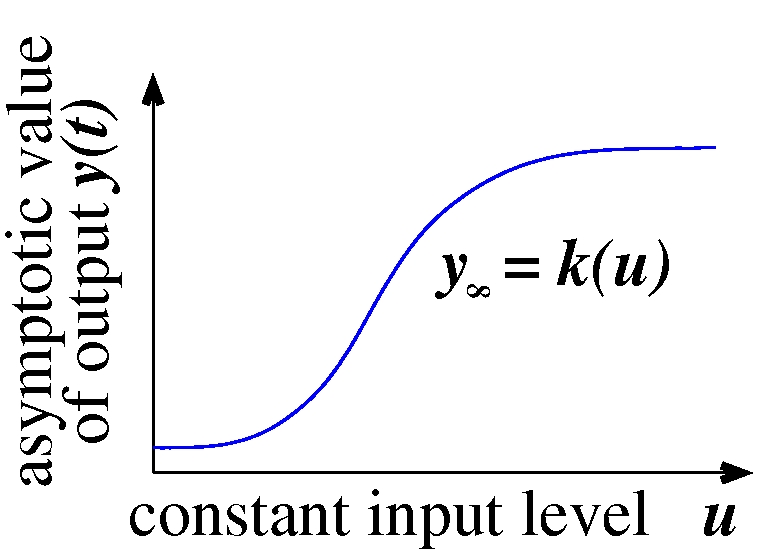}}
\caption{The I/O characteristic $k$: 
for each constant input $u$, there is a unique steady state $x_u$, which is
GAS, and $k(u):= h(x_u)$. A plot of a typical $k$ is shown, with
$\U=\Y=[0,\infty)$ endowed with the usual order from $\R$.}
\label{fig:dcgain}
\efigstar

We study a \textit{negative feedback} interconnection of two systems and
take the orders on states, inputs, and outputs of the first system
to be the usual order on $\R$, but impose the reverse order on the
outputs of the second system. For example, consider a system with
$\U=\X=\Y=(0,\infty)$, dynamics $\LDdiffdotI = 1-x+u$, and output
$y=\frac{1}{1+x}$.  This system is monotone when the reverse order is
used on $\Y$, because the output function is non-increasing.
In general, for monotone systems, the characteristic can be shown to
be monotonic with respect to the chosen orders. Thus, $k$ is a
non-decreasing function when the usual orders are used, and it is
non-increasing when a reversed order is imposed on outputs.
We now informally state one of the main theorems from~\cite{monotoneTAC}. That
paper should be consulted for full details and for the more general
formulation that allows arbitrary input and output spaces and
arbitrary orders on these spaces.  The theorem, illustrated in
Figure~\ref{fig:negative_feedback_monotones}, provides a sufficient
condition for global asymptotic stability (GAS) to a unique equilibrium of a
negative feedback configuration. We call $k$ the characteristic of the
first system and $g$ the characteristic of the second system.
We may assume in the theorem that the $g$ characteristic is just a
mapping, not necessarily arising from a nontrivial system, which allows
one to include static feedback laws.
The sufficient condition for GAS is that the discrete one-dimensional
``spider-web'' iteration:
\[
u_{i+1}=(g\circ k)(u_i)
\]
should globally converge to a unique fixed point (denoted $\bar u$ in
the figure). 
This condition can be interpreted as a contraction-type small-gain theorem for
monotone systems.
A key ingredient in the proof is a property that is also useful for
understanding transient behaviors: in monotone systems, bounded inputs
cause trajectories to eventually remain within suitable ``safe''
multidimensional order intervals in $\X\subset\R^n$, with the size of
these intervals determined by the bounds on the input.
\bfigstar
\minp{0.5}{\picc{0.12}{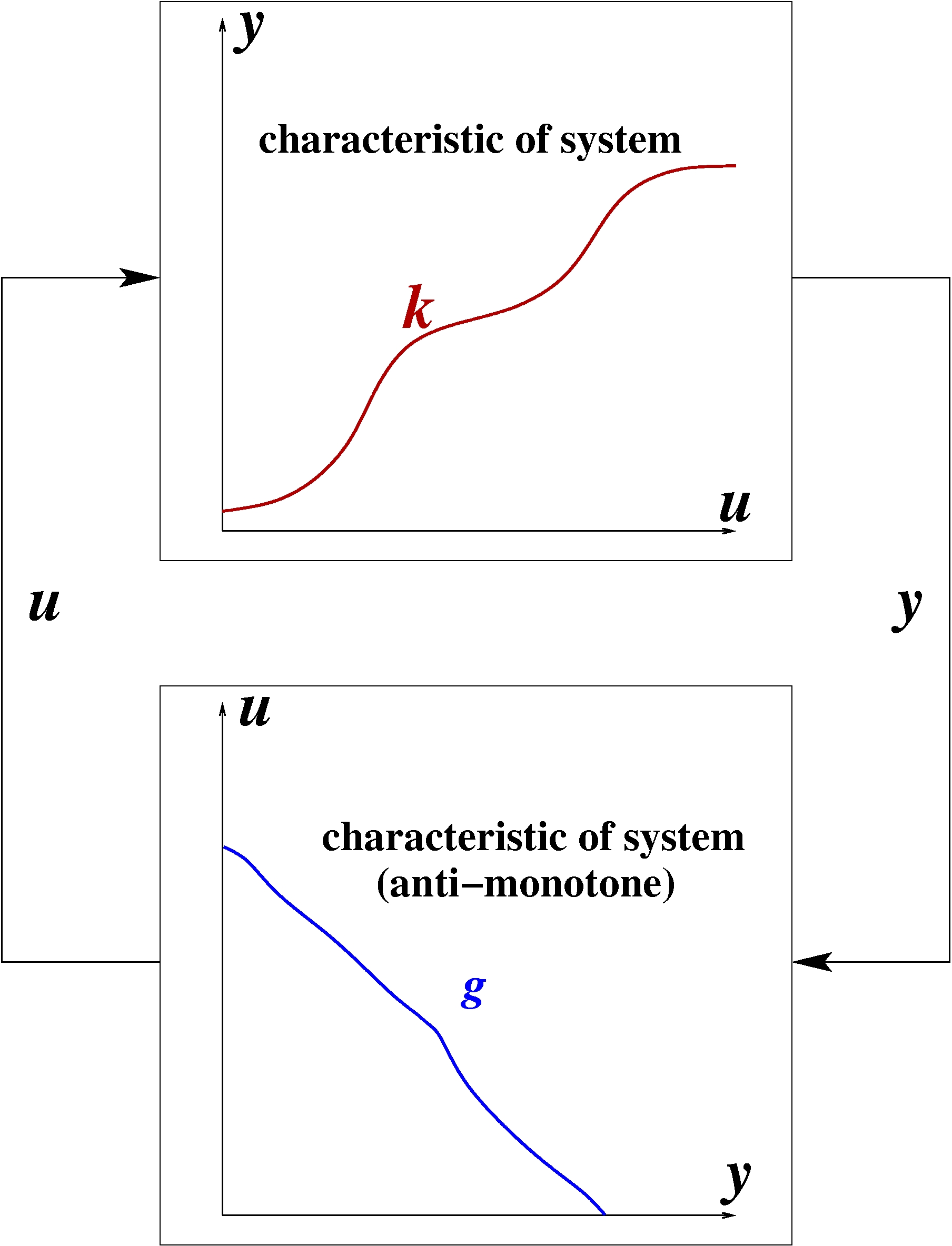}}%
\minp{0.5}{\picc{0.1}{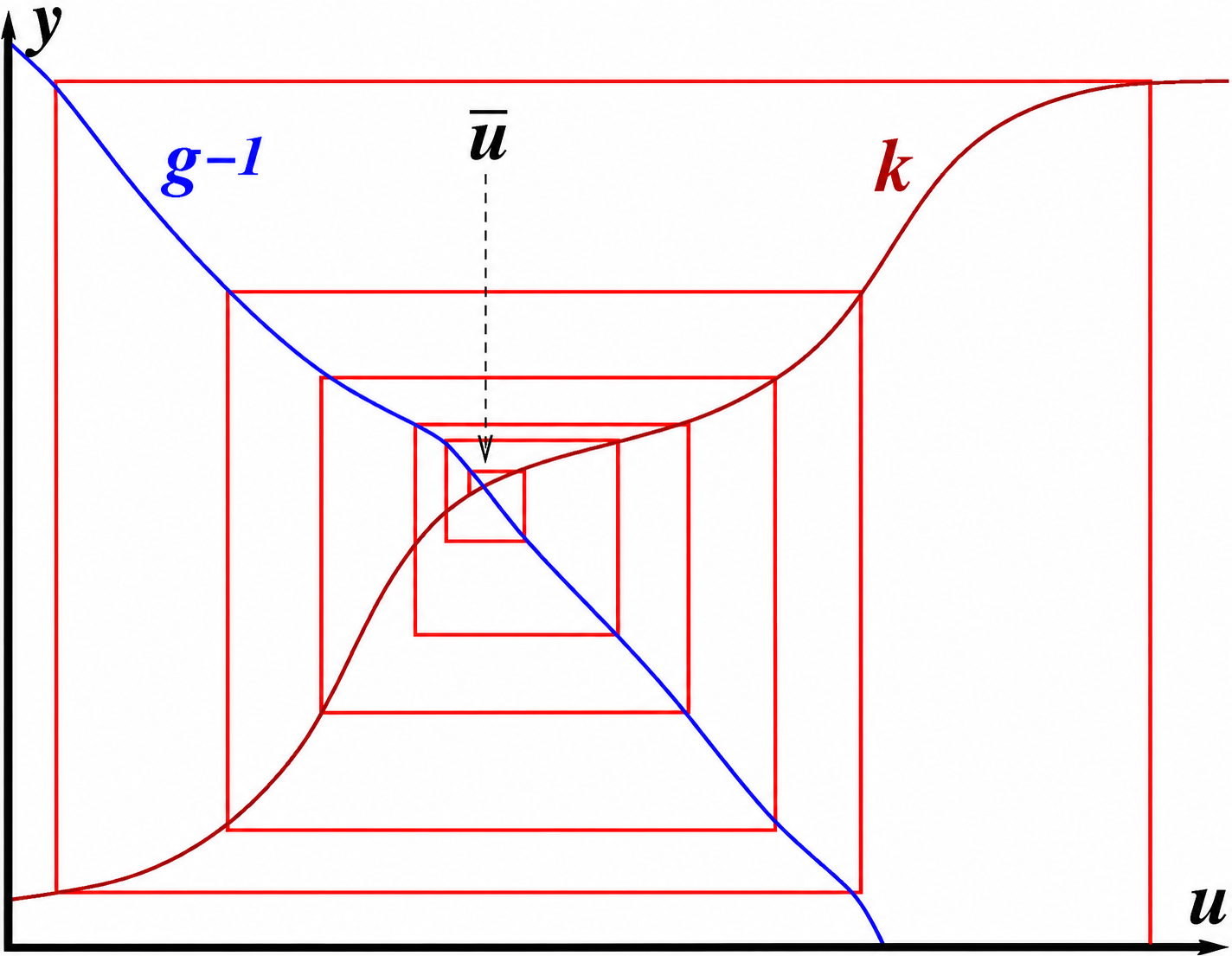}}
\caption{Two systems in a negative feedback configuration, and a
  convergent one-dimensional iteration.}
\label{fig:negative_feedback_monotones}
\efigstar

Stability
is guaranteed even in the presence of arbitrary delays (even
time-varying) in the feedback, so that the overall system becomes a
delay-differential system; see~\cite{dcds06}. The result can also be
generalized to certain partial differential equations that include
diffusion effects, as often encountered in biological
applications~\cite{enciso_smith_sontagJDE06}.

\myparagraph{Example of a negative-feedback interconnection:~a testosterone model}
                                
Circulating testosterone in healthy adult males exhibits multi-hour
rhythmic fluctuations driven by corresponding pulses of luteinizing
hormone (LH) from the pituitary and of luteinizing hormone-releasing
hormone (LHRH) from the hypothalamus
(see~\cite{Cart1986}, \cite{smith1980hormones}).
A classical model presented in~\cite{murray2002} accounts for these
oscillations using a system of delay-differential equations. In this
model, LHRH stimulates the production of LH, LH in turn
triggers testosterone synthesis in the testes (after a delay~$\tau$),
and testosterone feeds back to inhibit further release of LHRH.
Writing $R(t)$, $L(t)$ for the concentrations of LHRH and LH at time
$t$, and $T(t)$ for the concentration of testosterone at time
$t+\tau$, and assuming first-order degradation, the model takes the
following negative feedback form:
\beqn
\LDdiffdotJ&=&\alpha_2 L - \beta_3 T 
\quad\quad \mbox{\small testosterone}\\
\LDdiffdotK&=&\alpha_1 R - \beta_2 L 
\quad\quad \mbox{\small luteinising hormone}
\\
\LDdiffdotC&=&u-\beta_1 R 
\quad\quad\quad \mbox{\small $L$-releasing hormone}
\eeqn
where $u(t) = g(t-\tau)$ and $\tau>0$ is a delay. Here, $g$ is the function
$g(t)=\frac{A}{K+T(t)}$, thought of as a system with no dynamics.
We regard the system as a monotone linear subsystem with input $u$ and output $T$, whose characteristic is
\[
k(u)=\frac{\alpha_{1}\alpha_{2}}{\beta_{1}\beta_{2}\beta_{3}}\,u,
\]
placed in feedback with $g$. The composition $g\circ k$ is a simple fractional map of the form
\[
S(u)=\frac{p}{q+u}, \qquad p,q>0 .
\]
It is straightforward to verify that every map of this type has globally convergent iterates with a unique fixed point. Hence, by the theorem, oscillations cannot occur—regardless of the delay. This contradicts the result in~\cite{murray2002}, which asserts the existence of oscillations for certain delay values. 
Somewhat ironically, although monotone systems theory yields the non-oscillation result almost immediately, identifying the flaw in the proof of~\cite{murray2002} required substantial effort; see~\cite{testosterone}.

\section{Non-monotone systems: responses, motifs, and model discrimination}

A system is monotone with respect to an orthant order precisely when its
signed graph has no negative cycles after edge orientations are ignored.
In particular, any two directed paths connecting the same pair of distinct
nodes, a \textit{feedforward (FF)} motif, must carry the same sign, and
every directed cycle, a \textit{feedback (FB)} motif, must have a
positive sign. Feedforward motifs are especially common in biological
networks, including intracellular systems (metabolic, signaling, and
gene-regulatory networks) as well as intercellular interactions;
see~\cite{alonbook}.
We classify FF (respectively, FB) motifs as \textit{coherent} if the
two paths share the same sign (respectively, the loop has positive sign), and
\textit{incoherent} otherwise. Accordingly, we use the abbreviations CFF
and IFF for coherent and incoherent feedforward motifs, and CFB and IFB
for coherent and incoherent feedback motifs.
(We also use the term ``negative feedback loop'' instead of ``IFB.'')
Examples are shown in
Figure~\ref{fig:iffl_ifb}.

\bfigstar
\minp{0.16}{\picc{0.32}{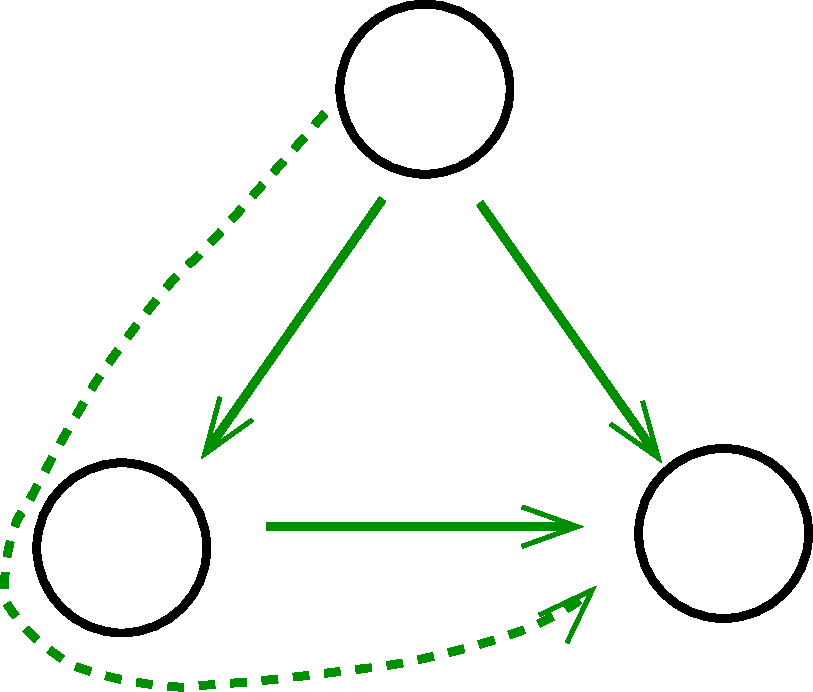}}%
\minp{0.16}{\picc{0.32}{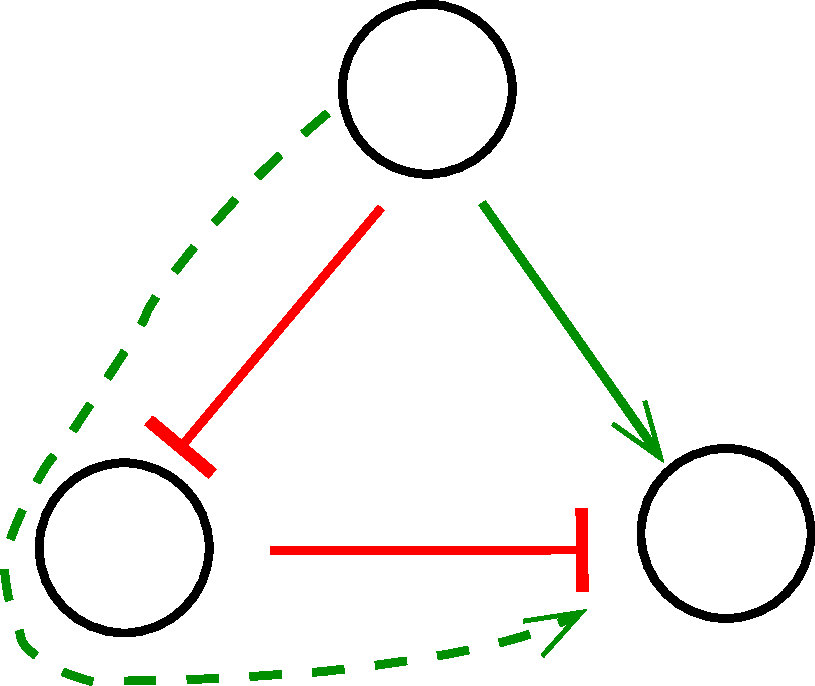}}%
\minp{0.16}{\picc{0.32}{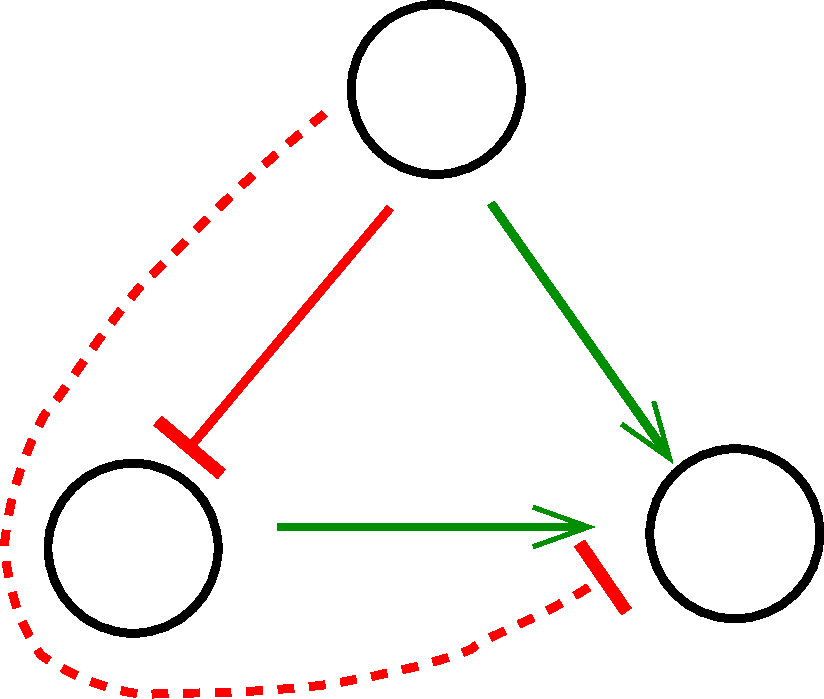}}%
\minp{0.50}{\picc{0.25}{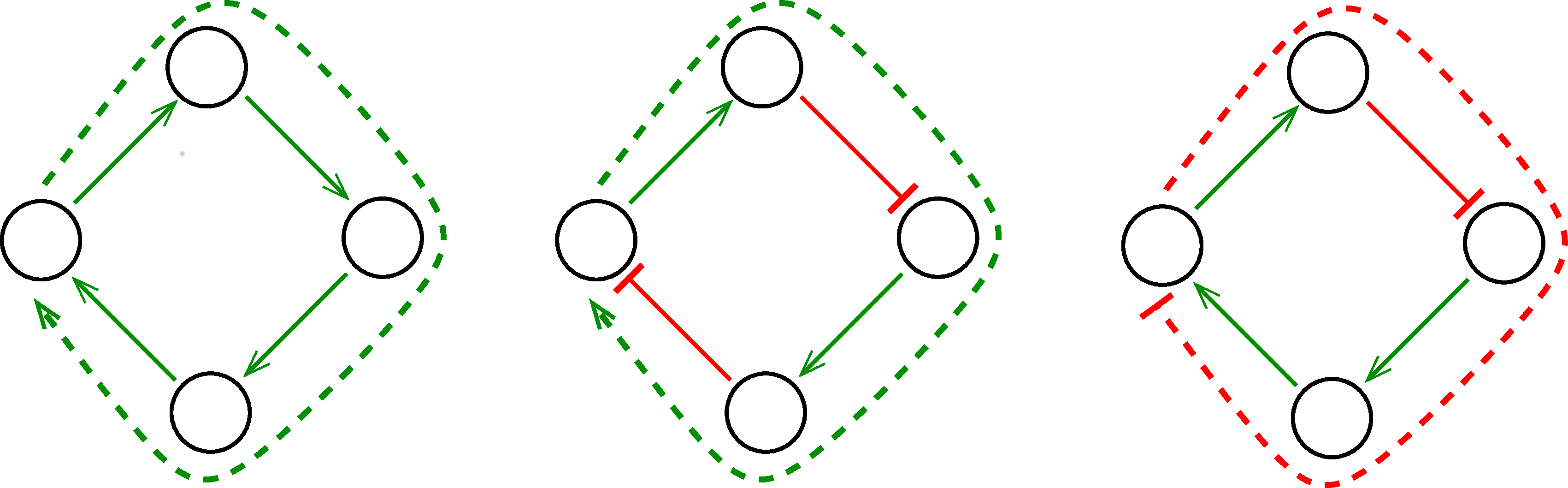}}

\medskip

\minp{0.16}{$\qquad$ CFF}%
\minp{0.16}{$\qquad$ CFF}%
\minp{0.16}{$\qquad$ IFF}%
\minp{0.50}{$\qquad\quad$ CFB $\qquad\qquad$ CFB $\qquad\qquad\;\;$ IFB}

\caption{Examples of Coherent and Incoherent FeedForward/FeedBack motifs}
\label{fig:iffl_ifb}
\efigstar

The following result is useful for invalidating putative
biological mechanisms. The proof is based on monotone systems theory,
but requires additional considerations because it only talks about
directed paths.  The precise statement is from
\cite{igoshin_et_al2015_monotone}, and relies heavily upon an earlier
version of the result given in~\cite{12cdc_angeli_sontag}.
We summarize below (for simplicity, only for scalar inputs and outputs):

\myparagraph{Biphasic responses as signatures of IFFs or IFBs}
Suppose that all the directed paths from the input node to the output
node have a positive (respectively negative) net sign.  Consider an
input $u$ that is monotonically increasing in time, and an initial
state that is an equilibrium state for the input value $u(0)$. Then
the output will monotonically increase (respectively decrease) in
time.  For inputs that are decreasing, the conclusions reverse.

A self-contained proof is not readily available, so we provide a
sketch in Appendix ``Sketch of proof of monotone-response theorem.''

Another way of stating the conclusion is that
a non-monotonic (biphasic, bell-shaped) response to a monotonic input
(for example, a step)
\emph{requires} a negative feedback and/or an incoherent feedforward motif,
see Fig.~\ref{fig:monotone_input}.
\bfigstar
\minp{0.2}{\picc{0.2}{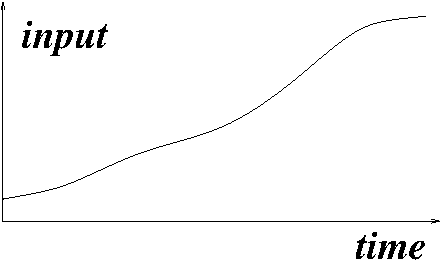}}%
\minp{0.2}{\picc{0.2}{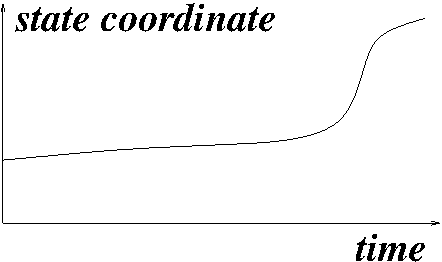}}%
\minp{0.2}{\picc{0.2}{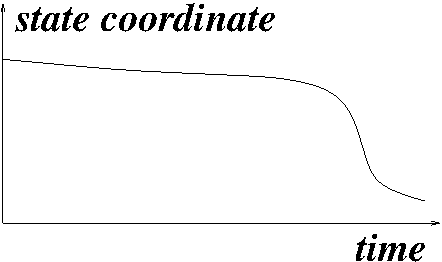}}%
\minp{0.2}{\picc{0.2}{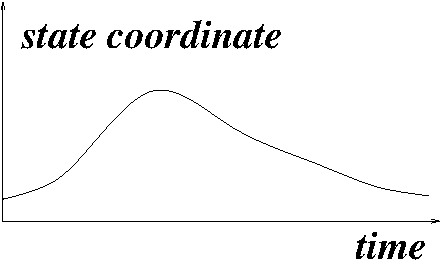}}%
\minp{0.2}{\picc{0.2}{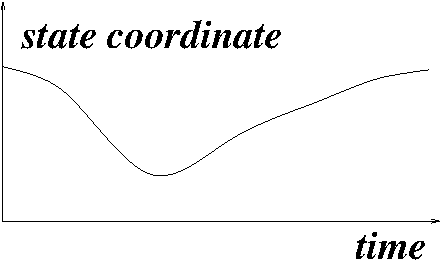}}%
\caption{Left to right: monotonic input; two possible state behaviors;
  two impossible behaviors, as the response is biphasic, unless there
  is a negative feedback and/or an incoherent feedforward motif in the
  graph.}
\label{fig:monotone_input}
\efigstar

\myparagraph{An example of model invalidation}

Regulation by the sigma factor $\sigma^E$ plays a key role in the
hypoxic stress-response pathway of \emph{M.~tuberculosis}, and in
particular modulates the expression of two central metabolism genes,
\emph{icl1} (Rv0467, glyoxylate shunt) and \emph{gltA1} (Rv1131,
methylcitrate cycle), both of which contribute to the persistence of
tubercle bacilli during infection. Transcription of \emph{icl1}
requires $\sigma^B$, itself transcribed under $\sigma^E$ control, as
well as the $\sigma^B$-dependent transcription factor \emph{lrpI}
(Rv0465c), a local regulator of \emph{icl1}. The resulting regulatory
architecture (see~\cite{datta11}) forms a coherent feedforward loop,
illustrated in the graph shown in Fig.~\ref{fig:sigmaE}. Because this
circuit contains neither feedback loops nor incoherent feedforward
branches, a monotone increase in $\sigma^E$ activity should, in
principle, produce a monotone increase in \emph{icl1} expression.
However, experiments in which oxygen levels are gradually depleted
over a three-day period show that \emph{icl1} expression is biphasic,
rather than monotone, despite the concomitant monotone rise in
$\sigma^E$ activity (plot in Fig.~\ref{fig:sigmaE}, adapted
from~\cite{Shi03}). This inconsistency between the predicted and
observed behavior motivated the search for alternative regulatory
architectures in~\cite{igoshin_et_al2015_monotone}. 
\bfig
\picc{0.15}{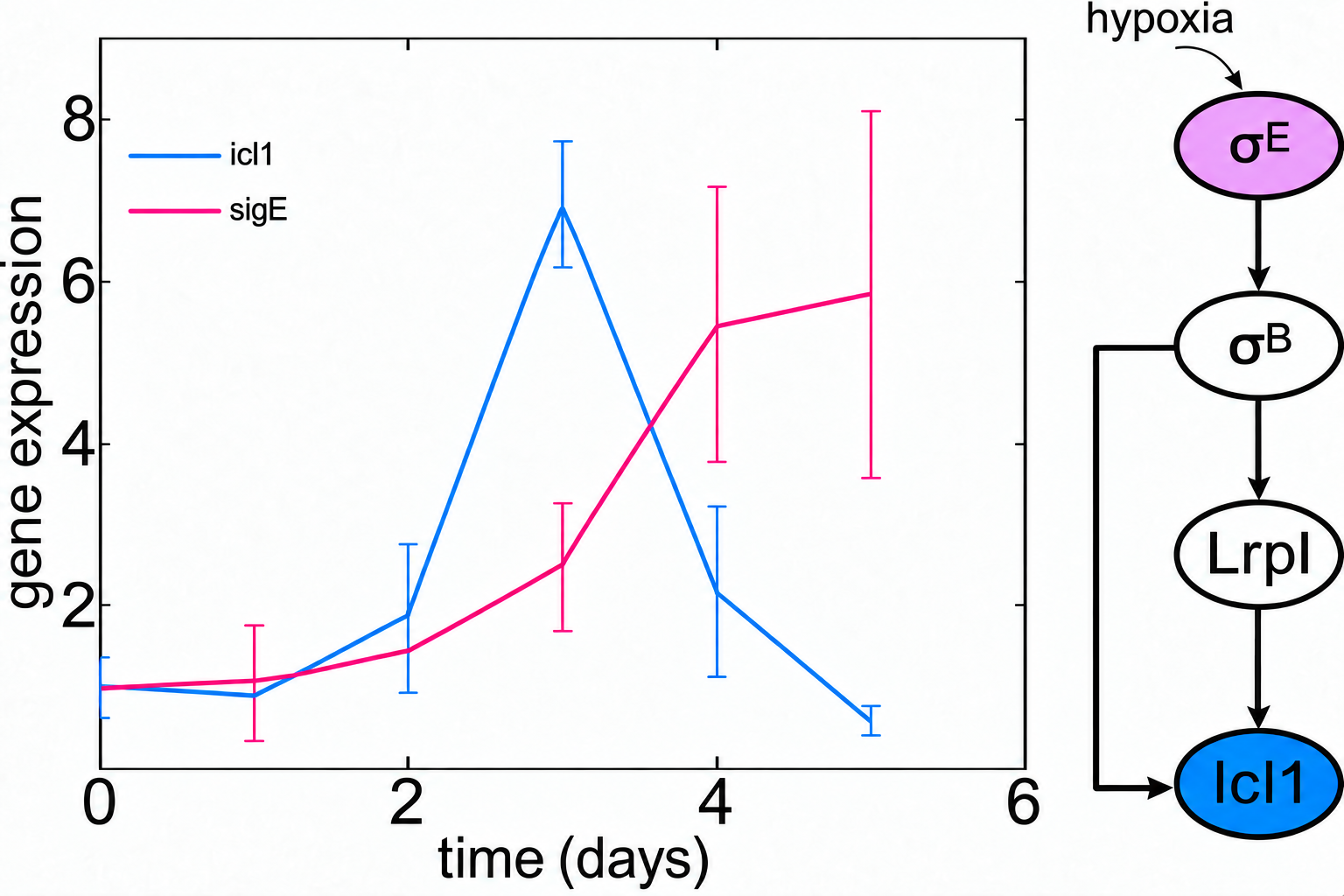}
\caption{Left: plot of experimental $\sigma ^E$ and resulting gene expression dynamics
  using data from~\cite{datta11}, showing a biphasic response. This is
  inconsistent with the network schematically shown on the right, which
  depicts the interactions described in~\cite{Shi03} .}
\label{fig:sigmaE}
\efig

\subsection{Transient responses followed by sensory ``perfect adaptation''}

A ubiquitous feature of biological sensory and regulatory systems is
\emph{(perfect) adaptation}: a sudden increase in an external stimulus
elicits a rapid transient response, such as activation of a signaling
cascade or induction of a gene, followed by a slower relaxation back
toward the original pre-stimulus level. Thus, even under a sustained
stimulus, the system ultimately ``resets'' its output. Adaptation is
crucial for maintaining key variables within physiological bounds and
for ensuring that cells and organisms remain sensitive to \emph{changes}
in their environment rather than being overwhelmed by constant
background signals.
Mathematically, we mean that, for constant (step) inputs, outputs are
not identically zero, but, after a transient recover, asymptotically
to their steady state value; see for example,~\cite{imp03} for a
precise definition.
Thus, adaptation to a step input, viewed as a nondecreasing input
starting from an equilibrium, necessarily entails a non-monotonic
response. We therefore have the following principle:

\myparagraph{Adaptation as a signature of IFFs and IFBs}
An adapting system requires a negative feedback and/or an incoherent feedforward motif.

Perfect adaptation is quite common in biology. In bacterial chemotaxis,
\emph{E.~coli} receptors transiently signal upon a step increase in
attractant, but biochemical modification systems restore the signaling
state to baseline. Sensory neurons in vision and olfaction adapt to
background light or odor so that new stimuli can still be detected.
Hormonal and immune systems similarly employ adaptive responses to
prevent chronic overstimulation. Gene-regulatory networks also display
adaptive pulse-like dynamics, often arising from characteristic circuit
motifs.
Figures \ref{fig:chemotaxis_adaptation},
\ref{fig:yeast_adaptation}, and
\ref{fig:trendel_adaptation} show respectively adaptation at several
biological levels: bacteria, yeast (a single-cell eukaryote), and
human immune cells.

\bfig
\picc{0.15}{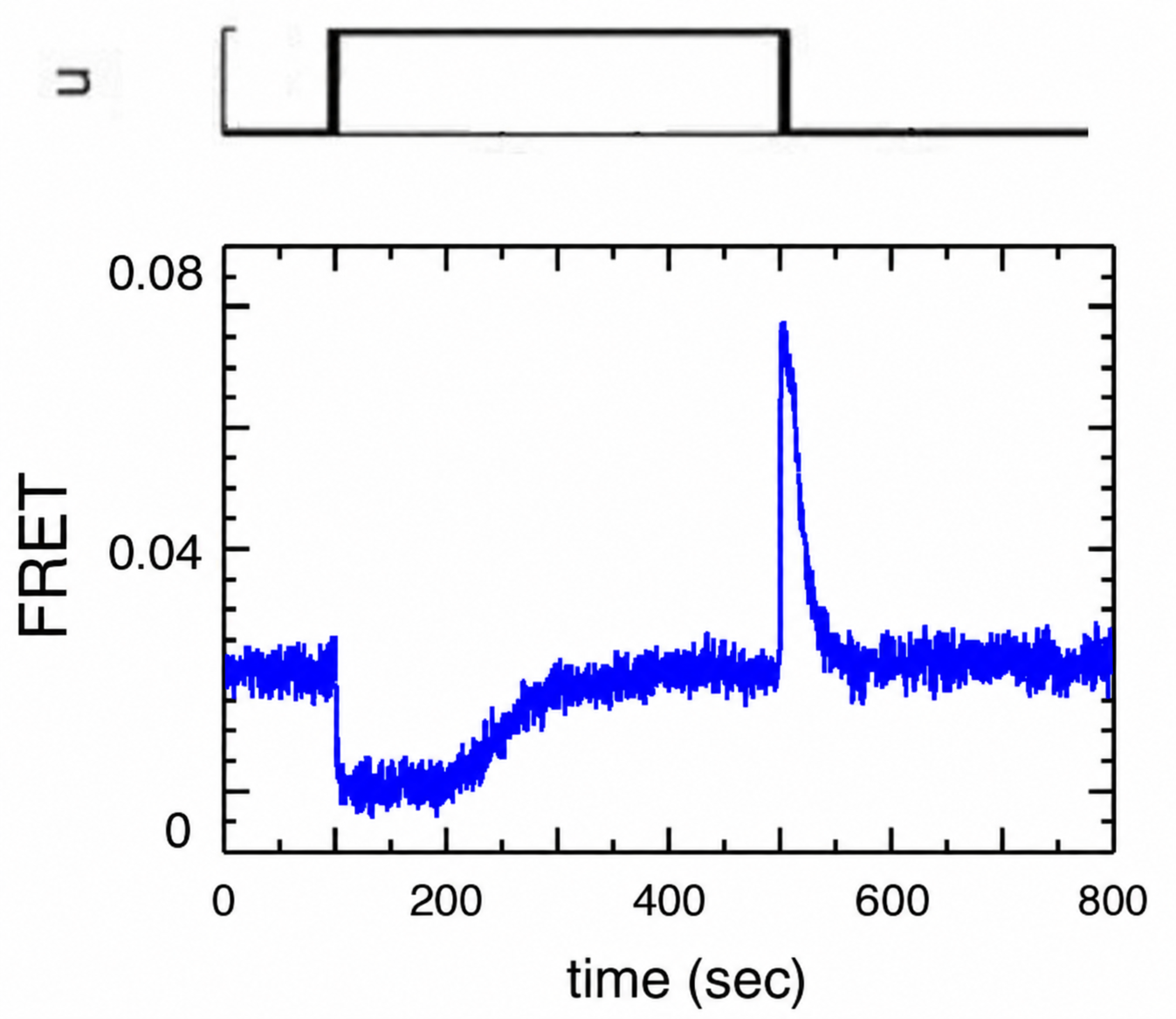}
\caption{Adaptive response in the \textit{E.\ coli} chemotaxis pathway,
measured \textit{in vivo} by fluorescence resonance energy transfer
(FRET). Stepwise addition and subsequent removal of attractant
(methyl-aspartate, MeAsp) are applied to cells expressing CFP--FliM and
CheY--YFP fusion proteins. CheY is the phosphorylated response regulator
that conveys chemotactic signals from the receptor kinase cascade to
the flagellar motor by binding to the motor switch protein FliM. YFP
and CFP (yellow and cyan fluorescent proteins) serve as the FRET donor
and acceptor pair, allowing real-time monitoring of the interaction
between CheY$\sim$P and FliM \textit{via} energy transfer efficiency.
The resulting FRET signal reports the fraction of motor-bound
CheY$\sim$P. Following each step change in stimulus, the signal exhibits
a rapid transient response and then relaxes back toward its prestimulus
level despite sustained input, demonstrating near-perfect adaptation
of the chemotaxis signaling output. Figure drawn using data
from~\cite{SourjikBerg2002}.}
\label{fig:chemotaxis_adaptation}
\efig

\bfig
\picc{0.5}{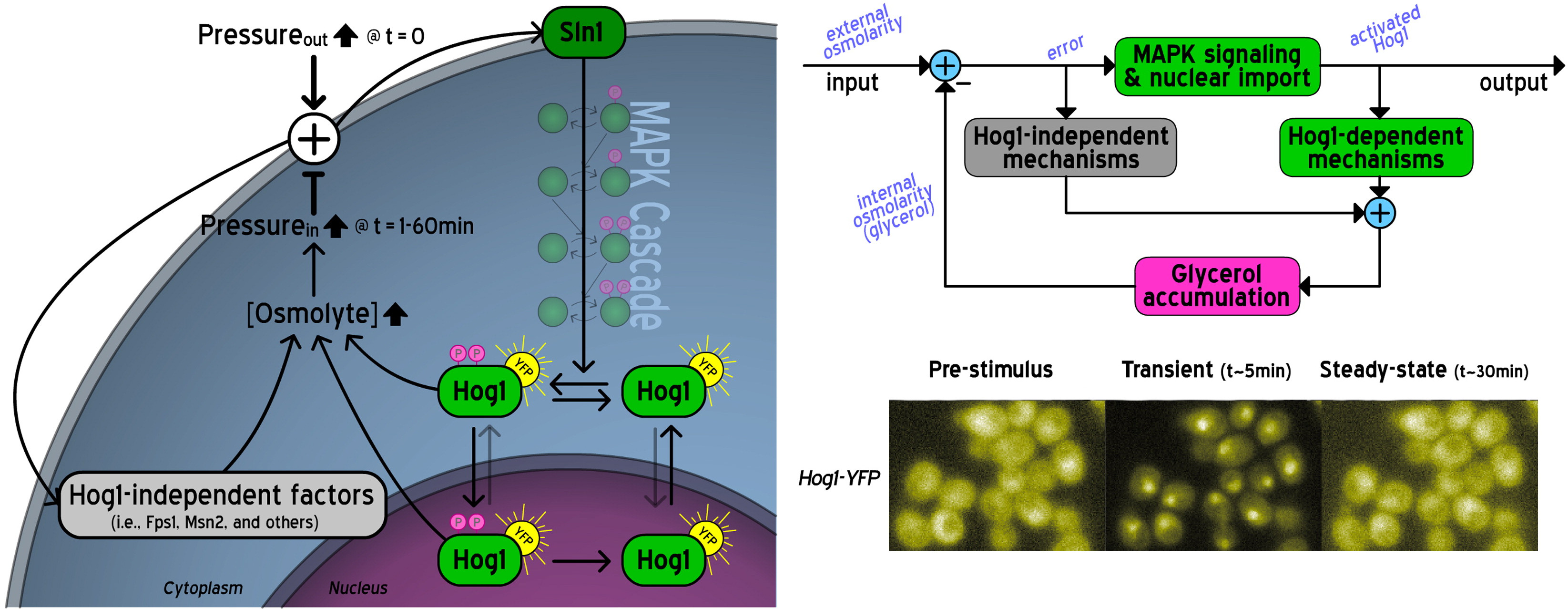}
\caption{Perfect adaptation in the yeast HOG pathway through an
integral-feedback module. Following hyperosmotic shock, the MAP kinase
Hog1 rapidly translocates to the nucleus, triggering transcriptional and
metabolic responses that promote glycerol production. As osmotic balance
is restored, Hog1 returns to its pre-stimulus cytosolic distribution
despite sustained elevated external osmolarity, demonstrating near-perfect
adaptation. The paper~\cite{Muzzey2009}
showed that this behavior requires a {single internal integrator}
located downstream of Hog1 and upstream of glycerol accumulation, forming
an effective {integral-feedback control loop} that restores turgor
pressure. Figure adapted from~\cite{Muzzey2009}, freely available through
Open Access Creative Commons Attribution License (CC BY).}
\label{fig:yeast_adaptation}
\efig

\bfigstar
\picc{0.18}{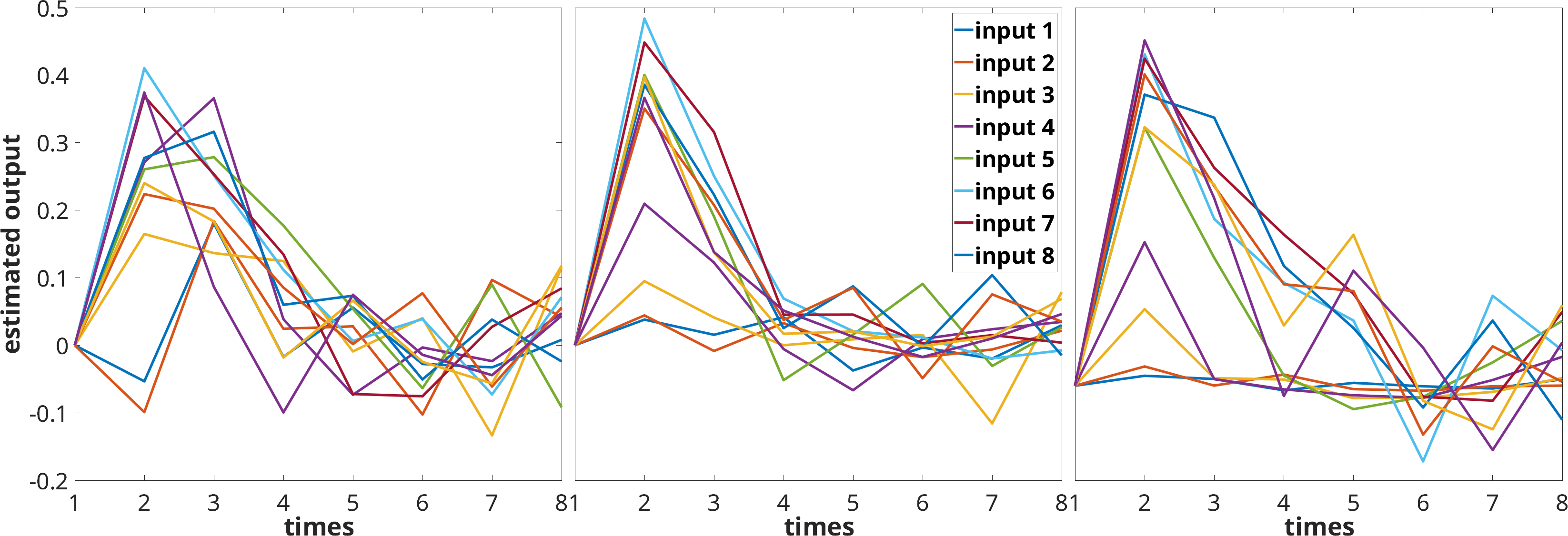}
\caption{An experimental demonstration of perfect (or near-perfect)
adaptation in a mammalian system. Human CD8$^+$ T cells were subjected to
constant antigen stimulation over many hours. Cytokine production
exhibited a transient activation pulse followed by a return to
near-baseline levels despite the continued presence of antigen. Shown
are three independent experiments, each performed over eight distinct
input magnitudes. The figure was generated from data reported
in~\cite{dushek2019}, and is discussed in more detail below.}
\label{fig:trendel_adaptation}
\efigstar

\myparagraph{Integral feedback for adaptation}

From a control-theoretic perspective, \emph{integral feedback} is a
classical and powerful mechanism for achieving perfect adaptation.
Widely used in engineering, it ensures that any deviation from a desired
baseline level is continuously accumulated and counteracted, eventually
driving the steady-state error to zero. Biological circuits such as
bacterial chemotaxis and certain homeostatic modules employ integral
feedback--like architectures, providing robust and precise restoration
to baseline even in the presence of parameter uncertainty or
environmental variability~\cite{Khammash2021}.

In synthetic biology, biomolecular integral controllers that precisely
regulate target gene expression so as to track externally imposed
reference levels, even in the presence of disturbances, have been
experimentally demonstrated both in living cells and in cell-free
\textit{in vitro} systems.
An approach that has been shown to be particularly powerful is based
on what Khammash and his coauthors~\cite{BriatGuptaKhammash2016}
called \emph{antithetic} controllers,
illustrated in~Figure~\ref{fig:antithetic}A.
Some theory is discussed in the Sidebar ``Some theoretical results for antithetic controllers.''
\bfigstar
\minp{0.25}{A}%
\minp{0.25}{B}%
\minp{0.25}{C}%
\minp{0.25}{D}

\minp{0.25}{\pic{0.12}{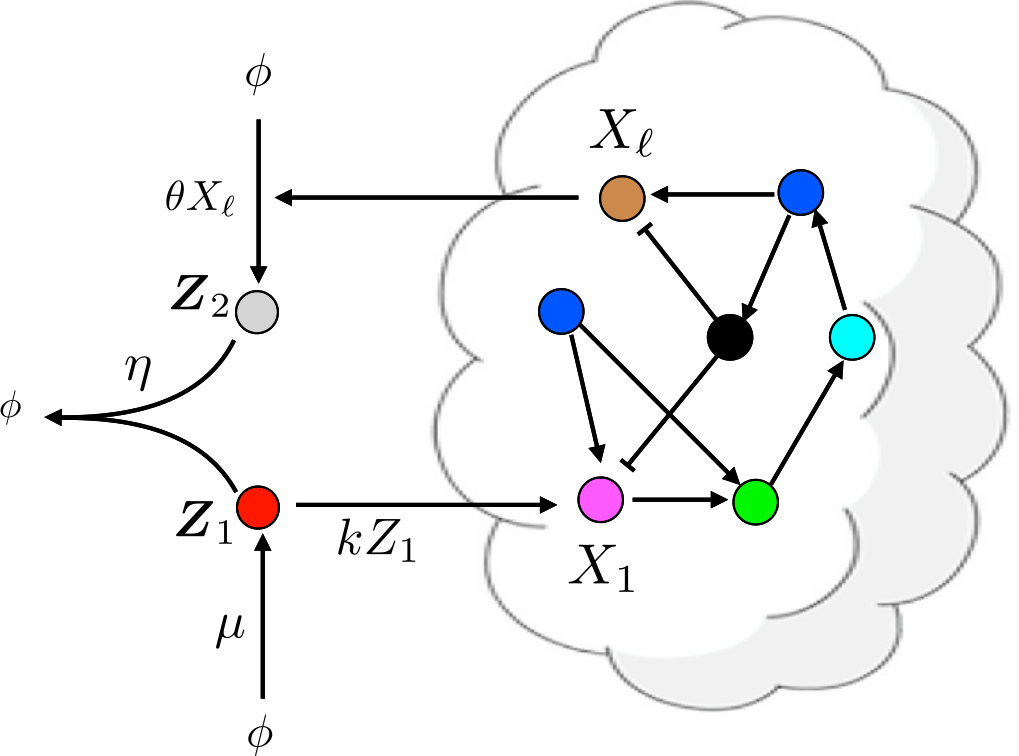}}%
\minp{0.25}{\picc{0.15}{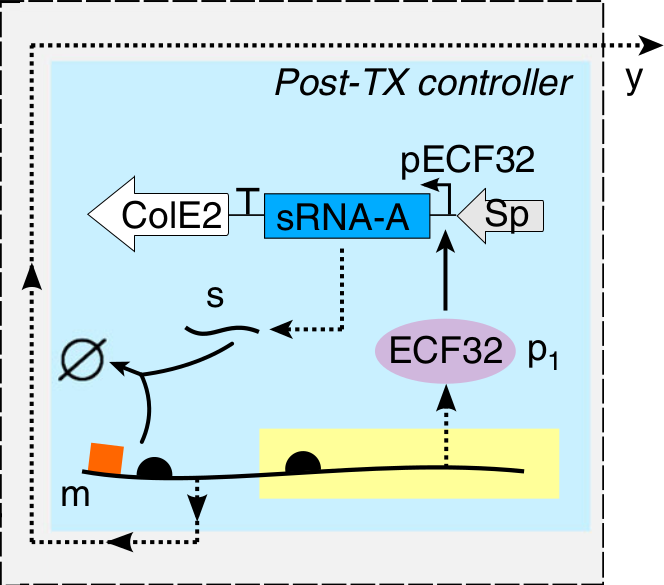}}
\minp{0.25}{\pic{0.07}{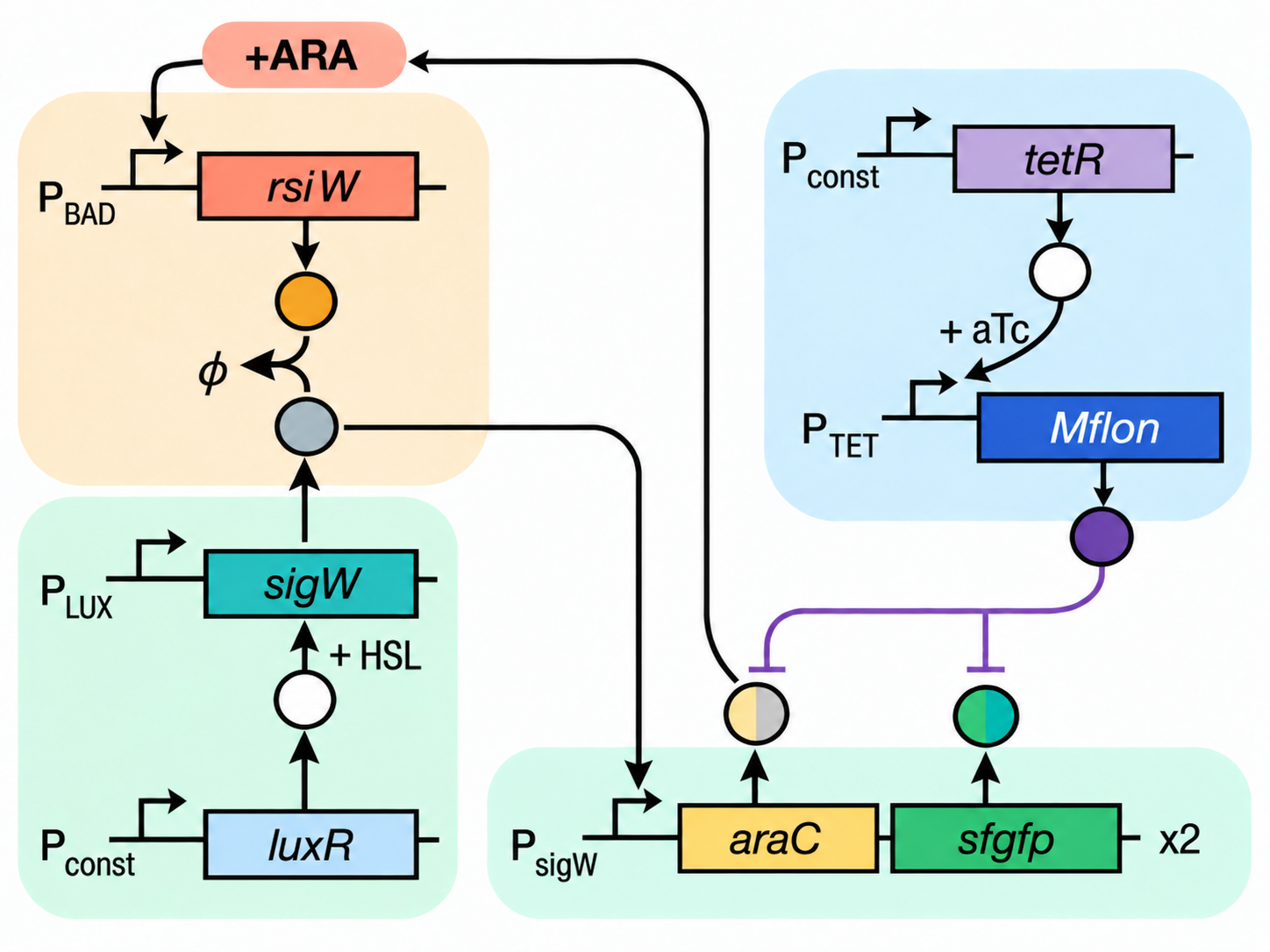}}%
\minp{0.25}{\pic{0.07}{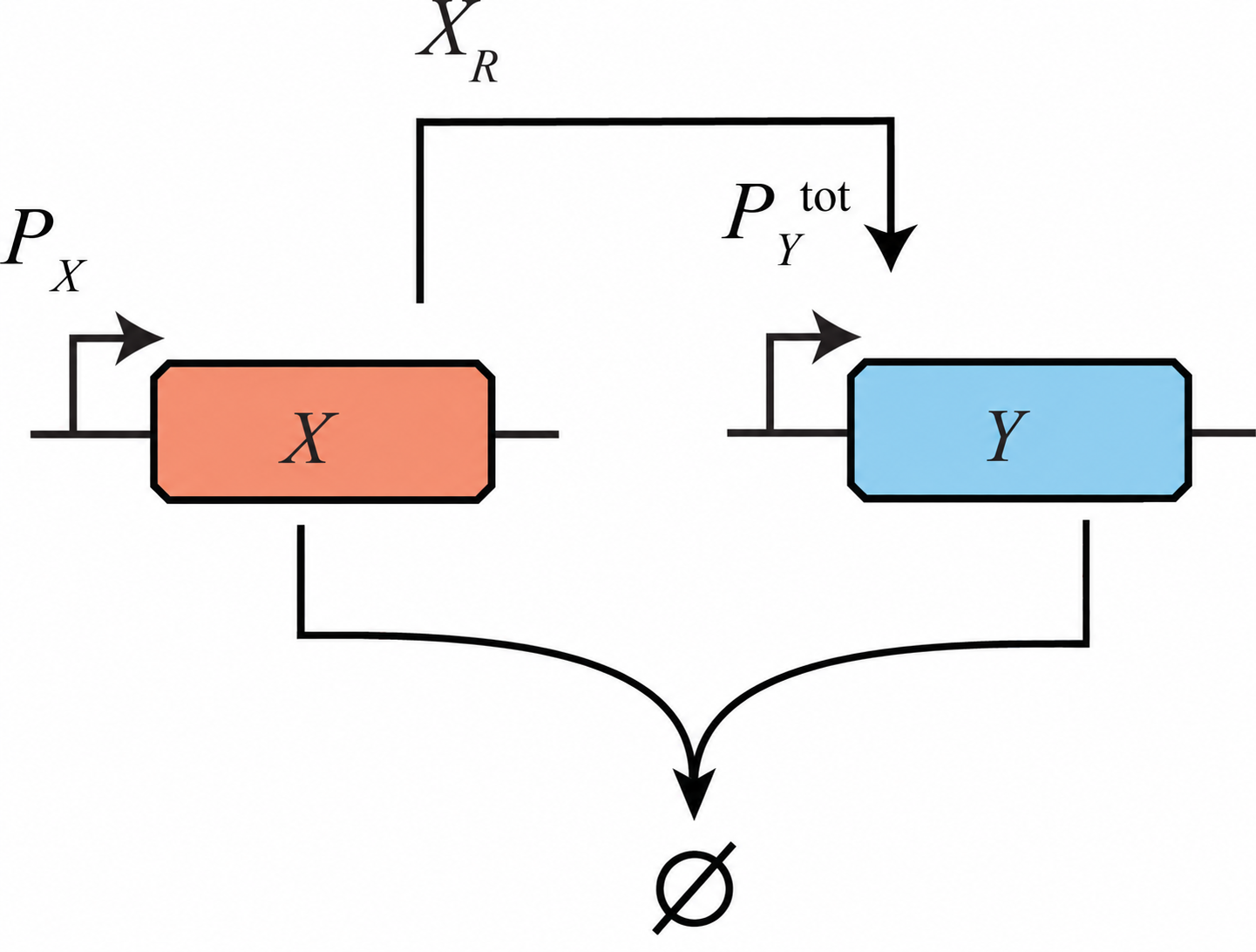}}

\caption{(A) The antithetic integral feedback controller.
Illustration adapted from~\cite{BriatGuptaKhammash2016} under
Open Access Creative Commons Attribution License (CC BY).
Figure represents
the antithetic integral-feedback architecture
consisting of two controller species, denoted $Z_1$ and $Z_2$.
Species $Z_1$ serves as the reference input and drives the production
of a network species $X_1$. Through an arbitrary downstream network,
which may be large and only partially characterized, changes in $X_1$
eventually influence an output species $X_\ell$. The second controller
species, $Z_2$, is produced at a rate proportional to the abundance of
$X_\ell$, thereby providing a measurement of the network output.
The key interaction is an \emph{antithetic} reaction between $Z_1$ and
$Z_2$, whereby the two controller species neutralize each other's
activity. This interaction need not correspond to their literal
degradation. For example, $Z_1$ and $Z_2$ may bind to form a
high-affinity complex that is biologically inactive. Through this
antithetic coupling, the controller compares the reference signal
encoded by $Z_1$ with the measured output encoded by $Z_2$, thereby
implementing an integral-feedback mechanism that achieves robust
perfect adaptation.
(B)
Antithetic integral feedback controller implemented
through anti-sense mRNA.
Illustration adapted from~\cite{HuangQiangDelVecchio2018},
available under Creative Commons Attribution 4.0 International License.
(C) Another implementation, through a sigma and an anti-sigma factor
in living \emph{E.~coli}.
Illustration drawn by chatGPT AI, based on the genetic diagram in~\cite{2019_nature_AokiKhammash_integral_feedback}.
(D) An implementation through an \emph{in vitro}
translation/transcription system, using also a sigma and an anti-sigma factor.
The diagram shows the main components of the construct in~\cite{2019_deepak_integral_controller}.}
\label{fig:antithetic}
\efigstar

\begin{sidebar}{Some theoretical results for antithetic controllers}
  
From a theoretical perspective, the antithetic architecture may be represented in
the general form
\beqn
\LDdiffdotZ &=& \alpha_1 - \alpha_2 z_1 z_2\\
\LDdiffdotAA &=& \alpha_3 y - \alpha_4 z_1 z_2\\
\LDdiffdotI   &=& f(x,z_1)\\
y &=& h(x).
\eeqn
Here, $z_1$ and $z_2$ are the antithetic controller variables, while
$y$ is the output of the plant, described by the state vector $x$,
that is to be regulated.
At steady state, and disregarding questions of existence and stability,
one immediately obtains
\[
y^*=\frac{\alpha_1\alpha_4}{\alpha_2\alpha_3},
\]
a value that is determined solely by the controller parameters and is
completely independent of the plant dynamics~$f$. Consequently, the
steady-state value of the regulated output is insensitive to uncertainty
in the plant, including disturbances and external inputs that affect
the vector field~$f$. This property is the hallmark of ``robust'' perfect
adaptation.
(As emphasized in~\cite{HuangQiangDelVecchio2018}, however, certain
implementations depart from this idealized behavior. In
particular, non-negligible degradation or dilution of the controller
species introduces additional terms of the form $-\gamma_i z_i$ into
the first two equations, thereby destroying the exact integral-action
property and leading to imperfect adaptation.)

Oscillations can occur in this deterministic dynamical system even when
the plant is the simple two-dimensional linear system
(see Figure~4 in~\cite{BriatGuptaKhammash2016}):
\beqn
\LDdiffdotE &=& kz_1 - x_1\\
\LDdiffdotS &=& x_1 - x_2\\
\LDdiffdotZ &=& \mu - \eta_1 z_1 z_2\\
\LDdiffdotAA &=&  y - \eta z_1 z_2\\
y &=& x_2,
\eeqn

This observation raises a natural dynamical-systems question: what can
be said, more generally, about the long-term behavior of solutions?

In the case considered here, the plant is a two-dimensional linear system,
resulting in a four-dimensional nonlinear closed-loop system, has been
studied in considerable detail. Although the system possesses a unique
equilibrium point~$e$, it may exhibit stable periodic orbits for
certain parameter values. This naturally leads to the question of
characterizing attractors.

A major step toward answering this question was obtained in~\cite{2025_margal}, where it was shown that, for arbitrary parameter values, every compact $\omega$-limit set that does not contain the equilibrium~$e$ must be a periodic orbit. The proof relies on the theory of \emph{$2$-cooperative systems} and the associated \emph{second additive compound} matrices, which generalize classical cooperative systems and, in an appropriate sense, ``behave like two-dimensional systems.'' 

This viewpoint can be understood through contractions in Hilbert's projective metric on positive cones, since $2$-cooperative systems may be viewed as cooperative systems on an exterior power; see~\cite{sontag2026kcooperativeintuition}.
This result left open an important issue: are all trajectories
precompact, and hence guaranteed to possess compact $\omega$-limit
sets?

A positive answer to this question was recently obtained
in~\cite{2026_05_wafi_oliveira_sontag_antithetic_arxiv}
(see also~\cite{deoliveira_wafi_sontag2026_pipo} for a nonlinear generalization). 
Combined with
the result from~\cite{2025_margal}, it follows that
every solution either keeps visiting arbitrarily small neighborhoods of
the equilibrium or else approaches a periodic orbit.
\end{sidebar}

There have been several experimental implementations of the antithetic
controller idea.
One of them can be found in~\cite{HuangQiangDelVecchio2018}.
In this construct, the objective is to regulate the product of the
gene of interest
(protein output $y$ in Figure~\ref{fig:antithetic}B) to a value that is
robust to ribosome competition (which arises from another gene being
expressed and is not shown in our simplified figure).
The variable $z_1$ in our model represents the concentration of an mRNA
($m$ in the figure), and $z_2$ represents the concentration of an sRNA (small
RNA, $s$ in the figure) which hybridizes with the mRNA of the regulated gene,
therefore inhibiting its translation and promoting degradation of both
$m$ and $s$.
The protein concentration $y$ can be thought of as produced by
a transcriptional system $\LDdiffdotH = \beta z_1 - \delta y$ in which
$\beta$ is proportional to ribosome availability and $\delta$ is a
degradation rate constant.
The concentration of $s$ is engineered to have a rate of
production proportional to $y$, through an ingenious co-expression
system involving the production of the targeting sequence for the sRNA
(ECF32 in the figure, binding to promoter pECF32).

Another implementation was provided
in~\cite{2019_nature_AokiKhammash_integral_feedback}.
In this case, two controller proteins stoichiometrically neutralize
one another through a sequestration reaction.
To implement this
reaction, the authors employed the \emph{Bacillus subtilis} sigma
factor SigW and its cognate anti-sigma factor RsiW, a protein pair
previously known to exhibit effective annihilation in vivo.
In their native biological role, RsiW binds SigW with high affinity,
sequestering it in an inactive complex. In the synthetic
controller, this strong sequestration interaction serves as the
antithetic reaction: SigW and RsiW bind and thereby mutually abolish
their biological activity, providing the molecular realization of the
integral-feedback mechanism.
See Figure~\ref{fig:antithetic}C.
In a hybrid implementation from the same lab, in which the antithetic
controller was implemented in software and connected to living yeast
cells through optogenetic feedback, pronounced but damped oscillatory transients
have been observed for certain parameter regimes, as predicted by
modeling (M. Khammash, personal communication). 

Yet another experimental realization of the antithetic-controller idea
is the \emph{in vitro} TXTL implementation of
Agrawal \textit{et al.}~\cite{2019_deepak_integral_controller}. In
contrast to implementations in living cells, this work used an
\emph{E.~coli} cell-free transcription--translation system, which
contains the molecular machinery required for gene expression over
several hours while allowing the experimenter to prescribe DNA
concentrations and stoichiometries directly. This feature is important
from a control-design perspective: the input, plant, and output modules
can be tuned quantitatively, disturbances can be introduced at specified
times, and the resulting dynamics can be compared with ordinary
differential equation models with much less uncertainty than in vivo.
The goal of the paper was not merely to obtain adaptation qualitatively,
but to build a controller for which the protein production rate of an
output gene tracks an externally imposed reference signal over a broad
range of input values.
The molecular implementation again uses sequestration as the antithetic
interaction; see Figure~\ref{fig:antithetic}D.
An input gene produces a protein $X$ (\emph{E.~coli} $\sigma^{28}$)
while a target gene produces a protein $Y$ (anti-$\sigma^{28}$ factor
FlgM), which
represents the regulated plant signal. When $X$ and $Y$ bind, both are
rendered inactive, so only the free amount of $X$ can activate
promoters driving production of $Y$.
The mathematical analysis was carried out for a model as above, where
$z_1$ and $z_2$ are the concentrations of $X$ and $Y$ respectively,
and the $f$-system represents a two-dimensional system in which $y$ is
the mRNA of the gene that codes for $Y$ (Figure 4C
in~\cite{2019_deepak_integral_controller} shows the precise equations, 
including an additional equation for a green fluorescent reporter not
shown in our figure).
The theory and experiments showed that the output production rate is
approximately linear in the promoter concentration $P_X$ of $X$,
providing reference tracking and rejecting local disturbances in the
DNA template concentrations $P_Y^{\mbox{tot}}$. In addition, the system
was found to be robust to a global perturbation produced by changing
the reaction temperature. 

We note that, in addition to antithetic integral feedback and IFFs (see
below), alternative biological architectures for achieving adaptation have
been proposed, notably ultrasensitive quasi-integral controllers
in~\cite{CubaSamaniegoFranco2021}.

\myparagraph{IFFs for adaptation}

A different mechanism for adaptation arises from \emph{incoherent
feedforward} (IFF) motifs, in which an input activates a target through
one pathway while simultaneously activating a second pathway that
represses the same target after a delay. This architecture produces an
initial activation pulse followed by a return toward baseline and is
common in transcriptional networks, allowing rapid and sharply timed
responses without an explicit ``integrator'' component.

Each mechanism has characteristic advantages and limitations. Integral
feedback is exceptionally robust and guarantees return to baseline despite
fluctuations or uncertainty, though it may respond more slowly and often
requires additional molecular components to implement the effective
integration. IFF motifs, in contrast, are compact and can generate fast,
transient responses, but their adaptation is generally less robust: the
degree of return to baseline depends on parameter balances and is not
structurally guaranteed. Both strategies are therefore pervasive in
biology, reflecting different trade-offs among robustness, speed,
metabolic cost, and biochemical simplicity.

Under suitable technical assumptions and for certain classes of
systems, the \emph{Internal Model Principle} (IMP) guarantees that any
system exhibiting perfect adaptation must incorporate, in some form,
an \emph{integral-feedback mechanism}, possibly after an appropriate
nonlinear change of coordinates. However, when interpreting biological
circuits in terms of \emph{network motifs}, such coordinate
transformations generally alter the network structure itself,
obscuring the direct correspondence between the abstract IMP guarantee
and specific biochemical architectures. As a result, the practical
significance of the IMP for motif-based biological analysis is not
always clear; see the discussion in~\cite{shoval_alon_sontag_2011,22_annual_reviews_tutorial_imp}.

\myparagraph{A synthetic biology example of IFF adaptation}

Not all negative feedback mechanisms yield perfect adaptation. In
linear systems, perfect adaptation to step inputs is equivalent to the
presence of an integrator, or, equivalently, a zero at the origin of
the transfer function. Purely static (memoryless) negative feedback,
even when applied with high gain, cannot generically enforce
adaptation. This distinction is illustrated experimentally
in~\cite{bleris10}, which contrasts synthetic gene circuits
implementing static negative feedback with incoherent feedforward
(IFF) architectures that effectively realize integral control.
The motivation in~\cite{bleris10} is the need for natural and
synthetic biological networks to operate reliably under fluctuations
in the stoichiometry of their molecular components, such as
variability in gene copy number or expression efficiency. From a
control-theoretic perspective, the question is whether the
steady-state output can be made insensitive to such constant
disturbances. The experiments show that while static negative feedback
reshapes the steady-state input--output map, only IFF-based circuits
achieve true adaptation, namely steady-state insensitivity of the
output to changes in input amplitude.

The negative-feedback genetic circuit studied in~\cite{bleris10} is
shown in Figure~\ref{fig:negfeedback_dna} (right), with corresponding
interaction and phenomenological diagrams in the middle and left
panels.  The phenomenological description assumes transcription is
fast relative to translation, so that the protein product represses
its own production rate. The measured output \(y\) (ZsGreen
fluorescence) is proportional to the concentration of the
self-repressing protein (LacI), while the input \(u\) is proportional
to plasmid copy number and is reported by DsRed fluorescence.
\bfigstar
\minp{0.1}{\picc{0.5}{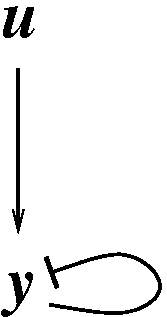}}%
\minp{0.9}{\picc{0.3}{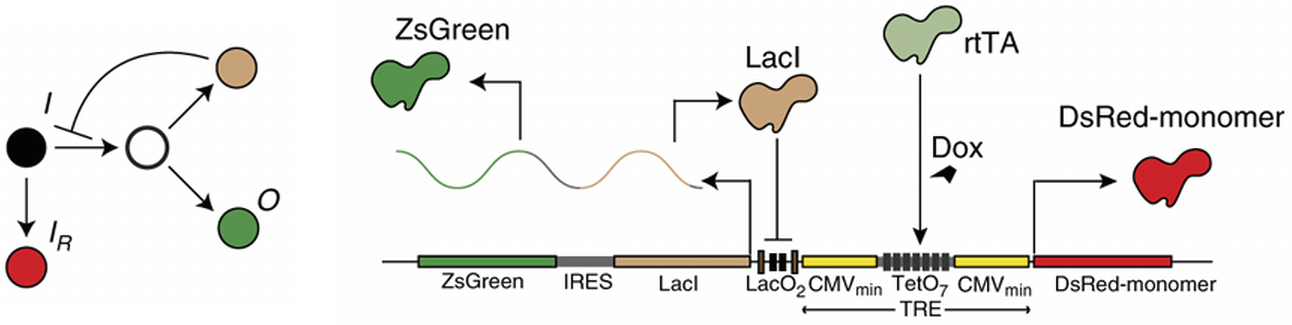}}
\caption{
Transcriptional negative autoregulation motif (tAM), adapted from
Figure~1D in~\cite{bleris10}, available under a CC-BY-NC 4.0 license.
Left: phenomenological static feedback representation.
Middle: interaction diagram of the genetic circuit ($I$ = input, 
$I_R$ = input reporter, $O$ = output).
Right: genetic implementation; LacI represses its own transcription
and that of the cotranslated ZsGreen reporter. The input is reported
by a divergently expressed DsRed protein. The transcription factor
rtTA is present at constant, non-limiting levels and is not treated as
a system input. 
}
\label{fig:negfeedback_dna}
\efigstar

A minimal phenomenological model for this circuit is the scalar nonlinear
system
\[
\LDdiffdotH = \frac{M u}{1+Ly} - \delta y,
\]
where \(u\) denotes the input, \(y\) the output, \(L\) the static feedback
gain, $M$ the efficiency of transcription and translation, and
\(\delta\) the degradation rate constant. While more detailed
dynamical models are analyzed in~\cite{bleris10}, this simple equation already
captures the essential steady-state behavior.
At equilibrium, $y$ satisfies
\[
 K y^2 + dy - u = 0
\]
where $d:=\delta/M$ and $K:=L d$.
For the subsequent analysis, we keep $d$ constant but think of $K$ as
a parameter, and write $y=f_K(u)$ for the positive solution of the quadratic.
A Taylor expansion for small \(K\) yields
\[
f_K(u)
= \frac{-d +d\left(1+\frac{2Ku}{d^2}+O(K^2)\right) }{2K}
= \frac{u}{d}+O(K).
\]
so that in the limit of vanishing feedback strength,
\[
\lim_{K\to 0^+} f_K(u)=\frac{u}{d}.
\]
Thus, in the absence of repression, the steady-state input--output map is
linear, and no adaptation is present.
In the opposite regime of strong static feedback (\(K\to +\infty\)),
asymptotic expansion gives
\[
f_K(u)
= \sqrt{\frac{u}{K}} - \frac{d}{2K}
+ O\!\left(\frac{1}{K^{3/2}}\right),
\]
so that
\[
f_K(u)\sim \sqrt{\frac{u}{K}}.
\]
Hence, increasing the feedback gain does not eliminate dependence on the
input; it merely compresses the static input--output nonlinearity from
linear to sublinear (square-root) scaling.
From a control perspective, this behavior is characteristic of high-gain
static feedback: sensitivity is reduced but cannot be driven to zero.
In contrast, integral feedback, as effectively implemented by the IFF
architectures in~\cite{bleris10}, introduces internal state
accumulation, enforcing zero steady-state sensitivity to constant
disturbances and thereby guaranteeing perfect adaptation, in accordance
with the internal model principle.

Equivalently, plotting \(\log y\) versus \(\log u\), the static-feedback
model predicts a slope close to \(1\) for weak repression and approaching
\(1/2\) for strong repression, but never zero. This prediction is
confirmed experimentally in Figure~\ref{fig:negfeedback_experiments},
where three repression levels yield slopes ranging from near \(1\) to
near \(0.5\).
About 200,000 cells were measured in each experiment, and the means and
standard deviations are shown for each of the three.
The key plot is the right one, showing a log-log plot of linear fits.
We see that, as predicted, the slopes range from near 1 to near 0.5.
\bfigstar
\minp{0.75}{\picc{0.23}{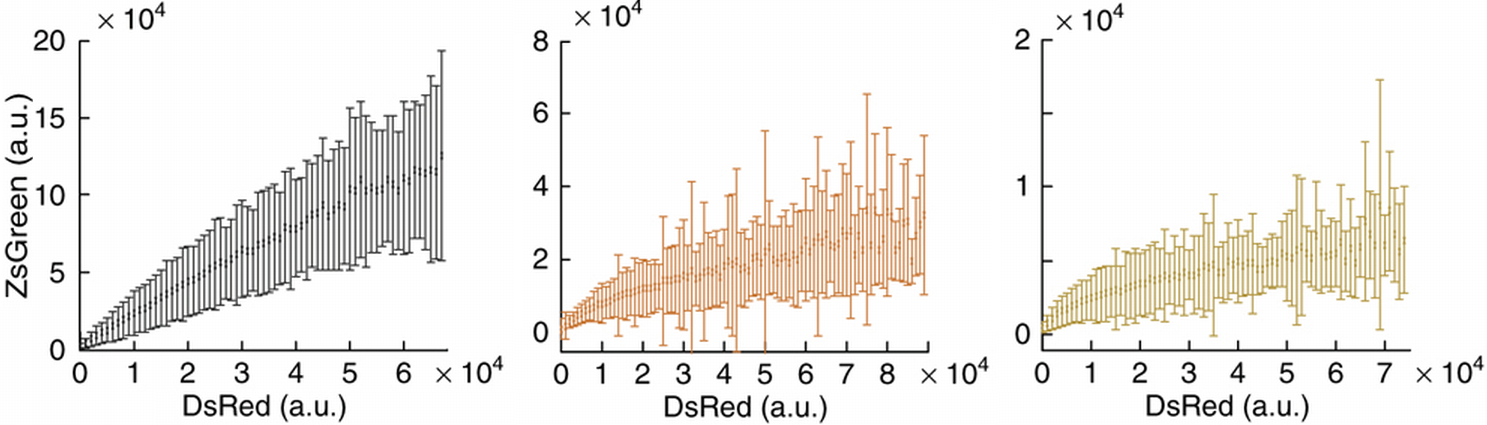}}%
\minp{0.25}{\pic{0.27}{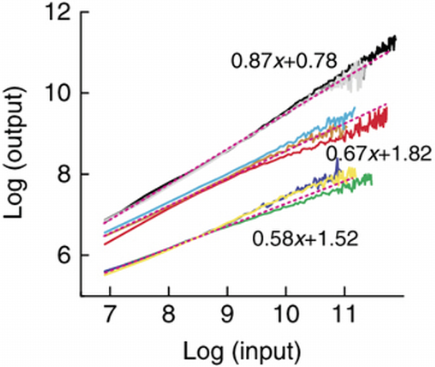}}
\caption{
Experimental characterization of transcriptional autoregulatory motifs
(tAM), adapted from Figure~5 in~\cite{bleris10}, available under a CC-BY-NC 4.0 license.
Flow cytometry measurements were obtained from at least 200{,}000 cells
per condition.
The first three panels correspond to increasing repression strength
(IPTG-interfered LacI binding, single LacO site, and double LacO repeat).
Right: log--log plots of mean output versus input, showing slopes
decreasing from near \(1\) to near \(0.5\) as repression strength
increases.}
\label{fig:negfeedback_experiments}
\efigstar

Several IFF genetic circuits were studied in~\cite{bleris10}. As an
illustrative example, we describe the post-transcriptional type-I
incoherent feedforward loop (ptI1-FFL) circuit from that reference,
shown in Figure~\ref{fig:iffl_dna}. The left panel shows a
phenomenological representation, while the second panel depicts the
interaction structure among the biological components. The two
rightmost panels show the corresponding genetic implementations: the
full circuit, followed by a variant in which the regulator
(``\(A\)'' in the interaction diagram, ``\(x\)'' in the
phenomenological model) does not repress the output (``\(O\)'' or ``\(y\)'').
\bfigstar
\minp{0.05}{\picc{0.25}{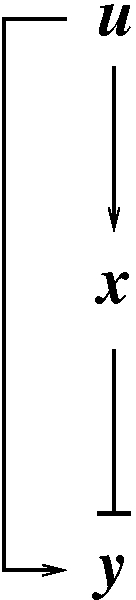}}%
\minp{0.95}{\picc{0.18}{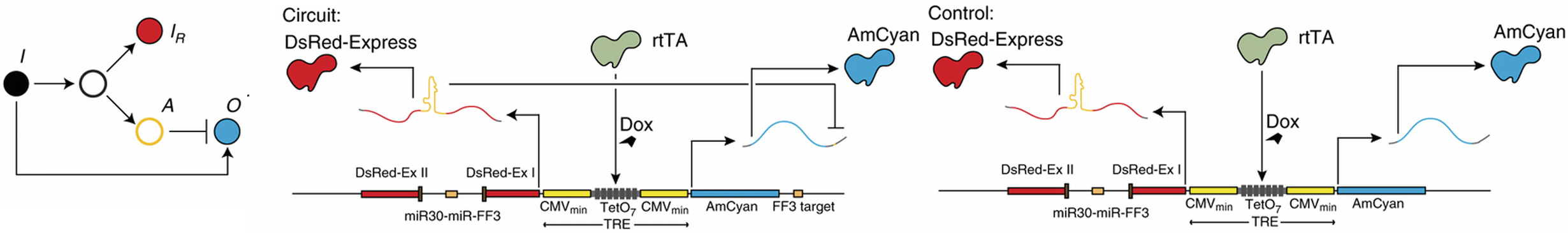}}
\caption{
Post-transcriptional type-I incoherent feedforward (ptI1-FFL)
circuit, adapted from Figure~1B in~\cite{bleris10}, available under a CC-BY-NC 4.0 license.
Left: phenomenological representation.
Middle: interaction diagram of the genetic circuit
(\(I\) = input, \(I_R\) = input reporter, \(A\) = auxiliary regulator,
\(O\) = output).
Right: genetic implementations.
The regulated output protein is AmCyan.
The auxiliary repressor (node \(x\) in the phenomenological diagram) is a
synthetic microRNA, miR-FF3, which targets a complementary RNA sequence
engineered into the 3$'$-UTR of the AmCyan mRNA, thereby repressing
AmCyan expression at the post-transcriptional level.
The microRNA miR-FF3 is co-expressed and processed post-splicing from an
intron inserted between two exons encoding the fluorescent protein
DsRed-Express.
DsRed-Express serves as an internal reporter of plasmid copy number and
is used as a proxy for the input \(u\).
As a biological control to test the necessity of repression, a variant
of the circuit omits the miR-FF3 target site in the AmCyan transcript,
thereby eliminating RNA-interference--mediated repression.
The transcription factor rtTA is present at constant, non-limiting
levels and is not treated as a system input.
}
\label{fig:iffl_dna}
\efigstar

A detailed model is analyzed in~\cite{bleris10}. For physiologically
relevant plasmid copy numbers, however, a suitable minimal
phenomenological description of the conceptual system in
Figure~\ref{fig:iffl_dna} is given by the two-dimensional nonlinear system
\beqn
\LDdiffdotI &=& -\delta x + M u \\
\LDdiffdotH &=& \frac{Nu}{1+Kx} - \varepsilon y,
\eeqn
where the parameters have their natural interpretations.
At steady state,
\[
 y = \frac{Vu}{1+Ku},
\]
where $V:=N/\varepsilon$ and $K:=LM/\delta$.
The parameter \(K\) thus acts as a proxy for the strength of the
repressive regulation of \(x\) on \(y\).
In the unrepressed case (\(K=0\)), the steady-state output \(y\) is
proportional to the input \(u\).
In contrast, in the strongly repressed regime (\(K \gg 1\)), the output
saturates at the constant value \(y \approx V/K\) for all but very small
inputs.
Thus, unlike static negative feedback, the ptI1-FFL architecture
achieves effective adaptation by rendering the steady-state output
insensitive to input magnitude over a wide dynamic range.
These predictions are confirmed experimentally in
Figure~\ref{fig:iffl_experiments}. 
\bfigstar
\picc{0.2}{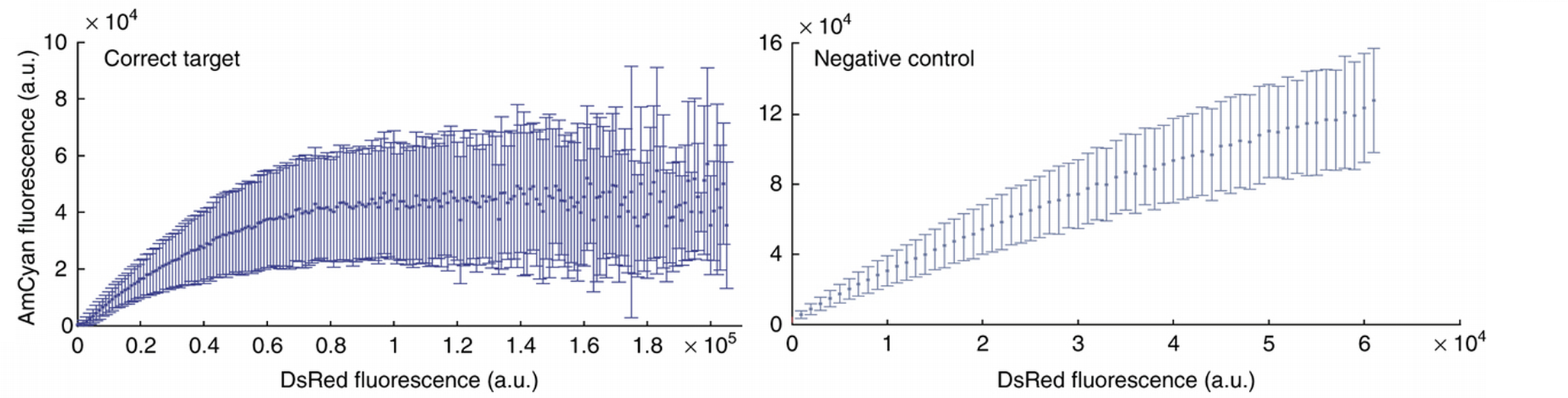}
\caption{Experimental results obtained with ptI1‐FFLI motif.
Figures adapted from Figures 4A and 4B in~\cite{bleris10}, available under a CC-BY-NC 4.0 license.
Left: original circuit with repression, exhibiting saturation of the
output at moderate input levels.
Right: comparison circuit without repression, showing an approximately
linear input--output relationship.
Flow cytometry measurements, including means and standard deviations,
are shown for approximately 200{,}000 cells in each condition.}
\label{fig:iffl_experiments}
\efigstar

\subsection{Distinguishing between adaptation topologies: Response to periodic inputs}

One challenging question in systems biology is that of comparing different
architectures for perfect adaptation.
We concluded that a system which perfectly adapts to step inputs must
contain at least one incoherent feedforward (IFF) or feedback (IFB)
motif. Beyond this, little can be inferred from step-response
adaptation alone, which naturally raises the question: how can one
distinguish between these two possibilities?  One possibility is to
use periodic signals to discriminate between these
models. The paper~\cite{rahi2017} provided the following principle:

\myparagraph{Subharmonics as a signature of IFBs}
Suppose that the input is $T$-periodic.
\bit
\item
For feedforward systems, solutions are eventually periodic with the
same period $T$ (``entrainment'').
\item
For monotone systems, if the solution starting at an initial state
$x(0)$ converges to a periodic solution $z(\cdot)$ of period $kT$, with $k$
a positive integer and $x(0)\preceq z(t)$ for all $t$, then necessarily $k=1$.
\eit 

We do not state a formal theorem, because there are several technical
points to be precise about. For example, feedforward systems are restricted to
those that can be written as a cascade of one-dimensional systems with
appropriate stability properties similarly to the systems analyzed
in~\cite{russo_dibernardo_sontag09}.
For details, see~\cite{rahi2017}
and Sidebar
``Sketch of proofs of entrainment.''

\begin{sidebar}{Sketch of proofs of entrainment}

\myparagraph{Feedforward systems}

Consider first purely feedforward systems, meaning
that the state variables $x_i$, $i=1,\ldots n$,
satisfy the property that if
$x_i$ influences $x_j$, then $x_j$ does not influence $x_i$.
One can always re-label variables in such a way that no element $x_i$
influences any element $x_j$ with $j<i$.
In this form the system has a cascade structure, that is to say
the Jacobian $\left(J_{ij}=\partial f_i/ \partial x_j\right)$ is lower
triangular. 
Let us make the following mild assumptions, which are reasonable for biological systems on trajectories, given an input $u$:
(i) solutions are bounded
($x_i(t) \leq  c_i$ for some $c_i>0$) for all $t$;
(ii) the diagonal elements of $J$ are negative ($\partial f_i/\partial x_i< 0$),
which means biologically that every species is degraded, typically in a
concentration-dependent manner such as
a linear degradation term like $- k_i x_i$ or a Michaelis-Menten
term like $ - k_i x_i^M / (1 + x_i^M/x_{i0}^M)$, where $M$ is a Hill
coefficient;
(iii) the off-diagonal elements of $J$ are bounded, i.e.,
all $\abs{J_{ij}} \leq p_\text{max}$ for some $p_\text{max}>0$.
Similar conditions are imposed in~\cite{russo_dibernardo_sontag09} for
analyzing cascades of enzymatic and gene transcription networks.

Suppose from now on that the input $u$ is $T$-periodic.
Then the system has a unique periodic solution with period $T$ (same as
stimulus), to which every other solution converges.
The proof consists of choosing a diagonal matrix $P$ with $P_{ii} = 1/p^i$, so
as to make the off-diagonal elements of $P J P^{-1}$ arbitrarily close to zero,
as $p \gg p_\text{max}$ .
Then, the matrix measures $\mu_1$, $\mu_2$, or $\mu_\infty$, associated with
the $L^1$, $L^2$, or $L^\infty$-norms, respectively, of $P J P^{-1}$ are all
approximately equal to the largest (i.e. least negative) diagonal element. 
Thus, the system is infinitesimally contracting and, by
results of~\cite{Loh_Slo_98} (see also proofs in
Theorem 2 in \cite{russo_dibernardo_sontag09}
or \cite{sontag10yamamoto}),
we conclude that all $x_i(t)$ are $T$-periodic, as claimed. 

\myparagraph{Cooperative systems}

For systems that are not feedforward, there are entrainment results as well,
as long as all loops are positive.  We now sketch the
case of monotone systems and equilibrium initial states.
We refer the reader to~\cite{rahi2017} for a sketch of proof that
cascades of monotone and feedforward systems preserve entrainment.

An important result for periodically forced monotone systems
$\LDdiffdotX=\vecf(\vecx(t),\vecu(t))$ is given as Theorem 5.26 in
\cite{hirsch_smith_2005}, which credits the unpublished 1997 Ph.D. thesis by
I. T\v{e}\v{s}\'{c}ak.  This result applies to systems that are
irreducible, meaning that all its Jacobian matrices are irreducible (that is,
every variable can indirectly affect every other variable, possibly through an
arbitrary number of intermediates; see also \cite{HirschSmith2003}).  The
result states that, assuming bounded trajectories,  $\vecx(t)$ converges to a solution with period $kT$,
where $k\geq 1$ is an integer, for almost all initial conditions if the
stimulus $\vecu(t)$ is periodic with period $T$
($\vecu(t)=\vecu(t+T)$) and if the system is dissipative (trajectories
are ultimately bounded).  It is important to note that, generally, there
may be stable periodic solutions with period $kT$ and $k>1$, as shown in
\cite{takac_procAMS1992}. Thus, if we are interested in entrainment (global
convergence to period-$T$ trajectories), we need to find additional conditions
which rule out $k>1$.
The simplest condition, which is enough for many applications, covers
systems evolving on the nonnegative orthant (all variables are
nonnegative) and starting from zero initial conditions.
Since it seems difficult to find a reference, we state and prove a simple result.

We assume given a closed, convex, pointed cone $K \subset \mathbb R^n$, and define
the partial order
$
x \le_K y \Leftrightarrow y-x \in K
$.
Let us write $f(t,x)$ for the time-varying vector field $f(x,u(t))$,
in which a periodic input has been specified.
Thus, we more generally consider a nonautonomous system
\begin{equation}\label{eq:system}
\LDdiffdotI = f(t,x),
\end{equation}
where $f:\R \times \R^n \to \R^n$ 
is continuous in $(t,x)$ and locally Lipschitz in $x$,
is $T$-periodic in time: $f(t+T,x)=f(t,x)$,
and the flow $\varphi(t,s,x)$ of~\eqref{eq:system} is monotone with
respect to $\le_K$, i.e.,
\[
x_1 \le_K x_2 \;\Rightarrow\;
\varphi(t,s,x_1) \le_K \varphi(t,s,x_2)
\quad \forall\, t \ge s .
\]

\textbf{Theorem.}
Let $x(t)=\varphi(t,0,x_0)$ be a solution of~\eqref{eq:system} that is
bounded for $t \ge 0$ and satisfies
$
x_0 \le_K x(t) \quad \text{for all } t \ge 0 .
$
Then $x(t)$ converges to a $T$-periodic solution of~\eqref{eq:system}.

We reduce the proof to a discrete-time convergence lemma:

\textbf{Lemma.}
Let $F:\R^n \to \R^n$ be continuous and monotone with
respect to $\le_K$. Suppose that a bounded sequence
$
x_{k+1} = F(x_k), \qquad k=0,1,2, \ldots
$
satisfies
$x_k \le_K x_{k+1}$ for all $k$.
Then $\{x_k\}$ converges to some $\bar x \in \R^n$, and $\bar x$
is a fixed point of $F$.

We prove this next.
Boundedness implies precompactness in $\R^n$, so $\{x_k\}$ has
convergent subsequences. Let
$
x_{k_j} \to a,
$, $
x_{\ell_j} \to b
$
be two convergent subsequences. Since both index sequences tend to
infinity, we may pass to further subsequences (without relabeling) such
that $k_j \le \ell_j$ for all $j$. Monotonicity of the sequence gives
$
x_{k_j} \le_K x_{\ell_j} \quad \text{for all } j.
$
Because $K$ is closed, taking limits yields $a \le_K b$. Reversing the
roles of the subsequences gives $b \le_K a$. Since $K$ is pointed, we
conclude $a=b$. Hence all convergent subsequences have the same limit,
and therefore $x_k \to \bar x$ for some $\bar x$.
Finally, continuity of $F$ and the relation $x_{k+1}=F(x_k)$ imply
\[
\bar x = \lim_{k\to\infty} x_{k+1}
       = \lim_{k\to\infty} F(x_k)
       = F(\bar x),
\]
so $\bar x$ is a fixed point of $F$.
This completes the proof of the lemma.

To prove the theorem, we start by sampling the trajectory.
Define
$
x_k := \varphi(kT,0,x_0)
$
for $k=0,1,2,\dots$, and note that the semigroup property of the flow says that
$
x_{k+1} = \varphi((k+1)T,kT,x_k).
$
Next we define the time-$T$ map (also called the period or
stroboscopic map) associated with the periodic system,
$
F(\xi) := \varphi(T,0,\xi).
$
Since
$
\varphi((k+1)T,kT,\xi) = \varphi(T,0,\xi)
\quad \forall\, \xi \in \R^n,\ \forall\, k,
$
it follows that $x_{k+1}=F(x_k)$ for all $k$.
Monotonicity of the flow implies that $F$ is monotone with respect to
$\le_K$, and continuity of the flow implies that $F$ is continuous. The
assumption $x_0 \le_K x(t)$ for all $t \ge 0$ implies
$
x_k \le_K x_{k+1}
$
for all $k$.
Boundedness of $x(t)$ implies boundedness of $\{x_k\}$.
Thus the lemma can be applied, and we know that
$x_k \to \bar x$ for some $\bar x \in \R^n$, and $F(\bar x)=\bar x$.
Define $\bar x(t):=\varphi(t,0,\bar x)$. Using time periodicity,
$
\varphi(t+T,0,\bar x)
=
\varphi\!\big(t,0,\varphi(T,0,\bar x)\big)
=
\varphi(t,0,\bar x),
$
so $\bar x(t)$ is $T$-periodic.

We must still show convergence of the full trajectory.
Now, for any $t \in [0,T]$,
$
x(kT+t)=\varphi(t,0,x_k)
$.
Continuity of the flow with respect to initial conditions and
$x_k \to \bar x$ thus imply
\[
x(kT+t) \to \bar x(t)
\quad \text{as } k \to \infty.
\]
Hence $x(t)$ converges to the $T$-periodic solution $\bar x(t)$.

We remark that periodicity of the vector field is essential for
identifying the discrete dynamics with iterations of a single map
$F$. Without periodicity, the step map $\varphi((k+1)T,kT,\cdot)$
would depend on $k$, and the Lemma would not apply.
\end{sidebar}

The above principle is often summarized as follows: if a system adapts to
constant inputs, but exhibits a \textit{subharmonic response}, namely,
a response with a frequency that is a submultiple of the forcing
frequency, then its underlying network must contain a negative-feedback
mechanism.
Observe that \textit{subharmonic responses are a strictly nonlinear phenomenon},
since linear (stable) systems always entrain.
In fact, entrainment is always guaranteed for \emph{contractive}
systems (in particular, stable linear systems are always contractive
in an appropriate weighted $L^2$ norm), as originally shown in
the work of Lohmiller and Slotine~\cite{Loh_Slo_98}; see also
\cite{russo_dibernardo_sontag09,sontag10yamamoto,margaliot_sontag_tuller_cast_2015}
for generalizations.

We conclude that non-entrainment to inputs is one ``signature'' of a
nonlinear negative feedback loop in the system.
In other words, \textit{testing with periodic inputs} provides a means
to distinguish IFF motifs from incoherent feedback (IFB) architectures,
or, more precisely, to identify situations in which some negative
feedback must be present.
It is well known in classical dynamical systems theory that negative
feedback can generate subharmonics, period-doubling bifurcations, and
even chaotic dynamics, as in the forced van der Pol oscillator.
What is perhaps surprising is that analogous phenomena can arise even
for very simple periodically forced biochemical networks~\cite{chaos2017}.
In fact, one can even give examples of systems with the property that
when the external input is constant, all solutions
converge to a steady state, yet with the input $u(t)= \sin t$ solutions
become chaotic~\cite{arxiv_chaos_2009}.
It is worth noting that monotone systems (at least under
a natural irreducibility assumption) may lead to subharmonics, as established by T\v{e}\v{s}\'{c}ak's Theorem that, when forced by
$T$-periodic inputs, solutions will generically converge to
$kT$-periodic solutions with $k$ possibly $>1$.
The additional condition that $x(0)\preceq z(t)$, which is satisfied if,
for example, we start at rest ($x(0)=0$) and solutions are positive for
$t>0$, rules out such subharmonic solutions.

As an application of this principle, consider the following experiment.
The worm \emph{C. elegans} can locate odor sources across a 100,000-fold
concentration range, and various sensory and interneurons participate in the
recognition pathway, see Fig.~\ref{fig:celegans_neurons}.
In~\cite{rahi2017} , we discussed how the above theorems can be used to rule
out IFF motifs as the sole mechanism responsible for adaptation in a local circuit controlling the
AWA sensory neuron, leading to the postulation of a negative feedback model.
This is because, at high input frequencies, one does not obtain
entrainment, see Fig.~\ref{fig:celegans_neurons} (top right panel and bottom).
\bfigstar
\minp{0.12}{\picc{0.18}{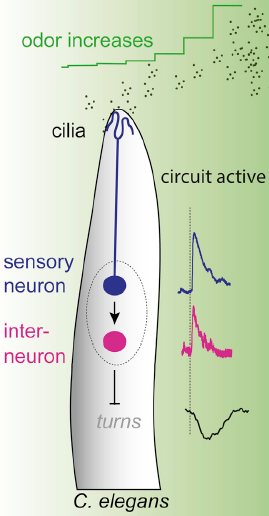}}%
\minp{0.24}{\picc{0.42}{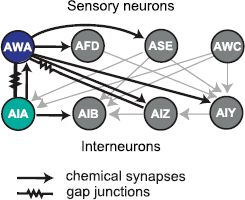}}%
\minp{0.18}{\picc{0.08}{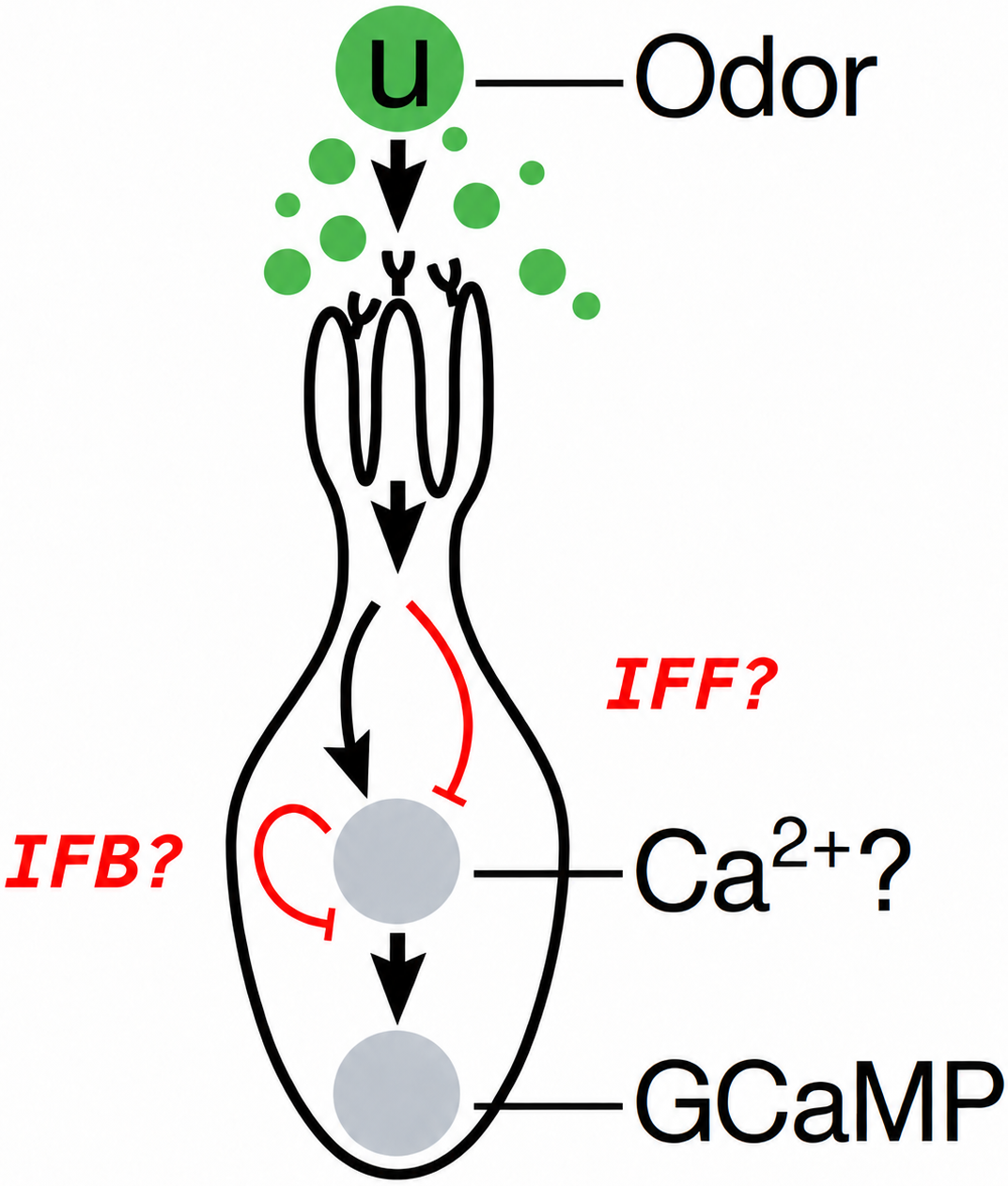}}%
\minp{0.22}{\picc{0.075}{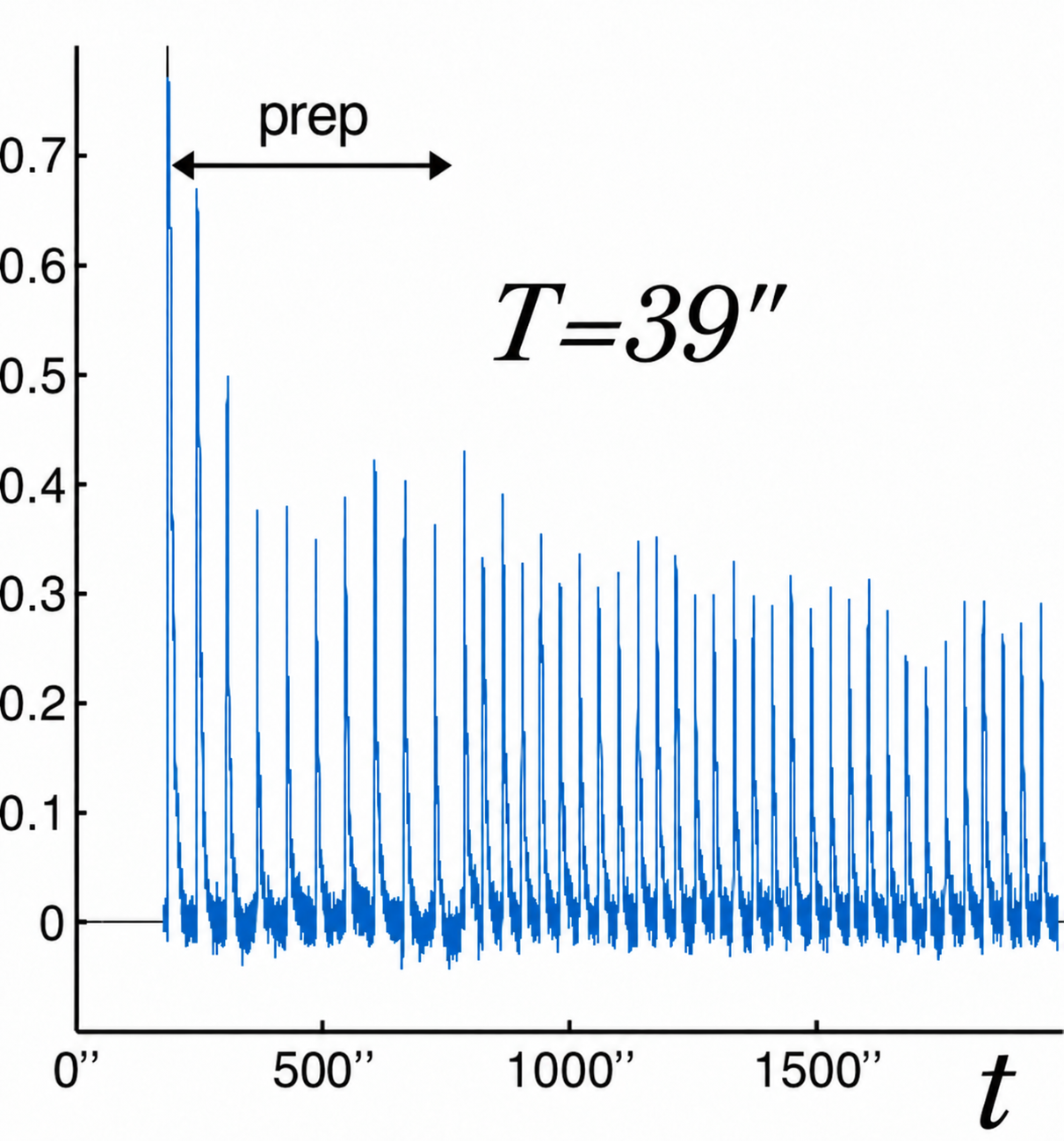}}%
\minp{0.22}{\picc{0.075}{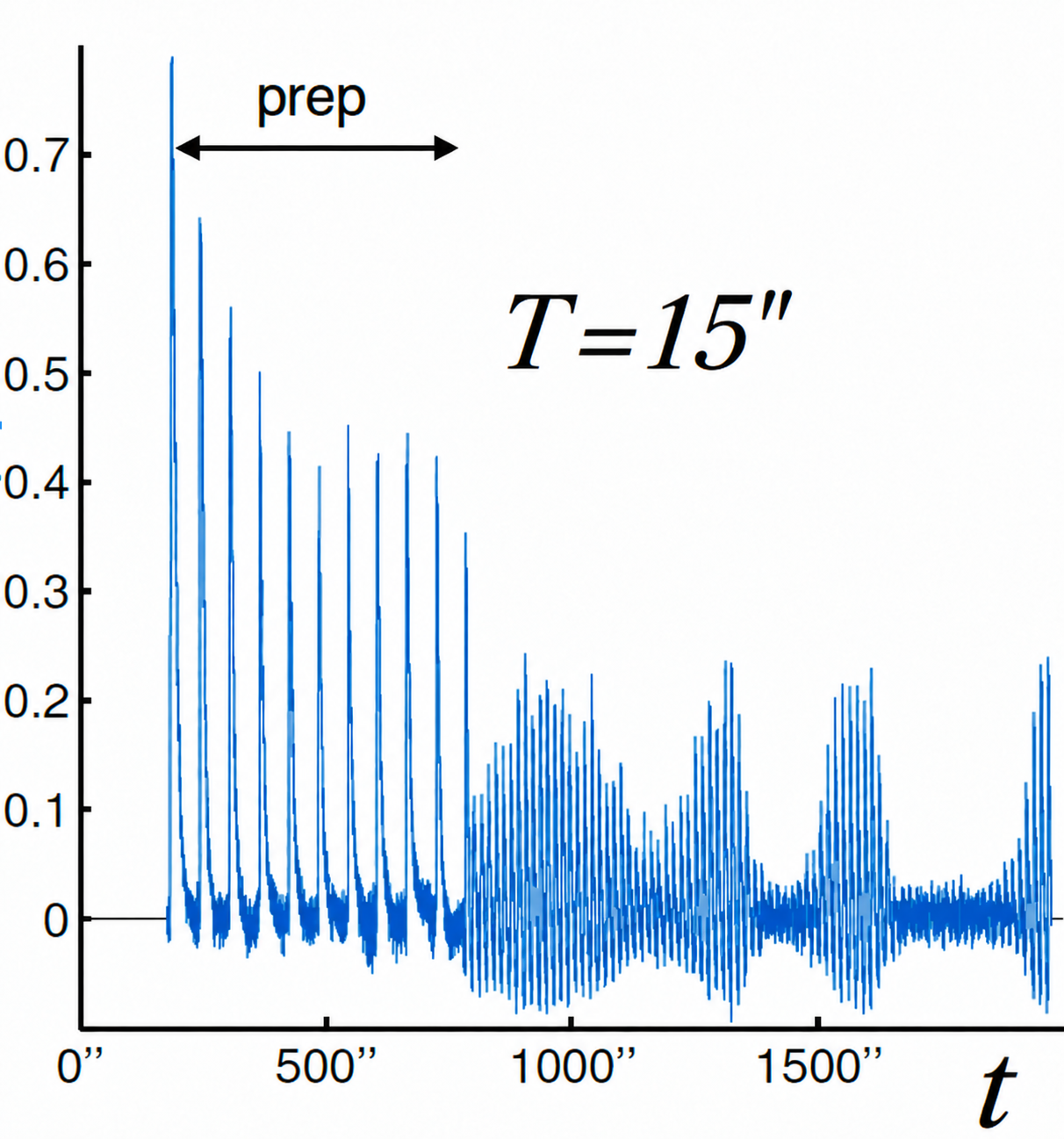}}

\picc{0.5}{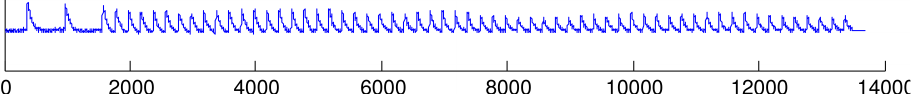}

\caption{Odor sensing in \emph{C.\ elegans} and sensory/interneurons.
Top, left to right: Location of sensory and interneurons, network
connections, and possible IFB or IFF mechanisms,
and testing for odor-evoked intracellular Ca$^{2+}$ response signature
via periodic on-off pulses of diacetyl. Population measurement shows
entrainment at 
low frequency ($1/39$Hz, period $T=39s$), but a higher
frequency ($1/15$Hz, period $T=15$s) results in subharmonic behavior.
Bottom: a sample single-neuron recording with higher frequency ($1/20$Hz, period
$T=20$s) gives subharmonic response (frequency $\sim 1/200$Hz, period $\sim200$s, suggesting that IFFs (or
positive feedback systems) cannot be the reason for the observed behavior.
First two figures adapted from~\protect{\cite{Larsch2015}}, available under a
CC-BY-NC-ND 4.0 license, and three plots done using data in~\protect{\cite{rahi2017}}.
}
\label{fig:celegans_neurons}
\efigstar

\section{Weber's Law, scale invariance, and log sensing}

\begin{wrapfigure}{r}{0.2\textwidth}
\ns
  \pic{0.2}{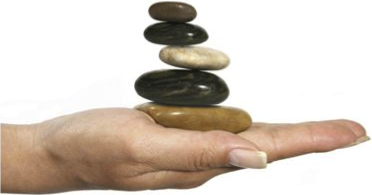}
\caption{Sensing.}
\label{fig:stones}
\vspace{-8pt}
\end{wrapfigure}

In the 1840s, Ernst Weber, a pioneer of psychophysics,
investigated the relationship between the physical magnitude of a
stimulus and its perceived intensity. He observed that the
\textit{just noticeable difference} (JND) in a stimulus variable $u$ is
not an absolute quantity but scales proportionally with the stimulus
itself, so that perception is sensitive to fractional changes
$\Delta u/u$. This empirical observation is now known as
\textit{Weber's law}.
Gustav Fechner subsequently proposed a mathematical interpretation of
Weber's law, arguing that perceived sensation is proportional to the
logarithm of stimulus intensity. This led to the classical
\textit{Weber--Fechner law}, which may be expressed as a logarithmic
relationship between physical and perceived quantities. In effect,
\textit{sensory systems respond to changes in $\log u$ rather than to changes
in $u$ itself (``logarithmic sensing'')}.
Typical Weber fractions reported in psychophysical experiments include:
$\approx \fractxt{1}{40}$ for lifted weight discrimination,
$\approx \fractxt{1}{10}$ for pitch perception,
$\approx \fractxt{1}{16}$ for light intensity,
$\approx \fractxt{1}{4}$ for odor intensity,
$\approx \fractxt{1}{30}$ for pain perception, and
$\approx \fractxt{1}{3}$ for taste.

Another way to express sensitivity to changes in $\log u$ is to view it
as an invariance with respect to multiplicative scaling of the input by
positive constants. Consider an input that switches from a value $u$
to a value $v$, and the system’s response to this change. Now suppose
that both values are multiplied by a factor $p>0$, so that the input
changes from $pu$ to $pv$. The corresponding change in the logarithmic
scale is
\[
\log(pv)-\log(pu)=\log v-\log u,
\]
which is identical to the original increment. In other words, sensing
on a logarithmic scale makes the response independent of the absolute
magnitude of the stimulus and sensitive only to relative changes.
This property is therefore referred to as \textit{scale-invariant}
sensing. An equivalent formulation is that the system is able to
detect at most
\emph{fold changes} of the input, since $pv/pu=v/u$. For this reason, the
phenomenon is also known as \textit{fold-change detection} (FCD).
Scale-invariant sensing is useful because it renders system responses
independent of measurement units and provides robustness to
multiplicative uncertainties in stimulus intensity.
Figure~\ref{fig:robustness}(left) illustrates a series connection in
which an ``upstream'' system is subject to multiplicative uncertainty and a
``downstream'' system has a response that is invariant to scale,
\bfigstar
\minp{0.7}{%
\setlength{\unitlength}{2000sp}%
\begin{picture}(9924,981)(1200,-3673)
\thicklines
\put(1396,-3211){\vector( 1, 0){1700}}
\put(3151,-2761){\line( 0,-1){900}}
\put(3151,-3661){\line( 1, 0){2250}}
\put(5401,-3661){\line( 0, 1){900}}
\put(5401,-2761){\line(-1, 0){2250}}
\put(5446,-3211){\vector( 1, 0){1700}}
\put(7201,-2761){\line( 0,-1){900}}
\put(7201,-3661){\line( 1, 0){2250}}
\put(9451,-3661){\line( 0, 1){900}}
\put(9451,-2761){\line(-1, 0){2250}}
\put(9496,-3211){\vector( 1, 0){1800}}
\put(3501,-3071){upstream}%
\put(3301,-3451){system \red{($p$?)}}%
\put(7431,-3071){downstream}%
\put(7761,-3451){system}
\put(1621,-2941){external}%
\put(1621,-3571){stimulus}%
\put(5861,-2941){$u(t)$}%
\put(9751,-2951){output robust}%
\put(9721,-3571){to \red{$p$} uncertainty?}%
\end{picture}
}%
\minp{0.3}{\pic{0.35}{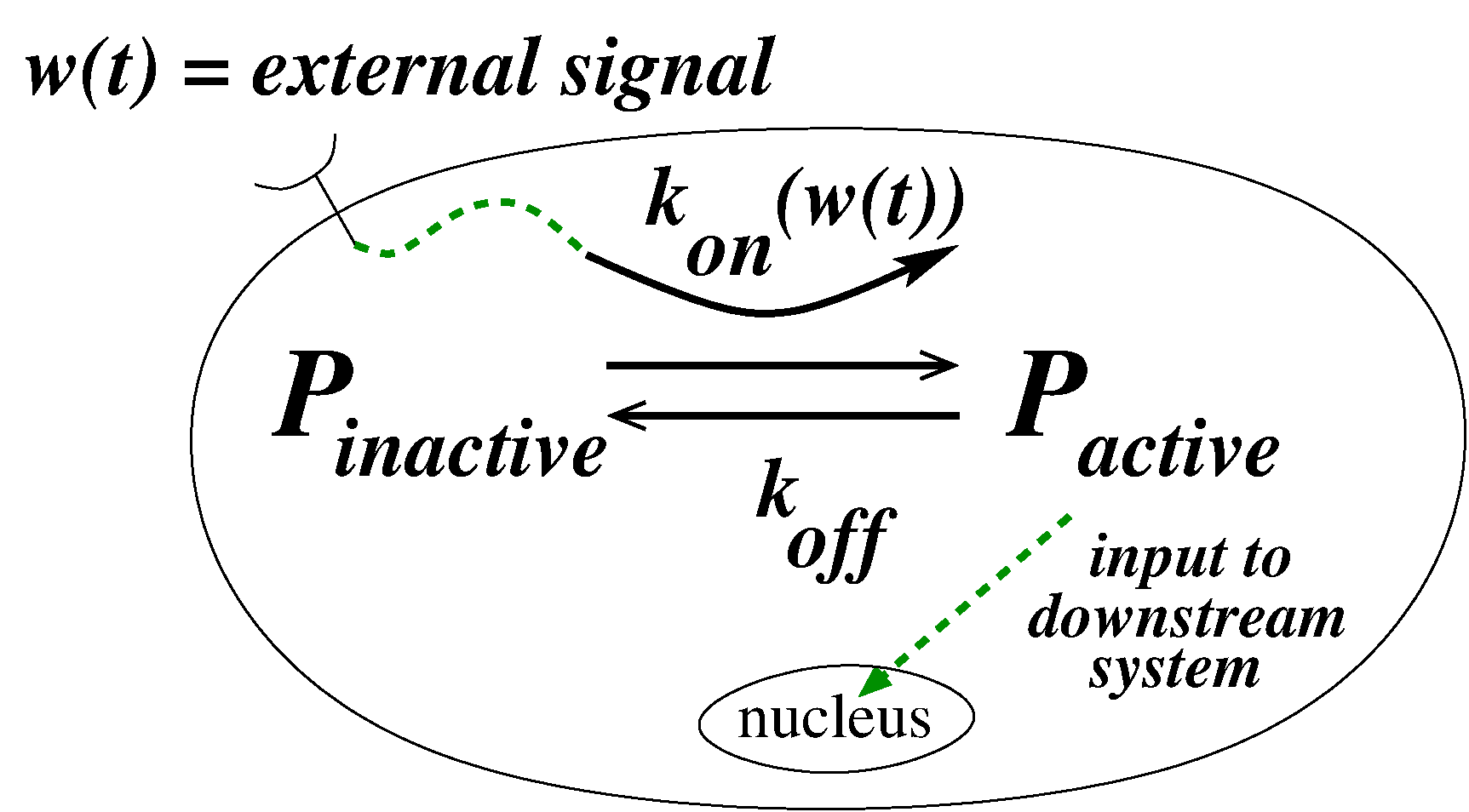}}
\caption{Robustness to multiplicative uncertainty}
\label{fig:robustness}
\efigstar

A typical example from cell signaling is the following. Suppose that a
signaling protein can switch between an inactive form
$P_{\textit{inactive}}$ and an active (for example, phosphorylated)
form $P_{\textit{active}}$. The rate of activation, mediated by kinase
action, is denoted
\[
a(t)=k_{\textit{on}}\!\big(w(t)\big),
\]
where $w(t)$ is an external signal. Inactivation occurs with constant
rate $k_{\textit{off}}$, independent of $w$.
See Figure~\ref{fig:robustness} (right).
Assume that the total protein concentration is conserved,
\[
P_{\textit{inactive}}(t)+P_{\textit{active}}(t)=p=P_T .
\]
Let
\[
u(t)=P_{\textit{active}}(t),
\qquad
P_{\textit{inactive}}(t)=p-u(t).
\]
Then the dynamics of the active form are described by the linear,
time-varying system
\[
\LDdiffdotL(t)=a(t)\big(p-u(t)\big)-k_{\textit{off}}u(t), 
\qquad u(0)=0 .
\]
This may be viewed as a linear system with constant input $p$:
\[
\LDdiffdotL(t)= -\big(k_{\textit{off}}+a(t)\big)u(t)+a(t)p .
\]
Therefore, the solution $u(t)$ scales proportionally with the total
protein concentration $p$.
An equivalent way to express this proportionality is in terms of
fold changes. Fix two times $t_1$ and $t_2$ and consider the ratio
$r:=u(t_2)/u(t_1)$. If the total protein level is changed from $p$ to
$\hat p$, and $\hat u(t)$ denotes the corresponding solution of the
ODE, then
\[
\hat r := \frac{\hat u(t_2)}{\hat u(t_1)} = r .
\]
Thus, this model predicts invariance of fold changes with respect to
scaling of $p$. If the initial condition is nonzero, this conclusion
requires that $\hat u(0)$ be scaled in proportion to $p$, as would
naturally occur when trajectories are initialized at steady state.

The active protein, which typically functions as a transcription
factor or signaling intermediate, often serves as the input to a
downstream genetic or signaling network. We conclude that the
downstream response to $w(t)$ will be robust to uncertainty in the
upstream protein abundance $p=P_T$ precisely when the downstream
network exhibits \emph{scale-invariant} behavior.
In practice, total protein concentrations such as $p$ often vary
substantially from cell to cell. Hence, scale invariance of the
downstream network is required in order for the overall system
response to remain independent of these fluctuations.
We summarize this informally:

\myparagraph{Robustness to multiplicative uncertainty as a signature of FCD}
If a downstream genetic or signaling system exhibits consistent
responses to a given external excitation despite variations in
upstream protein abundances, then the system likely possesses the
scale-invariance property, also known as fold-change detection (FCD).

Recent interest in fold-change detection (FCD) in signaling pathways was
largely stimulated by two influential studies published in
2009~\cite{Goentoro2009,Cellina2009}. These works reported that signaling
outputs may be insensitive to the absolute concentrations of pathway
components and instead respond primarily to \emph{relative (fold)
changes} in key signaling molecules. Such behavior was demonstrated
experimentally in the canonical Wnt/\(\beta\)-catenin
pathway~\cite{Goentoro2009} and in the EGF--ERK
pathway~\cite{Cellina2009}.
See Figure~\ref{fig:wnt}(left) for the Wnt pathway.
Both pathways are highly conserved in
eukaryotes and play essential roles in embryonic patterning, stem-cell
homeostasis, cell proliferation, and tissue maintenance. Notably,
dysregulation of Wnt signaling is strongly linked to multiple cancers,
particularly colorectal carcinoma.

In the Wnt study of Goentoro and Kirschner~\cite{Goentoro2009}, the authors
examined how cellular responses depend on nuclear \(\beta\)-catenin
(which we think of as $u=P_{\textit{active}}$),
the transcriptional effector of canonical Wnt signaling. Under Wnt
stimulation, \(\beta\)-catenin is stabilized through inhibition of the
destruction complex, enabling its accumulation and activation of
target-gene transcription. To vary absolute protein abundance
($p=P_{\textit{T}}$), they
applied independent perturbations, including pharmacological inhibition
of GSK3\(\beta\) using BIO or lithium chloride, RNA-mediated
overexpression of \(\beta\)-catenin, and manipulation of the
degradation machinery via overexpression of the scaffold protein
Axin1. These experiments were conducted in \emph{Xenopus} embryos and in
cultured RKO colorectal carcinoma cells, with pathway output assessed
through developmental phenotyping and quantitative RT-PCR measurements
of canonical Wnt target genes including \emph{siamois} and \emph{Xnr3}.

Across an approximate four-fold range of basal \(\beta\)-catenin
perturbations, the authors observed substantial variation in
\emph{absolute} protein levels, while the Wnt-induced
\emph{fold-change} in nuclear \(\beta\)-catenin remained nearly
constant. Correspondingly, downstream gene expression and developmental
phenotypes remained indistinguishable from wild type whenever fold
changes were preserved. For example, embryos treated with lithium or
injected with \(\beta\)-catenin RNA exhibited roughly two-fold increases
in absolute protein concentration, yet expression of
\emph{siamois} and \emph{Xnr3} remained at wild-type levels. In contrast,
perturbations that altered the magnitude of the Wnt-induced fold change,
such as strong modulation of Axin1 expression affecting
\(\beta\)-catenin degradation rates, led to aberrant signaling outputs
and abnormal phenotypes.
See Figure~\ref{fig:wnt}(right).
These findings provided compelling experimental
support for fold-change detection as a governing principle of Wnt
signal transduction.
\bfigstar
\minp{0.4}{%
\picc{0.55}{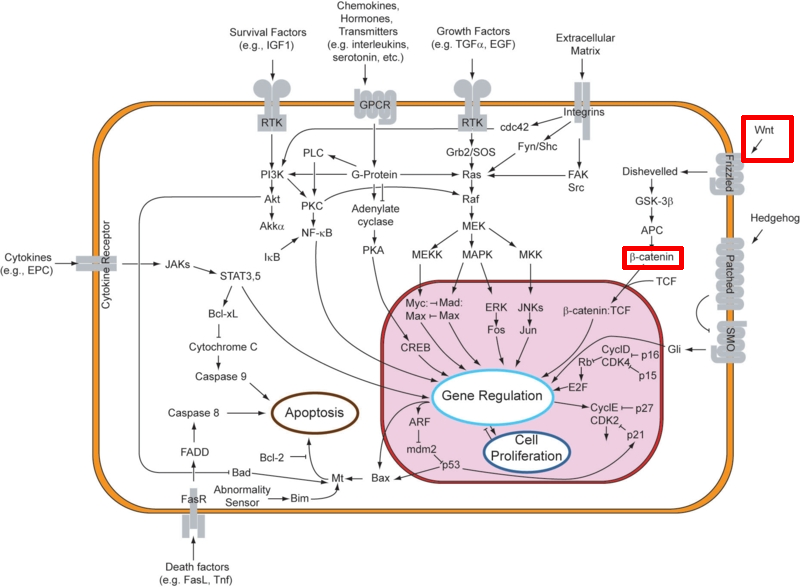}
}%
\minp{0.6}{\picc{0.35}{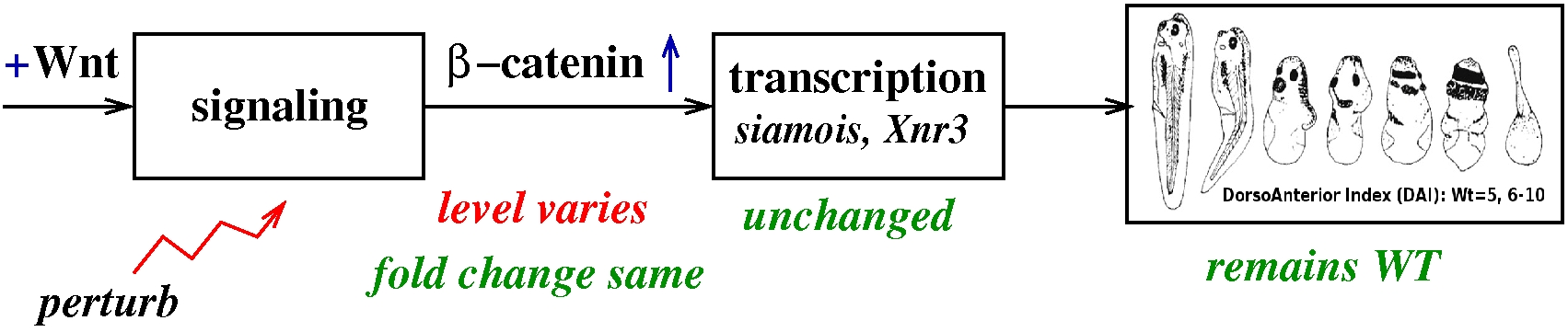}}
\caption{Left figure shows Wnt signaling pathway, adapted from
image file ``Signal transduction pathways.png'' (available in Public Domain) in
Wikipedia entry ``Paracrine Signaling.''
Red rectangles show Wnt
external ligand and $\beta$-catenin (whose concentration is highly
variable)
Right figure summarizes genetic and phenotypical observations in Goentoro and Kirschner's experiments.
}
\label{fig:wnt}
\efigstar


As illustrated in Figure~\ref{fig:steps}, perfect adaptation requires
only that the system output eventually return to its pre-stimulus value
after a change in input. The stronger property of
\emph{fold-change detection} (FCD) demands more: the entire temporal
response, including both amplitude and timing, must be preserved
whenever two inputs differ by a multiplicative constant. Thus, responses
depend only on relative changes in the input, not on its absolute
magnitude.
This distinction is conveniently illustrated using a sequence of step
inputs that transition from a baseline value $u_i$ to a new level
$u_{i+1}$. We assume that successive steps are sufficiently separated in
time so that the system has effectively reached steady state before
each new step occurs. Under perfect adaptation, all responses eventually
settle to the same steady value, but the transient dynamics may vary with
input magnitude. Under FCD, by contrast, the transient responses are
identical up to a time shift, reflecting invariance under input
scaling.
\bfigstar
\minp{0.20}{\picc{0.12}{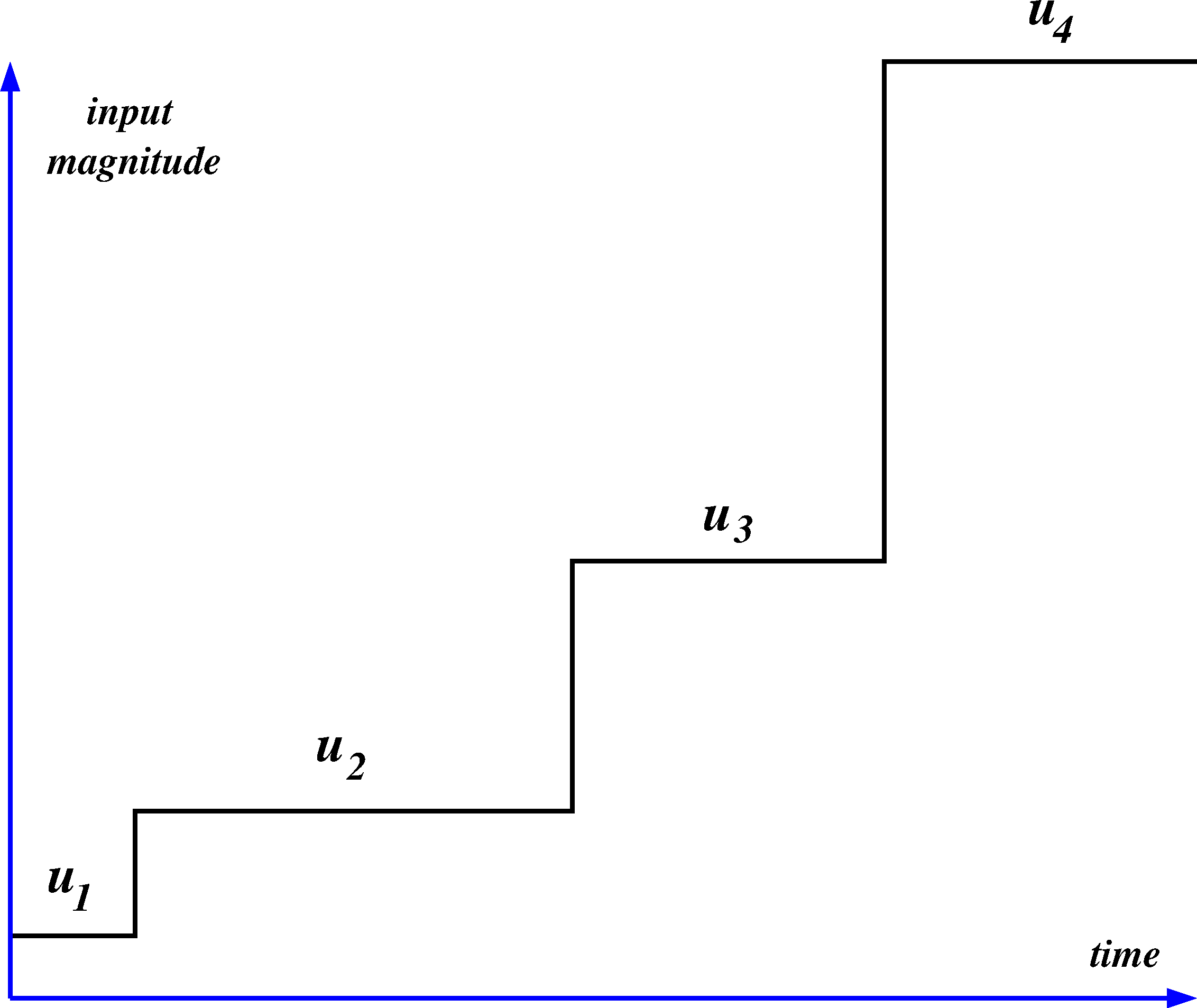}}%
\minp{0.30}{\picc{0.224}{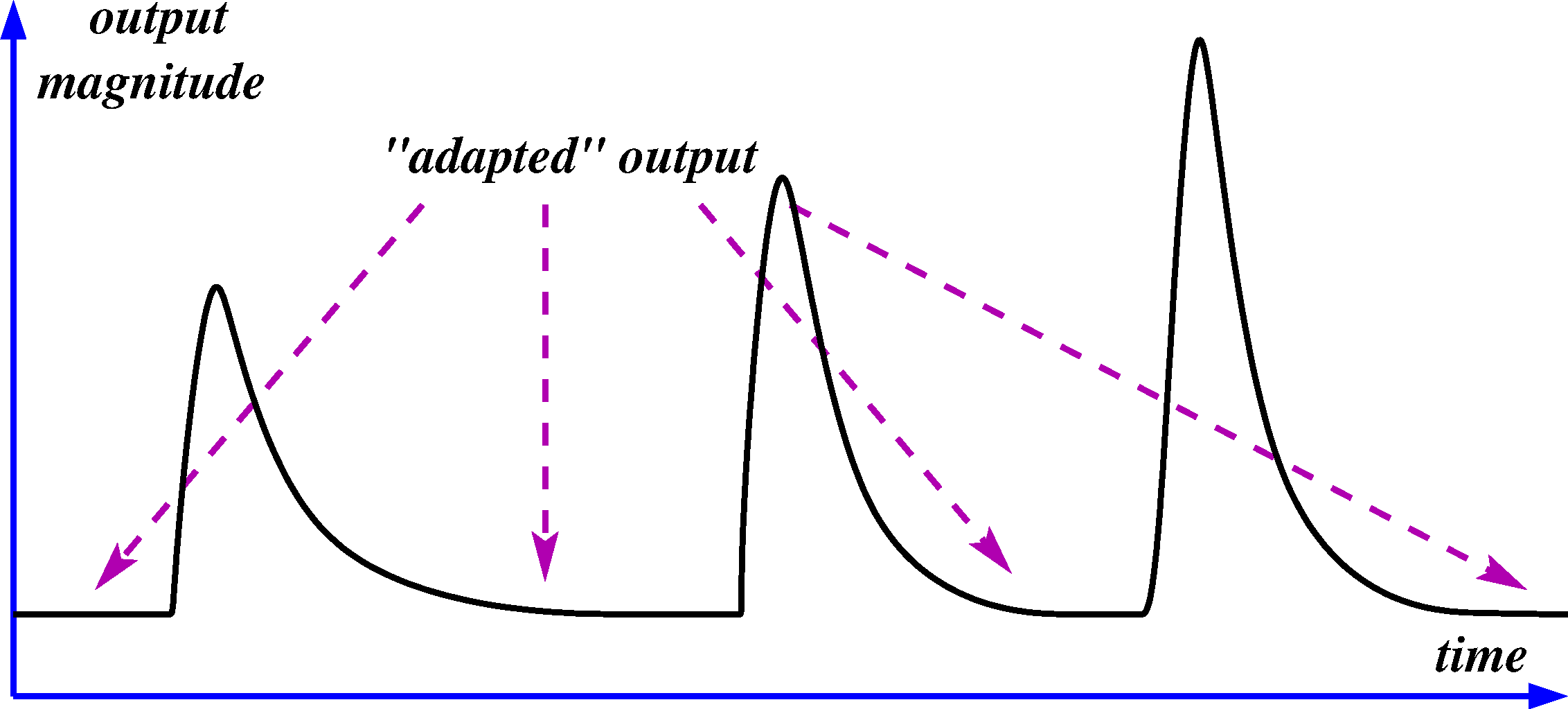}}%
\minp{0.30}{\picc{0.224}{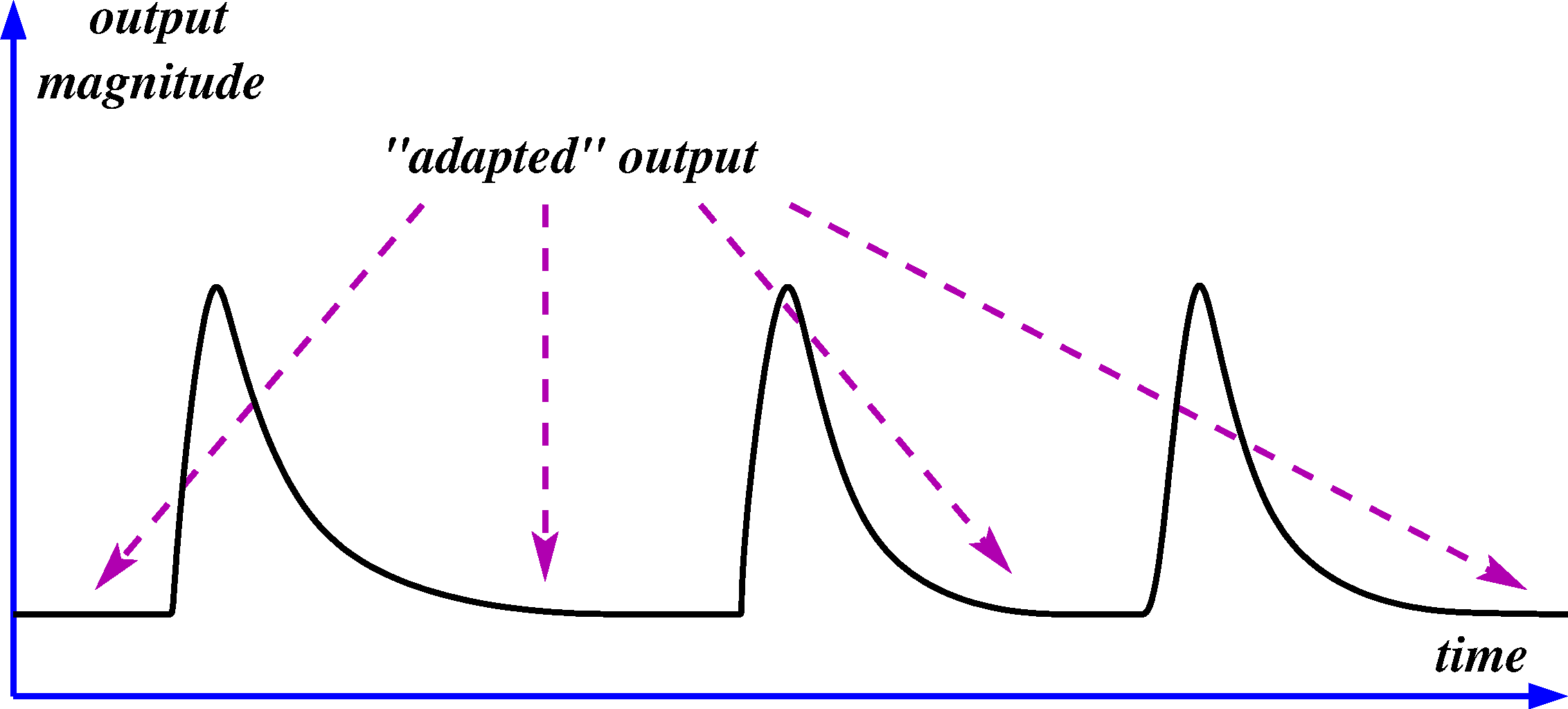}}%
\minp{0.20}{\picc{0.12}{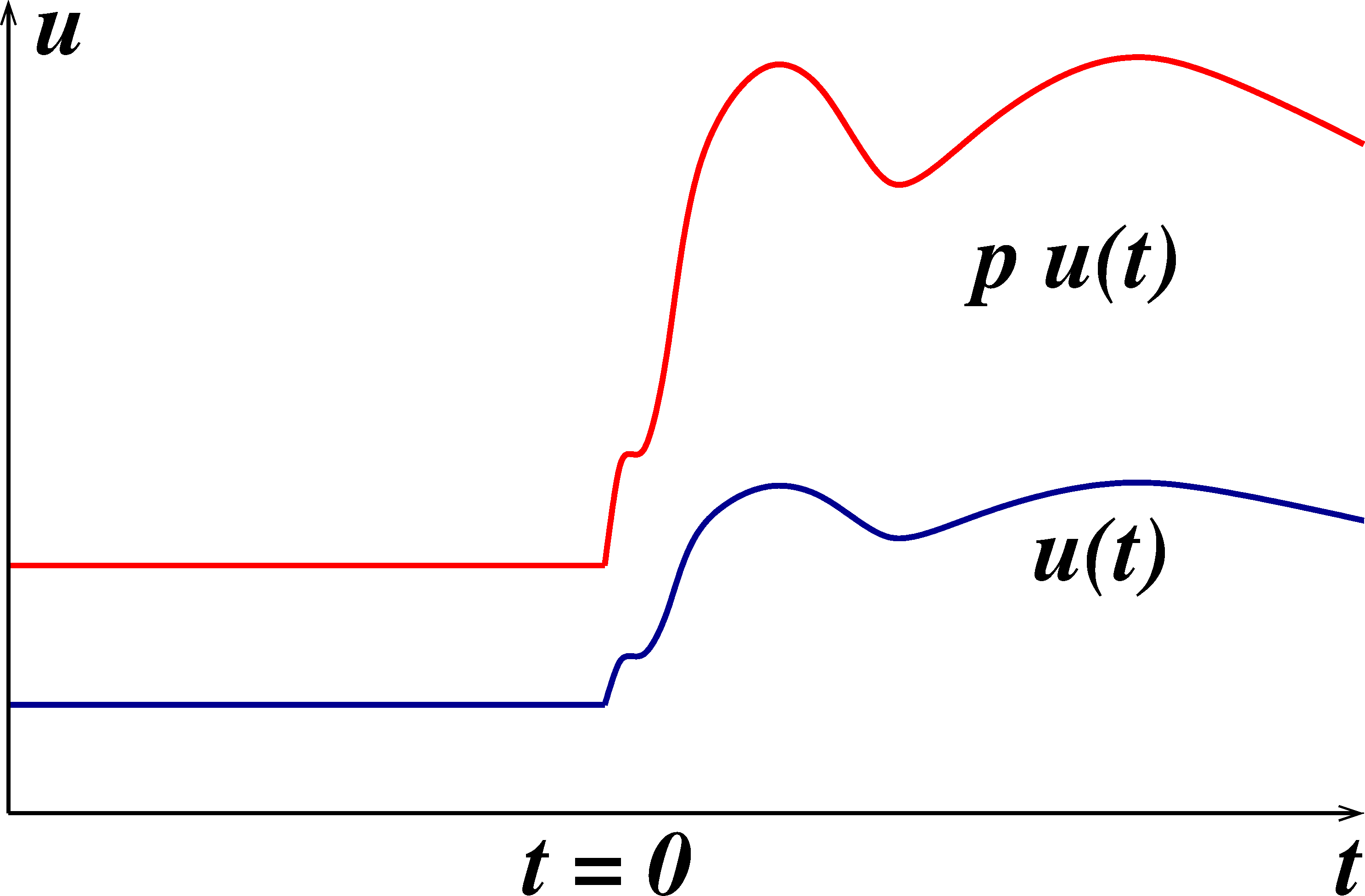}}
\caption{Left to Right: Step-wise input with equal fold-changes $\frac{u_{i+1}}{u_i}$.
Perfect adaptation requires that the response (eventually) returns to
pre-stimulus value, but the transient response amplitude and timing
generally depends on input magnitudes (for example, it is proportional
to magnitude, for any linear system).
Scale-invariance means that the response is exactly the same in
amplitude and timing. In other words, at best only fold-change can be
detected.
More generally, we study responses to non-step functions: response
should be the same for $u$ and $pu$.
}
\label{fig:steps}
\efigstar

\begin{sidebar}{Formalizing scale-invariance}

We now formalize this notion of scale invariance. The main references for
this material are~\cite{shoval10} and~\cite{shoval_alon_sontag_2011}.
More generally, the framework concerns invariance under the action of a
group of symmetries acting on inputs; FCD corresponds to the special case
in which the symmetry group is the multiplicative group of positive real
numbers.

We consider systems of the form
\[
\LDdiffdotI = f(x,u), \qquad y = h(x,u),
\]
where $x \in \X \subset \mathbb R^n$, $u \in \U$, and $y \in \Y$. We
assume that for each constant input $u(t)\equiv \baru$, there exists a
unique steady state $\barx=\sigma(\baru)$ satisfying
\[
f(\barx,\baru)=0.
\]
In many contexts one further assumes that this equilibrium is globally
asymptotically stable (GAS) for the system with constant input $\baru$,
so that
\[
\lim_{t\to\infty}\varphi(t,\xi,u)=\sigma(\baru)
\quad\text{for all initial conditions } \xi\in\X.
\]
Although this assumption is often natural in applications, it is not
required for the results presented here.

If $\X$ is an open subset of $\mathbb R^n$ (or the closure of such a
set), we say that the system is \emph{analytic} if the functions $f$ and
$h$ are real-analytic in $x$. The system is said to be
\emph{irreducible} if it is both accessible and observable.
Accessibility means that the accessibility rank condition holds,
\[
\mathcal L\mathcal A_{\mathcal F}(x)=\mathbb R^n
\quad \text{for all } x\in\X,
\]
so that no conservation laws restrict the dynamics to lower-dimensional
invariant manifolds. For analytic systems, this is equivalent to the
property that the set of states reachable from any given initial
condition has nonempty interior; see~\cite{mct} for details.
Observability means that no two distinct states produce identical
outputs for all possible inputs, that is,
\[
\psi(t,x,u)=\psi(t,\tilde x,u)\ \forall\,t,u
\quad\Rightarrow\quad x=\tilde x.
\]
For analytic input-affine systems, observability is equivalent to
separability of distinct states by the observation space; see~\cite{mct}.
In biomolecular applications, analyticity and irreducibility are
typically mild technical assumptions and are often satisfied.

We say that a system \emph{perfectly adapts to constant inputs} if the
steady-state output
\[
h(\sigma(\baru),\baru)
\]
is equal to a fixed value $y_0\in\Y$, independent of the particular
constant input value $\baru$. Thus, perfect adaptation requires
invariance of the steady-state output under constant inputs, but places
no restriction on transient behavior.

To define scale invariance, let $\mathcal P$ be a family of continuous,
onto transformations $\pi:\U\to\U$. For any input $u(t)$ and any
$\pi\in\mathcal P$, we write $\pi u$ for the time-dependent input
defined by $(\pi u)(t)=\pi(u(t))$. The continuity assumption ensures that
piecewise continuous inputs remain so after transformation. Ontoness is
assumed mainly to preserve irreducibility; weaker conditions are often
sufficient in practice.
The prototypical example is \emph{scale invariance}, where
$\U=\mathbb R_{>0}$ and
\[
\mathcal P=\{\,u\mapsto pu \mid p>0\,\}.
\]
In this case, invariance under $\mathcal P$ implies that responses
depend only on fold changes $u_{i+1}/u_i$, hence the alternative name
\emph{fold-change detection}.

We say that a system is \emph{$\mathcal P$-invariant} if, for all
$t\ge0$, all constant inputs $\baru$, all admissible inputs $u$, and all
transformations $\pi\in\mathcal P$,
\begin{equation}\label{eq:defFCD}
\psi\bigl(t,\sigma(\baru),u\bigr)
=
\psi\bigl(t,\sigma(\pi\baru),\pi u\bigr).
\end{equation}
In words, starting from the appropriate steady state, the response to an
input $u$ is identical to the response to the transformed input $\pi
u$.

If the action of $\mathcal P$ on $\U$ is transitive, meaning that any
two constant inputs can be mapped to one another by some
$\pi\in\mathcal P$, then $\mathcal P$-invariance implies perfect
adaptation. Indeed, the outputs in~\eqref{eq:defFCD} must coincide at
time $t=0$, and transitivity allows any constant input to be related to
any other.

Finally, we introduce the notion of equivariance. A family of
differentiable maps
\[
\{\rho_\pi:\X\to\X\}_{\pi\in\mathcal P}
\]
is called a \emph{$\mathcal P$-equivariance family} if the coordinate
change $x\mapsto\rho_\pi(x)$ maps solutions of the original system with
input $u$ to solutions of the transformed system with input $\pi u$,
while preserving outputs:
\begin{center}
\setlength{\unitlength}{1500sp}%
\begin{picture}(5385,2595)(3046,-4216)
\thicklines
\put(3601,-1861){\vector( 1, 0){2250}}
\put(3151,-2311){\vector( 0,-1){1350}}
\put(6301,-2311){\vector( 0,-1){1350}}
\put(3601,-4111){\vector( 1, 0){2250}}
\put(6751,-2086){\vector( 2,-1){1350}}
\put(6751,-3886){\vector( 2, 1){1350}}
\put(2906,-1906){$X$}
\put(2906,-4156){$X$}
\put(6056,-1906){$X$}
\put(6011,-4201){$X$}
\put(8416,-3031){$Y$}
\put(4276,-1726){$x\cdot u$}
\put(4231,-3931){$x\cdot \pi u$}
\put(3286,-2941){$\rho _{\pi }$}
\put(6436,-2986){$\rho _{\pi }$}
\put(7426,-2311){$h$}
\put(7426,-3951){$h$}
\end{picture}
\end{center}
where ``$x\cdot u$'' means the state that results from initial state
$x$ and a finite-length input $y$.
Equivalently, for all $x\in\X$, $u\in\U$, and $\pi\in\mathcal P$,
\begin{align*}
f(\rho_\pi(x),\pi u) &= (\rho_\pi)_*(x)\,f(x,u),\\
h(\rho_\pi(x),\pi u) &= h(x,u),
\end{align*}
where $(\rho_\pi)_*$ denotes the Jacobian of $\rho_\pi$. These relations
form a system of first-order quasilinear partial differential equations
that may, in principle, be solved by the method of characteristics.
Importantly, verification of equivariance does not require explicit
computation of system trajectories.

The main result of~\cite{shoval_alon_sontag_2011} provides a complete
characterization of scale invariance, a necessary and
sufficient ``certificate'' as follows.

\medskip
\noindent\textbf{Theorem.}
\emph{An analytic and irreducible system is $\mathcal P$-invariant if
and only if there exists a $\mathcal P$-equivariance family.}
\end{sidebar}

\myparagraph{Three typical adaptation motifs: two IFFs (one SI) \& a nonlinear IFB}

Scale invariance provides a powerful criterion for distinguishing
between network motifs that may appear equivalent at first glance.
Consider the two incoherent feedforward (IFF) motifs shown in
Figure~\ref{fig:three_motifs}(b,c). In both cases, the input activates
the intermediate variable \(x\) and the output \(y\), while \(x\)
represses \(y\). Despite this superficial similarity, the dynamical
behaviors of the two motifs differ fundamentally. The first motif
admits no nontrivial equivariance and therefore cannot exhibit scale
invariance.
(See Sidebar ``Formalizing scale-invariance.'')
By contrast, the second motif is scale invariant: scaling
the input by a factor \(p\) results in a corresponding scaling of the
intermediate variable, while leaving the output trajectory unchanged
(that is, if \(u(t)\) is replaced by \(p u(t)\), then
\((p x(t), y(t))\) satisfies the same equations with input \(p u(t)\)).
These two feedforward architectures will be revisited later in this
article. A third motif that naturally arises in the context of scale
invariance is the \emph{nonlinear integral feedback} structure shown in
Figure~\ref{fig:three_motifs}(d). This motif is likewise scale
invariant, by an argument analogous to that used for the second IFF
motif. All three motifs are analyzed in
\cite{shoval_alon_sontag_2011,gupta_sontag_cdr}, where proofs of global
asymptotic stability for each constant input \(u\) are also provided.
It is also interesting to note that linearizing IFF1 and IFF2 at
equilibria corresponding to any fixed constant input gives feedforward
linear systems with an exact
cancellation, hence a washout filter with zero DC gain. This holds
independently of the values of $\alpha,\beta,\gamma,\delta$.
In that sense, adaptation is robust.
\bfigstar
\minp{0.2}{%
\phantom{1}
\medskip

\setlength{\unitlength}{2200sp}%
\begin{picture}(598,2076)(1004,-3112)
\thicklines
\put(2476,-1302){\vector( 0,-1){519}}
\put(2475,-2242){\red{\line( 0,-1){519}}}
\put(2341,-1096){\line(-1, 0){315}}
\put(2026,-1096){\line( 0,-1){1890}}
\put(2026,-2986){\vector( 1, 0){315}}
\put(2355,-2764){\red{\line( 1, 0){225}}}
\put(2411,-1186){$u$}%
\put(2418,-2079){$x$}%
\put(2411,-3039){$y$}%
\end{picture}%
}%
\minp{0.25}{%
\thicklines
\setlength{\unitlength}{3400sp}%
\begin{picture}(2187,1125)(2326,-2866)
\put(2701,-1861){\vector( 1, 0){900}}
\put(3871,-2131){\red{\vector( 0,-1){450}}}
\put(2791,-1861){\line( 0,-1){900}}
\put(2791,-2761){\vector( 1, 0){360}}
\put(3601,-2761){\vector( 1, 0){900}}
\put(3781,-1901){$x$}%
\put(2441,-1901){$u$}%
\put(3286,-2801){$y$}%
\end{picture}%
}%
\minp{0.30}{%
\thicklines
\setlength{\unitlength}{3400sp}%
\begin{picture}(2190,1125)(2326,-2866)
\put(2701,-1861){\vector( 1, 0){900}}
\put(3421,-2761){\vector( 1, 0){900}}
\put(3871,-2131){\red{\line( 0,-1){450}}}
\put(3781,-2581){\red{\line( 1, 0){180}}}
\put(3421,-2761){\line(-1, 0){495}}
\put(2926,-2761){\line( 0, 1){900}}
\put(3781,-1901){$x$}%
\put(2441,-1901){$u$}%
\put(4501,-2801){$y$}%
\end{picture}%
}%
\minp{0.25}{%
\thicklines
\setlength{\unitlength}{3400sp}%
\begin{picture}(2190,1125)(2826,-2866)
\put(3521,-2761){\vector( 1, 0){700}}
\put(3871,-2131){\red{\line( 0,-1){450}}}
\put(3781,-2581){\red{\line( 1, 0){180}}}
\put(3781,-1951){$x$}%
\put(3041,-2831){$u$}%
\put(4501,-2831){$y$}%
\put(4551,-1881){\line( 0,-1){700}}
\put(4551,-1881){\vector( -1, 0){450}}
\end{picture}%
}

\null\vspace{-20pt}
\minp{0.2}{\phantom{1}}%
\minp{0.25}{%
\beqn
\LDdiffdotI &=& \alpha u - \delta x\\
\LDdiffdotH &=& \beta u - \gamma xy
\eeqn
}%
\minp{0.25}{%
\beqn
\LDdiffdotI &=& \alpha u - \delta x\\
\LDdiffdotH &=& \beta
\frac{u}{x}
- \gamma y
\eeqn
}%
\minp{0.3}{%
\beqn
\LDdiffdotI &=& x(\alpha y -\delta )\\
\LDdiffdotH &=& \beta
\frac{u}{x}
- \gamma  y
\eeqn
}
\medskip

\minp{0.2}{\centerline{(a) IFF}}%
\minp{0.25}{\centerline{(b) IFF1}}%
\minp{0.25}{\centerline{(c) IFF2}}%
\minp{0.3}{\centerline{(d) IFB}}
\caption{Two IFF and one IFB motifs (top) and respective two-dimensional models of dynamics (bottom).
In every case, $u$ is the input and $y$ is the output.
Variables are assumed positive; constants $\alpha,\beta,\gamma,\delta$
are positive.
(a) An IFF conceptual motif.
(b) One interpretation of $x$ repression of $y$: intermediate variable $x$ enhances degradation of $y$.
(c) Different interpretation: intermediate variable $x$ inhibits production of $y$.
(d) IFB (integral feedback).
}
\label{fig:three_motifs}
\efigstar

\begin{sidebar}{Scale-invariance and stochastic searches}

SI theory has an application to stochastic spatial search.
Consider a vehicle moving in space which continuously senses a space dependent
input $u=u(r,t)$, where $r$ is the space variable.
Suppose for concreteness that
the input represents a diffusive chemical being emitted from a point source,
\[
u(r,t) = \frac{\ds {u_s}}{(4\pi Dt)^{3/2}}e^{-\frac{(r-r_0)^2}{4Dt}}
\]
in three dimensions.
For example, in chemotaxis, discussed below, $u(r,t)$ represents a
diffusive chemoattractant field that scales with source intensity.
Since the field $u(\cdot ,t)$ at time $t$ 
scales with the source intensity $u_s$,
\emph{positional information about the source is encoded in the
shape of field, and not in its amplitude.}
Thus SI sensing should result in statistically the ``same'' stochastic
search (independent of strength $u_s$).
One way to formalize this is as follows.
We consider a Jump-Markov process
for $\pp(t,\s,v,x)$, the probability density at time $t$, location $\s$,
internal state $x$, and velocity $v$.
We assume that individuals performing the search have internal dynamics
$\LDdiffdotI = f(x,u)$ and that stochastic steering depends only on a
measured variable $y=h(x)$.
(In the \emph{E.~coli} chemotaxis application to be discussed below,
$x$ is a vector of concentrations of proteins in several modified
methylated/phosphorylated forms, and $y$ picks the phosphorylated form
of a protein called CheY.)
Reorientations (jumps in velocities) are guided by a Poisson process
with intensity $\lambda (y)$, according to the kernel
\newcommand{\vp}{\tilde{v}} 
\[
T_y(v,\vp)= \mbox{Prob} \{\vp \leadsto v \,|\, \mbox{input is\ }y\}.
\]
The density $c$ satisfies the transport
(Fokker-Planck, forward Kolmogorov, Smoluchowski) equation
\[
\frac{\partial \pp}{\partial t}
+
\nabla_{\s}\cdot \pp v
+
\nabla_{x}\cdot \pp f
=
-\lambda (y)\pp
+
\int_V
\lambda (y)T_y(v,\vp)\pp(t,\s,\vp,x)d\vp
\]
The input at location $\s$ and time $t$ is $U(t,s)$, and it appears in these
equations through the vector field $f$.
The SI property implies the invariance of marginal distributions for time and
position, defined as:
\[
n(t,\s) =\, \int_\X \int_V \pp(t,\s,v,x) \,d\mu _{\X}(x)\,d\mu _{V}(v)\,.
\]
This is proved
in~\cite{fcd_pde_arxiv_aug2011},
using an equivariance provided by the necessary part of the theorem,
and showing that then
the density with respect to the transformed field $\pi U$ is
\[
\widetilde \pp(t,\s,v,x) = \pp(t,\s,v,\rho^{-1}(x))
  \, \det \left[\rho^{-1}_*(z)\right].
\]  
\end{sidebar}

\myparagraph{Scale invariance in chemotaxis}

Chemotaxis refers to the directed movement of organisms in response to
chemical cues. Chemical stimuli, or chemoeffectors, may function as
attractants or repellents, giving rise to chemoattraction or
chemorepulsion, respectively. These cues can be environmental nutrients
or toxins, or they may be signaling molecules secreted by organisms to
coordinate long-range interactions within a population.

Chemotaxis plays a role in behavior across many biological scales, from
bacteria and single cells in multicellular organisms to amoebae and
social insects~\cite{chemotaxisbook_armitage}. It underlies a wide range
of biological phenomena, including nutrient foraging by \emph{E.\ coli},
aggregation in slime molds, morphogenesis and tissue organization during
embryonic development, tumor angiogenesis, wound healing, and numerous
human diseases. The mathematical study of chemotaxis has therefore been
an active and expansive area of research; see, for example,
\cite{TindallPorterMainiGagliaArmitage2008} for a survey of bacterial
chemotaxis models, or~\cite{keshet_alzheimers2002} for an application to
Alzheimer’s disease based on chemotactic mechanisms. Although our focus
here is on single-cell movement, it is worth noting that a complementary
and equally important branch of chemotaxis research concerns pattern
formation in spatially distributed systems, such as the emergence of
fruiting bodies or animal coat patterns. This line of work was initiated
in large part by the seminal Keller--Segel model~\cite{keller-segel}.

Among known signal transduction pathways, the chemotaxis networks of
\emph{E.\ coli} and othermotile bacteria are arguably the most thoroughly studied. This system
enables bacteria to navigate gradients of chemical attractants or
repellents by sensing \emph{temporal} changes in chemoeffector
concentrations rather than spatial differences. Motile chemotactic
bacteria alternate between relatively straight swimming episodes
(\emph{runs}) and brief reorientation events (\emph{tumbles}). These
behaviors correspond to counterclockwise (CCW) and clockwise (CW)
rotation of the flagellar motors, respectively. While tumbles lead to
random reorientation, motion in favorable directions suppresses tumbling
and results in extended runs, thereby biasing the random walk toward
regions of higher attractant concentration~\cite{BergBrown1972}.
See Figure~\ref{fig:ecoli}.
\bfigstar
\minp{0.14}{\picc{0.22}{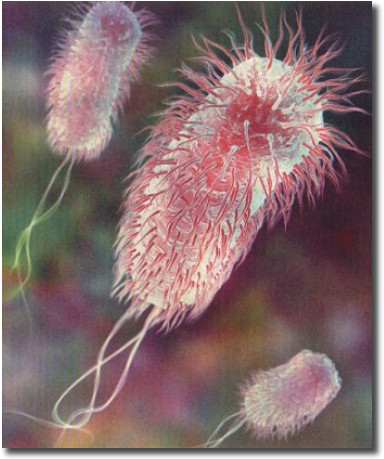}}
\minp{0.14}{\picc{1.6}{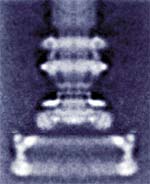}}
\minp{0.14}{\picc{0.06}{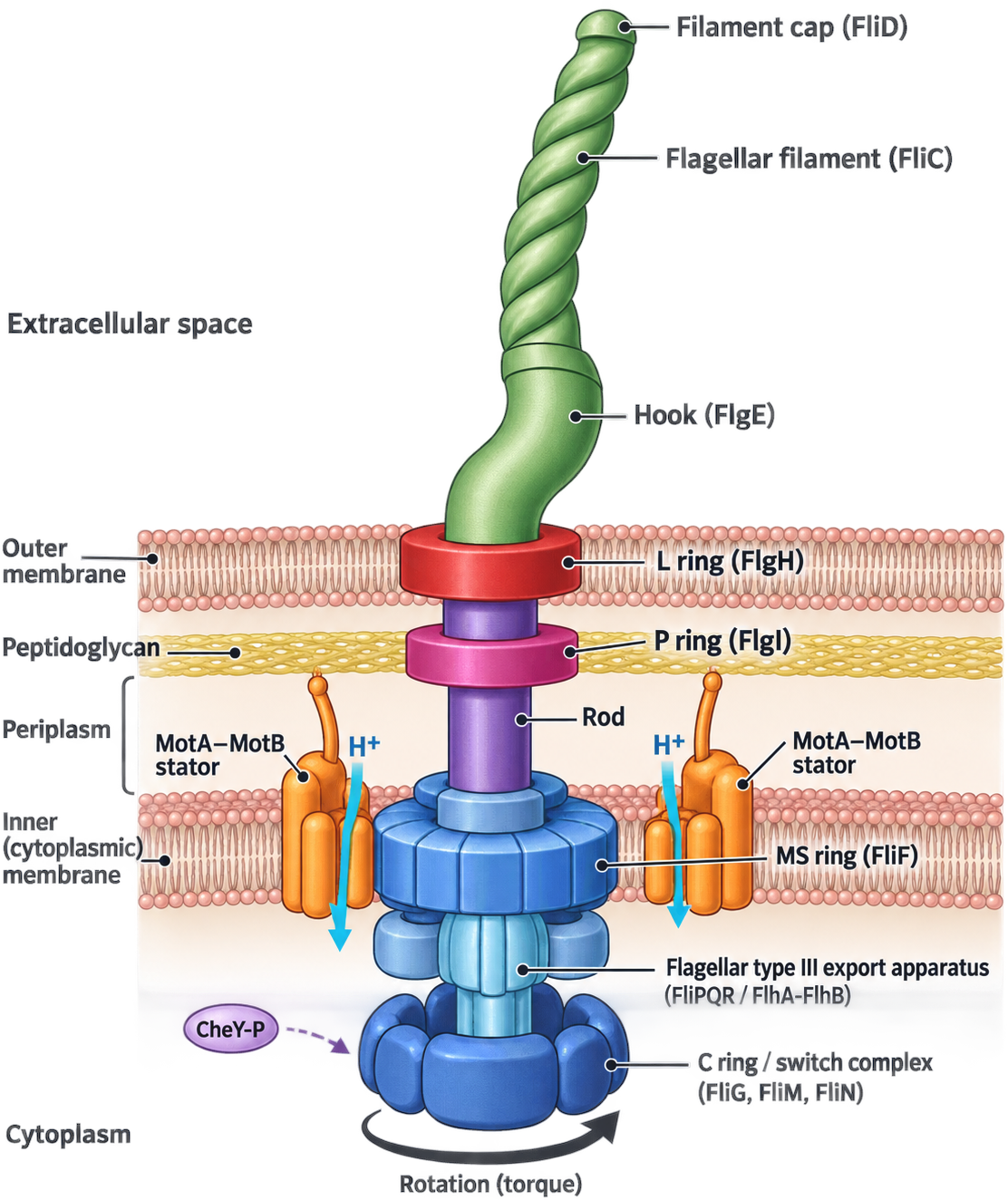}}
\minp{0.40}{\picc{0.12}{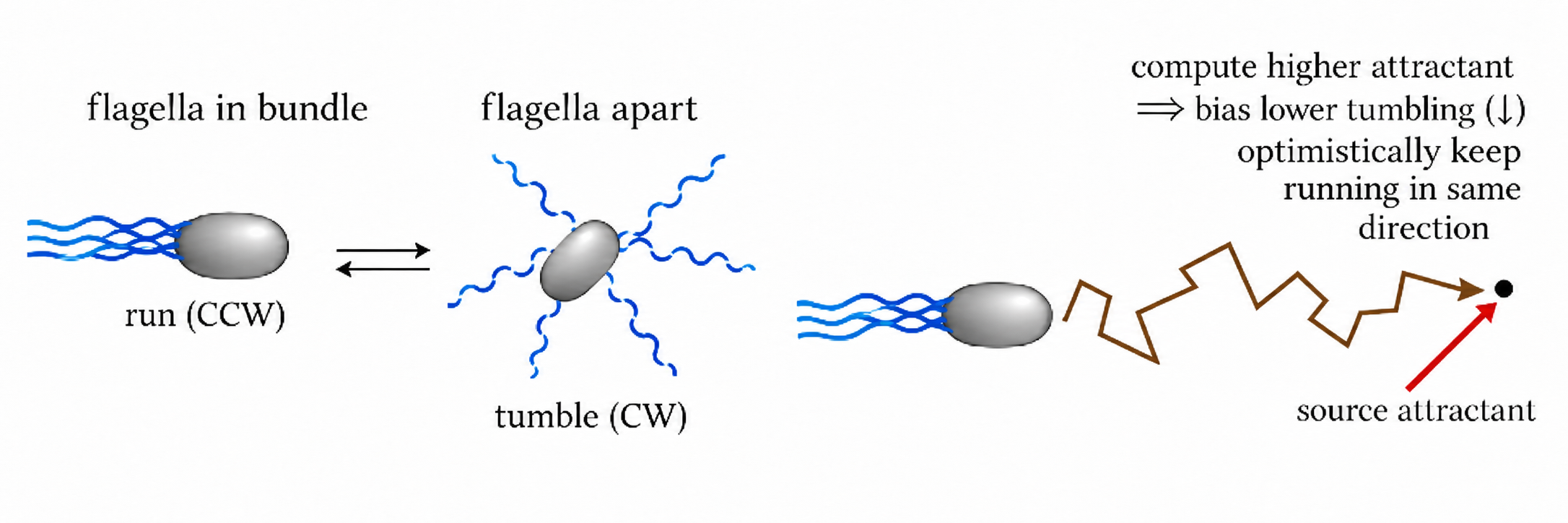}}
\caption{\emph{Escherichia coli} is a single-celled bacterium with a typical length of approximately $2\,\mu$m
(photo in public domain from \cite{usgs_ecoli_fig}).
Motile bacteria swim using several helical flagella driven by a proton-powered
rotary motor, illustrated here by a micrograph from \emph{Salmonella}
from~\cite{Francis1994} (digital print courtesy of David DeRosier, Brandeis University),
and a schematic of a bacterial flagellar motor, drawn by
chatGPT based on the structural organization reviewed in~\cite{NakamuraMinamino(2024}.
Chemotaxis in \emph{E.\ coli}
implements a biased stochastic search in a nutrient concentration
field. When the flagellar motors rotate clockwise, the flagella
disrupt their bundle and the cell undergoes a reorientation
(\emph{tumble}); when the motors rotate counterclockwise, the flagella
form a coherent bundle and the cell moves approximately straight
(\emph{run}). If the bacterium senses an increase in attractant
concentration along its trajectory, the tumbling rate is reduced,
prolonging runs in favorable directions; otherwise, tumbling resumes,
allowing the cell to reorient and sample new directions.}
\label{fig:ecoli}
\efigstar

\emph{E.\ coli} can detect a wide range of chemical cues, including amino
acids, sugars, and dipeptides, as well as physical stimuli such as pH,
temperature, and redox state. The most abundant chemoreceptors, present
in thousands of copies per cell, are Tar and Tsr, which primarily sense
aspartate and serine, respectively. Additional lower-abundance receptors
detect dipeptides (Tap) and sugars such as ribose and galactose
(Trg)~\cite{Sourjik2004}. The chemotaxis system exhibits remarkably high
sensitivity, responding over several orders of magnitude in ambient
concentration to small relative changes in attractant or repellent
levels~\cite{bray2002}.

A major advantage of the \emph{E.\ coli} chemotaxis pathway as a model
signal-transduction system is its relative insulation from other
cellular processes, such as metabolism, allowing it to be treated as a
largely autonomous functional module~\cite{kollmann2005}. 
Most
biochemical rate constants and average protein concentrations in this
pathway have been measured experimentally~\cite{LiHazelbauer2004}. The
core signaling network is well characterized: chemoreceptors assemble
with the adaptor protein CheW and the histidine kinase CheA into large
clusters at the cell poles. Attractant binding inhibits CheA
autophosphorylation, whereas repellent binding enhances it. Activated
CheA transfers phosphoryl groups to the diffusible response regulator
CheY, which in turn biases flagellar motor rotation toward clockwise
(CW) rotation and tumbling. CheY is rapidly dephosphorylated by the
phosphatase CheZ, enabling fast responses to changes in input signals.

The dynamics of the chemotaxis network have been characterized using two
main experimental approaches. Phenotypic assays measure the fraction of
time tethered cells spend in CW versus counterclockwise (CCW) rotation
in response to step changes in chemoeffector concentration
\cite{LarsenReaderKortTsoAdler1974}. Complementary biochemical assays
employ fluorescence resonance energy transfer (FRET) to monitor CheY
phosphorylation levels in vivo~\cite{SourjikBerg2002}. In particular,
FRET between CheY--YFP and CheZ--CFP provides a quantitative readout of
CheA kinase activity, since the FRET signal is proportional to the
concentration of the CheY-P--CheZ complex and thus reflects the steady-
state balance between CheY phosphorylation and dephosphorylation
\cite{ShimizuTuBerg2010}.
Figure~\ref{fig:sourjik2004} schematically summarizes the principal
components of the \emph{E.\ coli} chemotaxis signaling network.
\bfigstar
\picc{0.2}{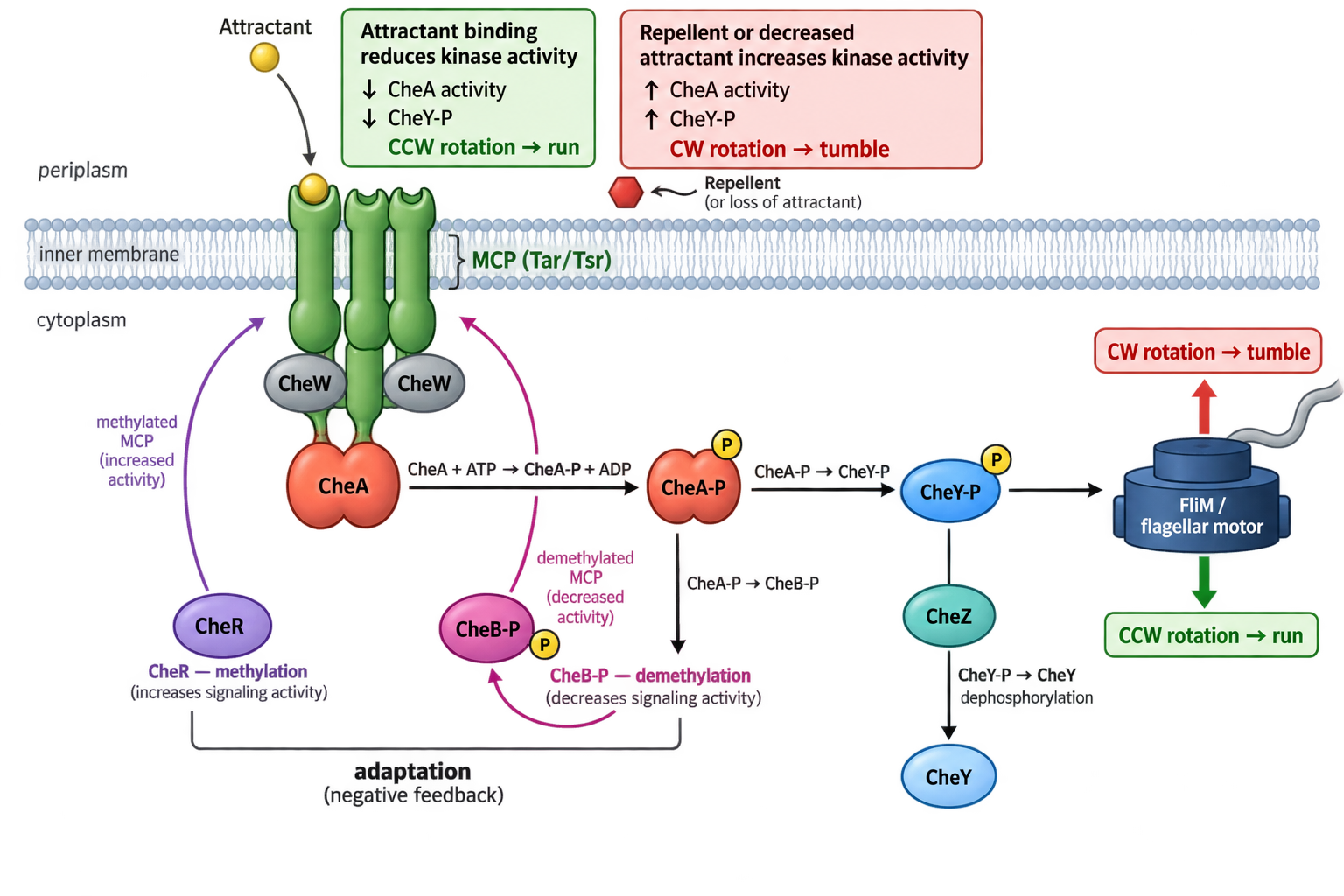}
\caption{Schematic of \emph{E.\ coli} chemotaxis biochemistry, drawn using ChatGPT.}
\label{fig:sourjik2004} 
\efigstar

A defining feature of the chemotaxis network is \emph{adaptation}:
following a sustained change in chemoeffector concentration, tumbling
activity returns to its pre-stimulus level. Adaptation is mediated by
receptor methylation and demethylation, catalyzed by the enzymes CheR
and CheB, respectively, at multiple modification sites on each receptor.
This methylation system provides a molecular memory that effectively
compares past and present inputs. Integral feedback arises through the
phosphorylation of CheB by CheA, which couples pathway output to the
adaptation machinery~\cite{asbl99,YiHUangSimonDoyle2000}.

\begin{sidebar}{Mathematical model of chemotaxis signaling network}
Mathematically, the intracellular biochemistry of the chemotaxis
signaling network is well described by a system of ordinary differential
equations which capture the temporal evolution of the relevant protein
concentrations and modification states.
A representative reduced model is
\beqn
\LDdiffdotM &=& V_R \,r\, \frac{1-a}{1-a+K_R}\,-\,V_B\, b_p\, \frac{a}{a+K_B}\\
\LDdiffdotO &=& k_4 \,a\, (B - b) \,-\, k_5\, b\\
\LDdiffdotH &=& k_1 \,a\, (Y - y)\, -\, k_2\, y\, z \,- \,k_3 \,y,
\eeqn
where:
\begin{itemize}
\item \(m\) is the receptor methylation level,
\item \(b\) is the concentration of phosphorylated CheB (CheB-P),
\item \(y\) is the concentration of phosphorylated CheY (CheY-P), which
      serves as the pathway output,
\item \(B\) and \(Y\) are the total concentrations of CheB and CheY,
\item \(r\) and \(z\) denote the concentrations (or effective activities)
      of CheR  (methyltransferase) and CheZ (phosphatase), respectively.
\end{itemize}
The kinase activity \(a\) is given by the Monod--Wyman--Changeux model
of allosteric regulation
\[
a(m,u)
=
\left(
1
+
c\, e^{-\gamma N m}
\left(\frac{1+u/K_I}{1+u/K_A}\right)^N
\right)^{-1},
\]
where \(u\) is the external ligand concentration, \(N\) is the number of
receptor homodimers in a functional signaling complex, and
\(V_R, V_B, K_R, K_B, k_i, c, \gamma, K_I, K_A\) are positive constants.
The parameter values are typically chosen to represent the Tar receptor
responding to \(\alpha\)-methyl-DL-aspartate (MeAsp).
\end{sidebar}

The resulting cell motion can be modeled separately as a stochastic
process: tumbles occur at random times according to a Poisson process
whose instantaneous rate depends on the concentration of
phosphorylated CheY (CheY-P). In this way, the deterministic signaling
dynamics modulate the statistics of the run-and-tumble motion.
Figure~\ref{fig:ecoli_chemotaxis_circuit}
shows the wiring diagram of this system.
\bfig
\picc{0.32}{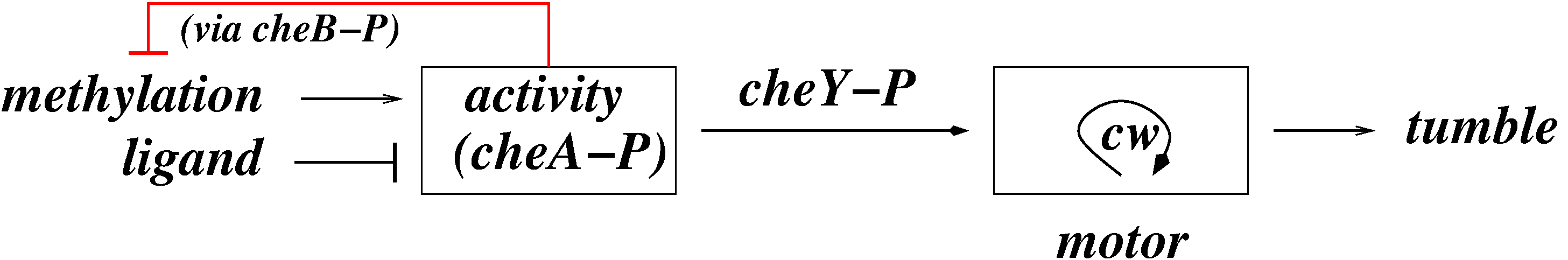}
\caption{Phenomenology of \emph{E.\ coli} adaptation circuit}
\label{fig:ecoli_chemotaxis_circuit}
\efig

The paper~\cite{shoval10} carried out this analysis, using the model
described in Sidebar ``Mathematical model of chemotaxis signaling network,''
and pointed out that
for inputs $u$ (ligand level) in the experimental range
$18\mu M\!\approx 0.006K_A \approx K_I\ll u \ll K_A$, one has the
simplified form
\[
a(m,u)\;\approx\; F(u  e^{-\gamma  m})
\]
for some function $F$.
Then equivariance given by
\[
u\mapsto pu, \quad y\mapsto y, \quad b\mapsto b, \quad m\mapsto m+\frac{1}{\gamma }\ln p
\]
establishes that \textit{E.\ coli} chemotaxis should have scale-invariant behavior.
Thus, in view of the discussion in Sidebar ``Scale-invariance and stochastic searches,''
the stochastic search algorithm as performed by bacteria should result
in equivalent stochastic search processes, 
independently of the strength $u_s$ of a diffusive field.
The paper~\cite{shoval10} predicted this spatial behavior as well.

\myparagraph{Experimental verification of scale-invariance for \textit{E.\ coli} chemotaxis}

To test the scale-invariance predictions from~\cite{shoval10},
the authors of \cite{ShimizuStocker2011} carried out
fluorescence resonance energy transfer (FRET) experiments on tethered cells and
microfluidic experiments on freely swimming cells.
In the FRET experiments, the bacteria were adapted to a given concentration of
MeAsp, $u_0$, and then exposed to a smoothly varying temporal waveform of
stimulus, $u(t)$.
The FRET response, $\Delta FRET(t)$, was recorded continuously
throughout many such experiments in which $u_0$ was varied over a broad range,
while keeping constant the ratio $u(t)/u_0$. Comparisons of the temporal
response profiles over a $>10$ fold range of $u_0$
(Fig.~\ref{fig:stocker_experiments})
show that both the amplitude and waveform of response are invariant. The
concentration range over which this invariance is observed is consistent with
expectations from the FCD analysis in~\cite{shoval10}, i.e. between the
two dissociation constants, $K_I$ and $K_A$, of the Tu-Shimizu-Berg model.
The
waveform of MeAsp concentration in the FRET experiments
was $u(t) \seq u_0\exp(Ae^{-\beta ^2(t-t_c)^2} \sin2\pi \nu t)$,
a function whose logarithm is a baseline $\ln(u_0)$ plus a sinusoidal waveform.
\bfigstar
\minp{0.35}{\picc{0.13}{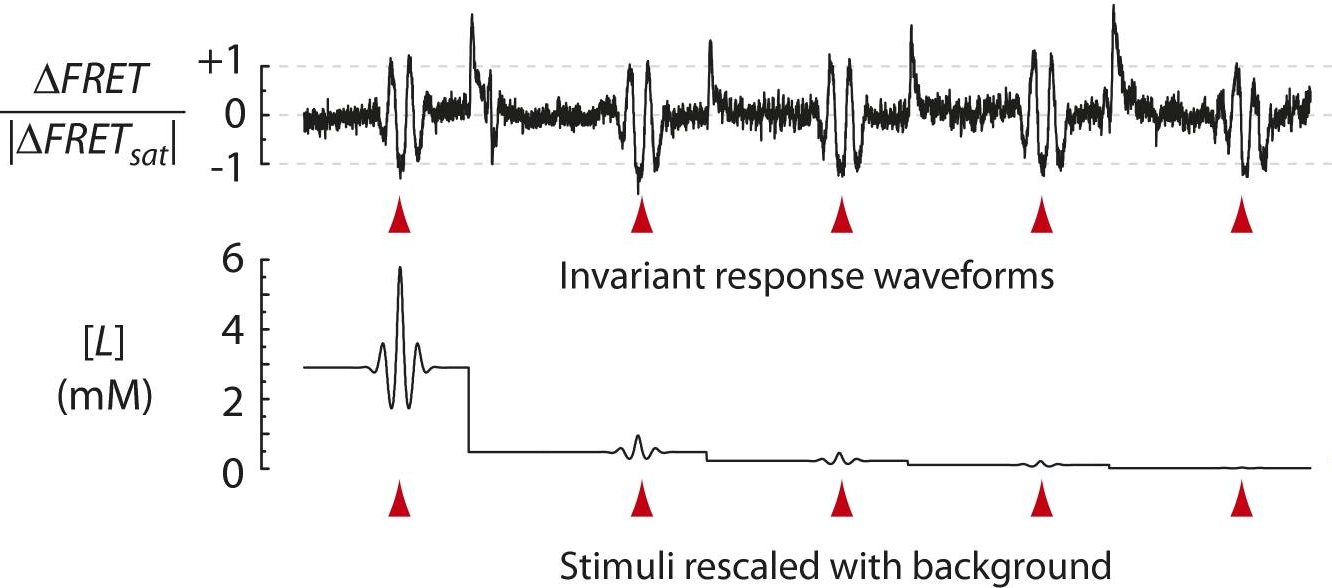}}%
\minp{0.30}{\picc{0.09}{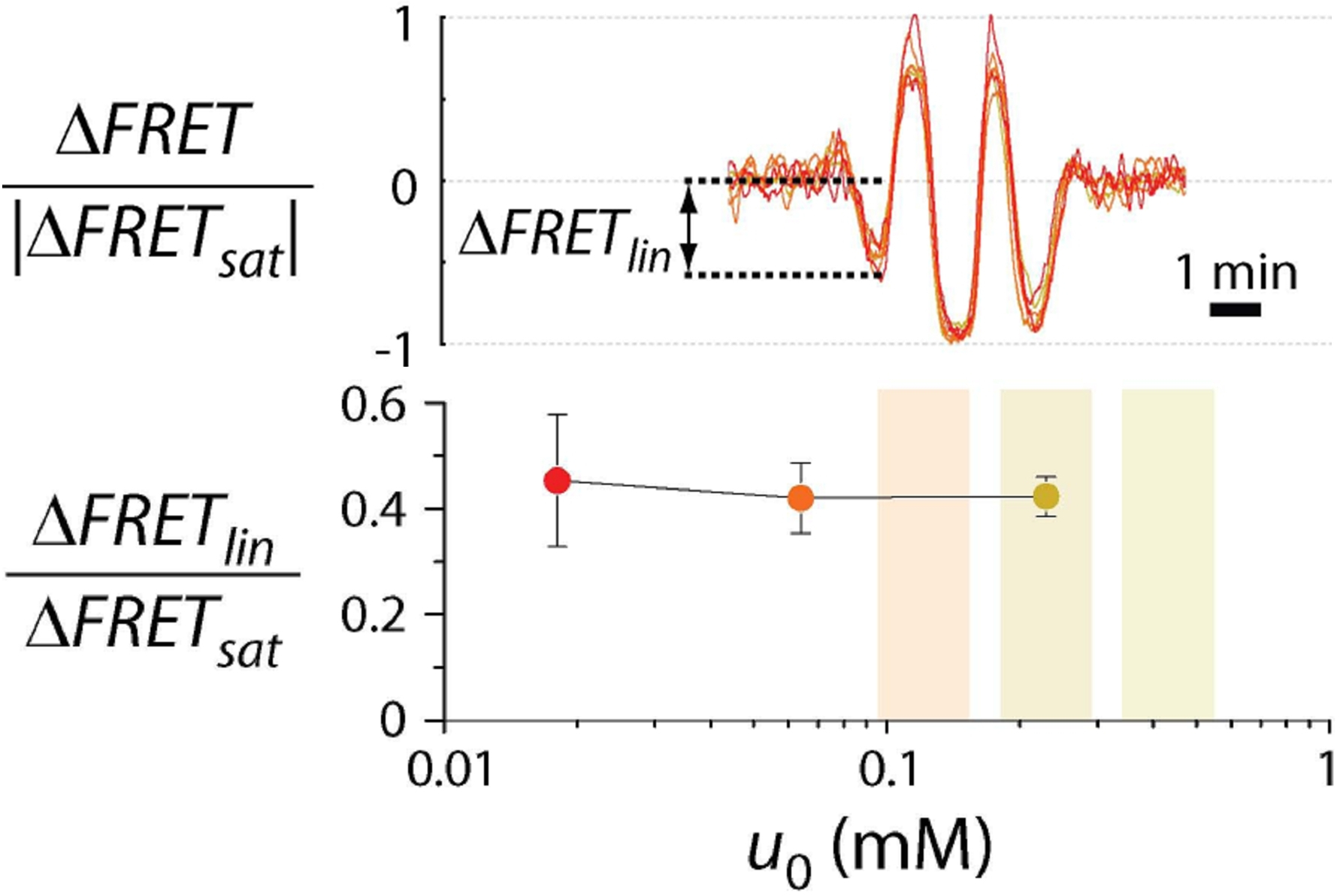}}
\minp{0.25}{\picc{0.08}{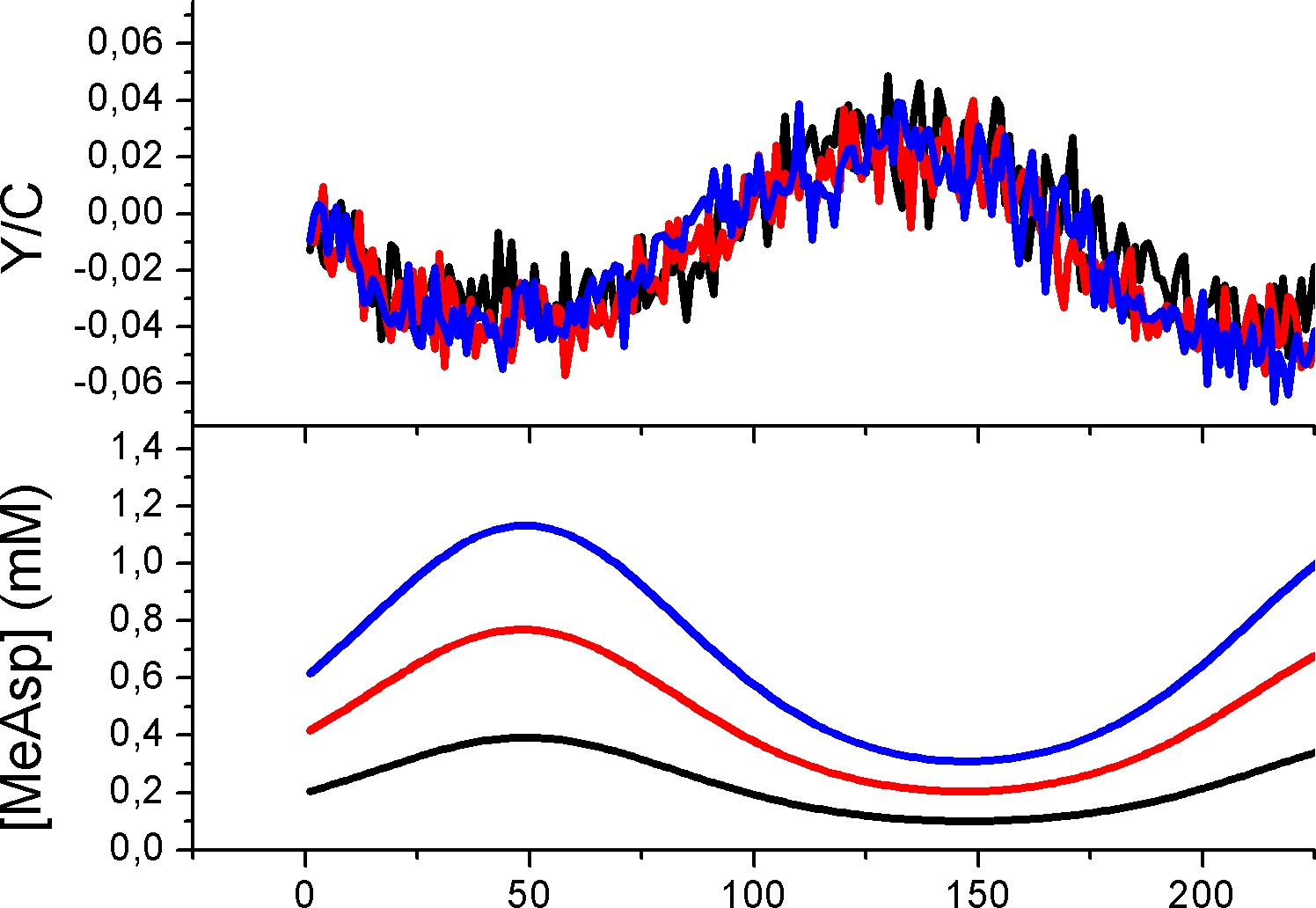}}
\caption{Demonstration of FCD in \emph{E.\ coli}'s chemotaxis network
\cite{ShimizuStocker2011}, shown by invariance of FRET response
waveforms and amplitudes upon rescaling of background
concentration, using tethered cells.
Left:
responses to time-varying ligand with means in range $0.018$-$2.9$ mM.
Center: Overlapped time series of FRET responses to the
stimulus waveform. Each curve shows the deviation, $\Delta FRET$, from
the adapted state FRET signal, normalized by the deviation upon a
saturating stimulus, $|\Delta FRET_{\mbox{sat}}|$; lower panel
shows FRET response amplitudes $|\Delta FRET_{\mbox{\small lin}}|$ of the
waveforms shown in the upper panel, computed from the small-amplitude
peak within the linear regime of the response.
Right: Another experiment, overlapped FRET traces with different
signal (personal communication from T.\ Shimizu).
The label $Y/C$ represents the change in the ratio CheY-P to total CheY, and is
roughly proportional to $\Delta FRET(t)/\abs{\Delta FRET_{\mbox{\small sat}}}$.
Illustrations adapted from Figures 1 and 2 in~\cite{ShimizuStocker2011}, available under a CC-BY-NC-ND 4.0 license.
}
\label{fig:stocker_experiments}
\efigstar

\myparagraph{Experimental verification of scale-invariant spatial search for
  \textit{E.\ coli} chemotaxis}

An experimental confirmation of invariance of search statistics was 
also provided in~\cite{ShimizuStocker2011}.
If the chemotaxis signaling pathway exhibits invariance under
simultaneous rescaling of temporal stimulus gradients and background
ligand levels, then an individual swimming \emph{E.\ coli} cell should
be equally effective at ascending spatial gradients that differ only
by an overall multiplicative factor. Although tracking the trajectory
of a single bacterium over long time intervals is experimentally
challenging, the collective behavior of a large population can be
monitored reliably over extended periods.
One thinks of each bacterium's motion as an independent realization
of the same stochastic process. 
If fold-change detection persists at the behavioral level, one expects
that the spatial evolution of bacterial populations will be
indistinguishable across appropriately rescaled chemoeffector
gradients, provided that the cells are initially adapted to background
concentrations scaled in the same proportion as the gradient
magnitude. This hypothesis was tested experimentally
in~\cite{ShimizuStocker2011}, which examined population-level
migration in controlled spatial gradients of the attractant MeAsp
using a microfluidic setup.

In these experiments, a linear concentration gradient was generated by
flowing buffer solutions with different MeAsp concentrations through
two parallel side channels, referred to as the source and sink
channels, with higher and lower concentrations, respectively (see
Figure~\ref{fig:shimizu_experiments}). Diffusive exchange across a
thin agarose layer established a corresponding gradient within the
central observation channel on a rapid time scale. The motion of
bacterial populations within this channel was recorded using video
microscopy. Initially, cells were distributed approximately uniformly,
but over time they preferentially accumulated toward the side adjacent
to the source channel. Quantitative analysis of the recorded image
sequences yielded the spatial cell density profile $B(x)$ along the
gradient direction.
The concentration profile $B(x)$ of bacteria along the linear gradient shown in
Figure~\ref{fig:shimizu_experiments} shows a clear chemotactic
response, evident as an accumulation towards higher concentrations
(i.e. $x=0$). The nearly identical response despite the large
variation in input stimuli supports FCD. 
\bfigstar
\minp{0.32}{\picc{0.34}{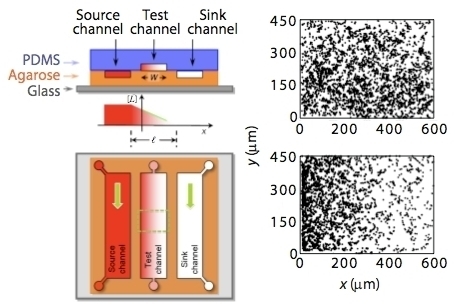}}%
\minp{0.32}{\picc{0.33}{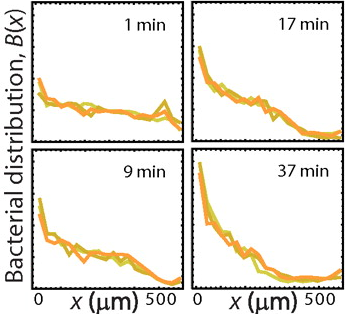}}%
\minp{0.34}{\picc{0.4}{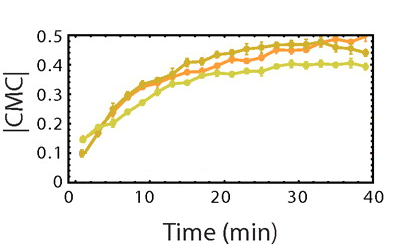}}
\caption{
Free-swimming cells in a linear microfluidic gradient, supporting scale-invariant search.
Left: Microfluidics setup and representative microscopy images, showing
spatial distribution of cells at the start of an experiment
(upper) and 19 min after establishment of a 0.15 mM/mm gradient
along the space dimension (lower), demonstrating chemotaxis.
Center: Three superimposed histograms with $\fractext{\nabla u}{u}$
fixed at 2/3 mm$^{-1}$ and midchannel
concentrations:~0.018/0.064/0.23mM.
Right: chemotactic migration coefficient (CMC) measuring the mean
displacement of the population from the center of the channel, showing
scale invariance. 
Illustrations adapted from Figure 3 in~\cite{ShimizuStocker2011}, available under a CC-BY-NC-ND 4.0 license.}
\label{fig:shimizu_experiments}
\efigstar

\myparagraph{An application to an IFF model invalidation}

The paper~\cite{takeda_et_al2012} examined adaptive signaling in a
eukaryotic chemotaxis pathway by exposing
\emph{Dictyostelium discoideum} cells to controlled temporal profiles of
the chemoattractant cyclic adenosine monophosphate (cAMP) using a
microfluidic device. Pathway activity was monitored via activated Ras
(Ras--GTP), reported by RBD--GFP, a fluorescent probe based on the
Ras-binding domain of human Raf1. The experiments revealed near-perfect
adaptation of Ras activity over a broad range of cAMP concentrations,
from $10^{-2}\,\mathrm{nM}$ to $1\,\mu\mathrm{M}$.
To account for these observations, the authors compared several
mechanistic models and identified an incoherent feedforward architecture
as providing the best fit. The resulting model consists of six coupled
ordinary differential equations describing receptor activation, Ras
regulation through RasGEF and RasGAP, and reporter dynamics. Two receptor
populations with distinct affinities for cAMP contribute additively to
a pooled upstream signal, which drives both Ras activation and
inactivation.

In~\cite{12skataric_fcd}, we revisited this model from the perspective
of fold-change detection, in a regime where the levels of
constitutive activation of the two types of receptors are much lower
than the cAMP input level.
In this regime,
the receptor dynamics
reduce to approximately linear filters, so that the pooled signal
$u(t)$ is proportional to the external cAMP concentration. The
downstream subsystem governing RasGEF, RasGAP, and Ras--GTP is therefore
effectively driven by an input $u(t)$
(see~\cite{12skataric_fcd} for details). 
The fitted parameters exhibit a clear separation of time scales, with
Ras activation and deactivation occurring much faster than the dynamics
of RasGEF and RasGAP. Exploiting this separation yields a reduced system
in which the slow variables satisfy linear dynamics driven by $u(t)$,
and the Ras--GTP level is given by a rational function of these slow
variables:
\[
\begin{aligned}
\LDdiffdotE &= -a_1 x_1 + b_1 u,\\
\LDdiffdotS &= -a_2 x_2 + b_2 u,\\
y &= \frac{K b_3 x_1}{a_3 x_2 + b_3 x_1},
\end{aligned}
\]
where $x_1=\GEF$, $x_2=\GAP$, and $y$ represents the quasi-steady-state
level of Ras--GTP. The constants $a_i$, $b_i$, and $K$ are determined by
the original kinetic parameters.
This reduced system is equivariant under the scaling
transformation $x_i \mapsto p x_i$, implying scale invariance and hence
fold-change detection.
Simulations of the full six-dimensional model confirm this prediction:
responses to cAMP step inputs differing by a constant factor (for
example, $1\!\rightarrow\!2\,\mathrm{nM}$ versus
$2\!\rightarrow\!4\,\mathrm{nM}$) collapse onto essentially identical
trajectories, see
Fig.~\ref{fig:dicty_fcd}.
However, this behavior appears to be inconsistent with
experimental observations. According to a personal communication from the
authors of~\cite{takeda_et_al2012}, experimentally doubling the cAMP
input in the relevant ranges produced responses that were not
fold-invariant.
\bfig
\picc{0.15}{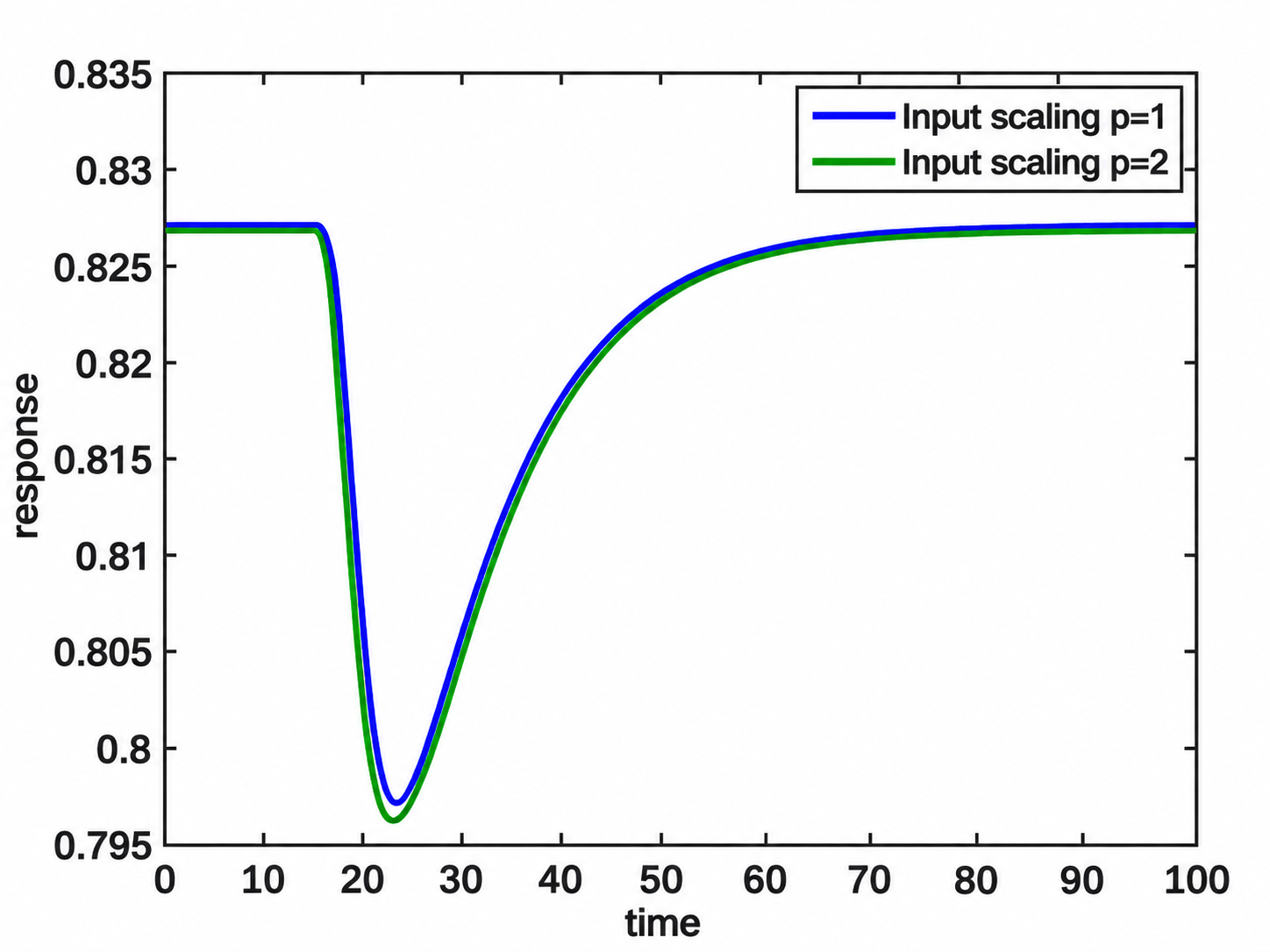}
\caption{Scale-invariance for model from
  \protect{\cite{takeda_et_al2012}}: responses to steps
  $1$$\rightarrow $$2$ and $2$$\rightarrow $$4$ coincide.
Figure reproduced from~\cite{12skataric_fcd}.}
\label{fig:dicty_fcd}
\efig

This discrepancy is instructive. It suggests that the published model,
or at least the fitted parameter regime, does not fully capture the
input-scaling properties of the biological system. More broadly, it
illustrates how fold-change detection can serve as a stringent
\emph{dynamic phenotype}: a simple test comparing responses to inputs
related by a constant factor can strongly constrain admissible network
structures and parameter values, without requiring direct access to
internal molecular states.

\section{Another transient phenotype: cumulative dose responses}

\newcommand{\DRuT}{\mbox{DR}(u, T)} 
\newcommand{\cDRuT}{\mbox{cDR}(u, T)}
\newcommand{\bary}{{\bar y}}

In many areas of pharmacology and biomedical research, outcomes depend
not only on instantaneous concentrations but on the \emph{time-integrated
exposure} to a substance. A standard way to quantify such exposure is
through the area under the concentration--time curve (AUC), which
captures the cumulative abundance of a molecule over a specified
interval. In~\cite{gupta_sontag_cdr} we introduced the term
\emph{cumulative dose response} (cDR) to emphasize this perspective in a
dynamical-systems context. Measurements of this type arise naturally
across a wide range of experimental and clinical settings, illustrating
the broad relevance of cDR-based analysis.

One prominent class of examples is provided by \emph{cytokine release
assays}, which quantify the total amount of cytokines secreted by immune
cells in response to drugs, pathogens, or other stimuli. Cytokines are
key regulators of immune and inflammatory processes, as well as cell
growth, differentiation, and tissue repair. Commonly measured species
include pro-inflammatory cytokines such as IL-6, IL-1$\beta$,
TNF-$\alpha$, and IFN-$\gamma$; anti-inflammatory cytokines such as IL-10
and IL-4; and growth factors including GM-CSF and VEGF. Experimental
techniques such as ELISA and multiplex immunoassays (for example,
Luminex\textsuperscript{\tiny\textregistered}) are routinely used to
quantify cumulative cytokine secretion. As an illustration,
\cite{2023_frank_bratta_perspective_neural_regeneration_cytokine_release}
describes an \emph{in vivo} immunosensing approach in which an
antibody-coated optical fiber is implanted in the rodent brain to capture
cytokines released over time within a localized region; the accumulated
cytokine load is then quantified by ELISA. In a clinical context,
cumulative cytokine measurements are also central to the assessment of
cytokine release syndrome (CRS), a potentially severe adverse effect of
T-cell bispecific antibody therapies. The study
\cite{leclercq2023dissecting}, for instance, employed
Luminex\textsuperscript{\tiny\textregistered}-based assays and AUC
analyses of cytokine profiles to compare strategies aimed at mitigating
CRS.

A second major application of cumulative measurements arises in the
study of \emph{drug metabolism and clearance}. Following administration
of a drug or prodrug, plasma concentrations are typically sampled over
time, and the resulting concentration--time curve is integrated to
obtain the AUC~\cite{cho2018understanding}. This quantity provides a
compact summary of systemic drug exposure and plays a central role in
pharmacokinetic modeling, dose selection, and safety assessment,
including the evaluation of hepatic and renal clearance. A comprehensive
discussion of cumulative drug excretion measures and their clinical
relevance is given in~\cite{li2013systematic}, which highlights the use
of AUC-based metrics in dosing and toxicity studies.

Finally, a familiar example from routine clinical practice is the
\emph{HbA1c (A1c) blood test}, widely used in the diagnosis and
management of diabetes. The A1c value reflects the average blood glucose
level over the preceding two to three months and can be interpreted as
being proportional to the time integral of glucose concentration over
that interval. This effective integration arises from the relatively
long lifespan of red blood cells and the irreversible binding of glucose
to hemoglobin, which smooths short-term fluctuations and yields a
cumulative measure of glycemic exposure.

\subsection{Dose response and cumulative dose response}

Throughout this section we restrict attention to constant inputs
$u(t)\equiv u$, and we write $y_u(t)$ to emphasize the dependence of the
output on both the input value and time. Typical temporal responses to
several input levels are illustrated in the left panel of
Figure~\ref{fig:example_time}, where, for simplicity, the notation
$y_i(t)$ is used in place of $y_{u_i}(t)$. Dependence on initial
conditions will not be made explicit; unless otherwise stated, all
responses are assumed to start from a fixed reference state.

As discussed earlier, a system is said to exhibit \emph{perfect adaptation} to constant
inputs if the long-time output converges to a common value,
independent of the particular input level. That is, there exists a
constant $\hat y$ such that
\[
\lim_{t\to\infty} y_u(t) = \hat y
\quad \text{for all } u .
\]
This asymptotic value represents a habituated or baseline state, as in
sensory systems that adapt to constant background stimuli such as
steady illumination or noise. From an engineering viewpoint, such
systems behave as high-pass filters, responding primarily to changes
in the input rather than to its sustained level. By its very nature,
adaptation is an asymptotic notion and does not constrain the behavior
of the system at finite times.

In many applications, however, the transient response is of primary
interest. For a fixed observation time $T$, one may ask how the output
$y_u(T)$ varies with the input level $u$. Examples include the drug
concentration in a tumor microenvironment at a prescribed time after
administration, tumor size at a given stage of therapy, or the number
of infected individuals at a specified time during an epidemic. We
refer to the mapping
\[
u \;\mapsto\; y_u(T)
\]
as the \emph{dose response} of the system at time $T$, and denote it by
$\DRuT$. Experimentally, this curve is obtained by applying different
constant input levels and recording the corresponding output at the
chosen time point. The center panel of
Figure~\ref{fig:example_time} illustrates dose-response curves derived
from the time courses shown on the left.

In many biological and pharmacological settings, the output variable
turns over slowly or accumulates in a compartment such as tissue or
bloodstream. In such cases, the experimentally accessible quantity,
and often the relevant phenotypic response, depends not on the
instantaneous output but on its time integral. This motivates the
definition of the \emph{cumulative dose response}
\[
\cDRuT := \int_0^T y_u(t)\,dt ,
\]
that is, the area under the output curve up to time $T$. The right
panel of Figure~\ref{fig:example_time} shows cumulative dose-response
curves corresponding to the same set of temporal responses. Unlike
perfect adaptation, which constrains only steady-state behavior, dose
response and cumulative dose response capture distinct aspects of
finite-time dynamics that are often central to experimental
observation and interpretation.
\bfigstar
\minp{0.4}{\picc{0.35}{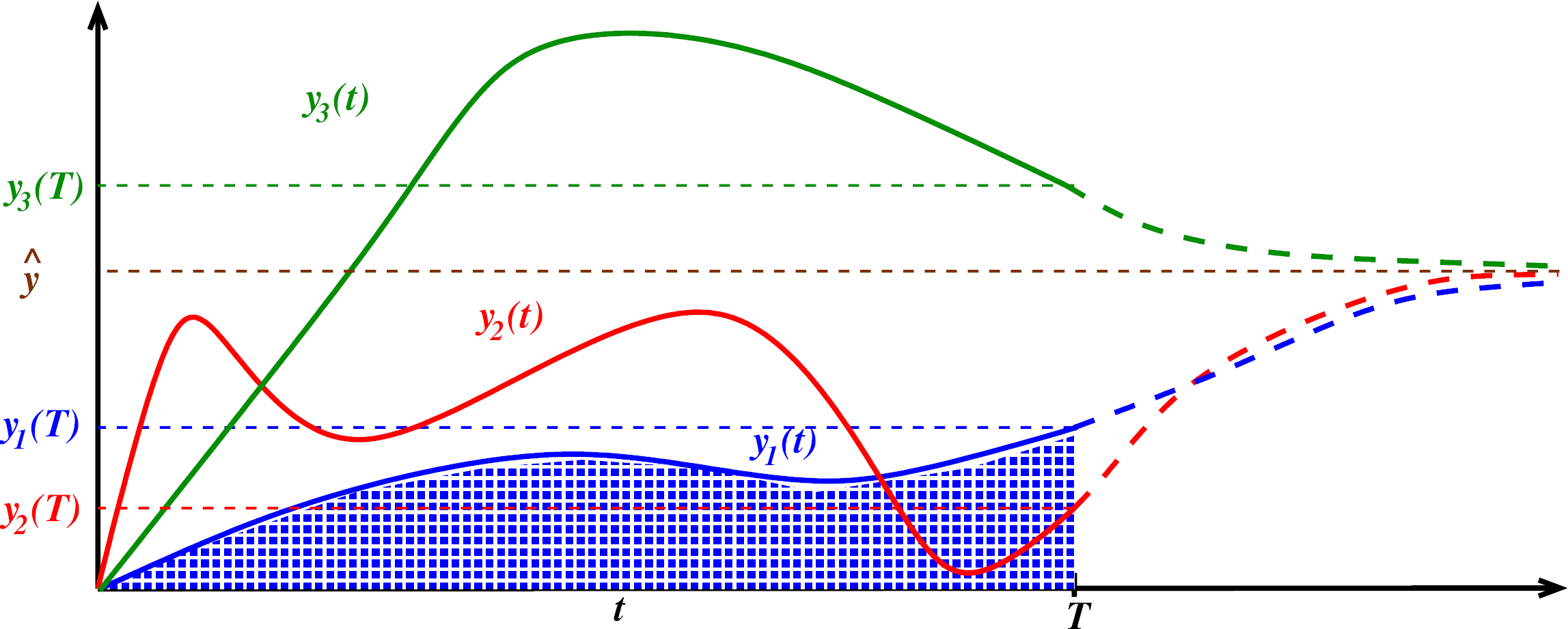}}%
\minp{0.6}{\picc{0.35}{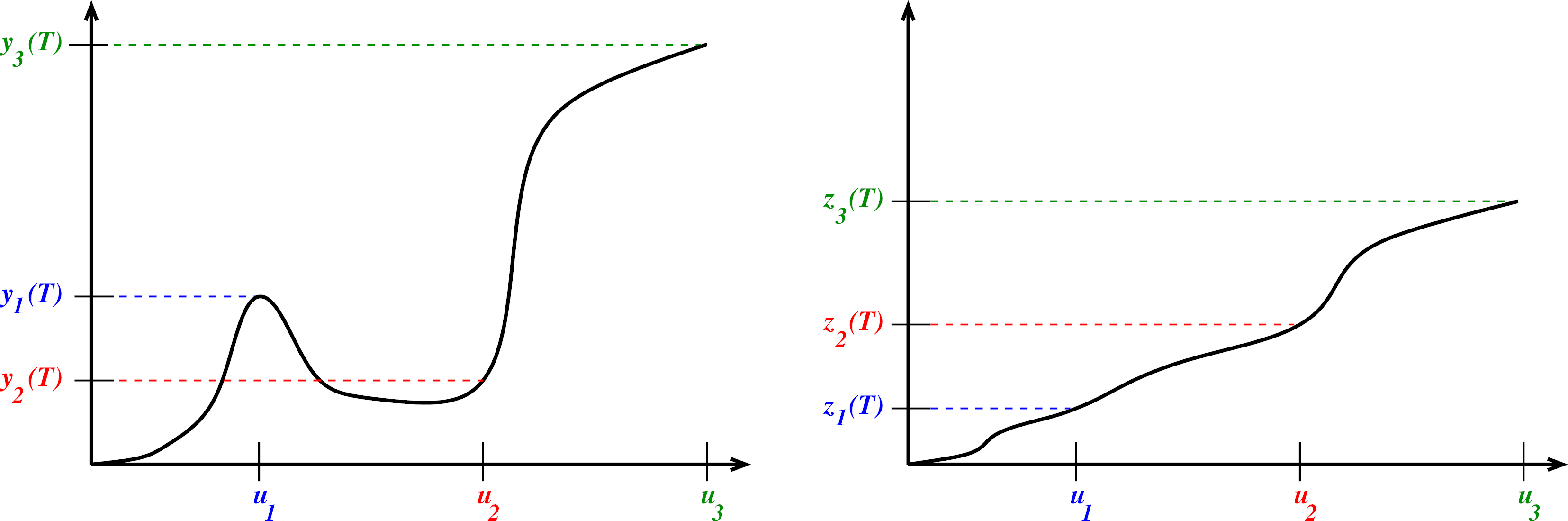}}

\caption{
\textbf{Left:} Time courses of the system output \(y(t)\). Three representative
responses \(y_1(t)\), \(y_2(t)\), and \(y_3(t)\) are shown, corresponding
to constant input levels \(u_1\), \(u_2\), and \(u_3\), respectively. The
values of the responses at a prescribed observation time \(T\) define
their dose--response (DR) values, indicated on the vertical axis. In an
adaptive system, all responses converge to the same steady-state value
\(\hat y\) as \(t \to \infty\), independently of the input magnitude.
Solid curves depict the responses up to time \(T\), while dashed curves
indicate their continuation to steady state. The shaded region
illustrates the integral
\(z_1(T)=\int_0^T y_1(t)\,dt\), which underlies the definition of the
cumulative dose response (cDR).
\textbf{Center:} Dose response (DR) at time \(T\), obtained from the
temporal responses shown in the left panel. The vertical axis represents
the output evaluated at \(t=T\), now plotted as a function of the input
level \(u\). In this example, the DR is nonmonotonic; for instance,
\(u_1<u_2\) while \(y_1(T)>y_2(T)\).
\textbf{Right:} Cumulative dose response (cDR) at time \(T\). The
vertical axis now shows the integrated output
\(z(T)=\int_0^T y(t)\,dt\), plotted against the input \(u\). The value
\(z_1(T)\) corresponds to the shaded area in the left panel. In this
example, the cDR is monotonic: although \(y_2(T)<y_1(T)\), the area
under \(y_2(t)\) exceeds that under \(y_1(t)\), leading to
\(z_1(T)<z_2(T)<z_3(T)\) for \(u_1<u_2<u_3\).
}
\label{fig:example_time}
\efigstar

\myparagraph{Motivation: T cell recognition}

As discussed earlier, cumulative dose responses (cDRs) arise naturally
in many biological contexts. The specific motivation for the study
in~\cite{gupta_sontag_cdr}, however, originated in experimental work on
T cell signaling, particularly the measurements reported in
\cite{dushek2019}. We briefly summarize that biological setting and its
implications here.

Adaptation is a fundamental requirement of the immune system. T cells
must mount strong responses to pathogens and malignant cells while at
the same time maintaining tolerance to self-antigens in order to avoid
autoimmunity. T cell activation is initiated when T cell receptors
(TCRs) bind to peptide major histocompatibility complex (pMHC)
molecules displayed on antigen-presenting cells (see
Figure~\ref{fig:Tcell}). This recognition event triggers intracellular
signaling cascades that ultimately lead to the secretion of cytokines,
which coordinate downstream immune responses by recruiting and
activating additional immune cells.
\bfig
\picc{0.1}{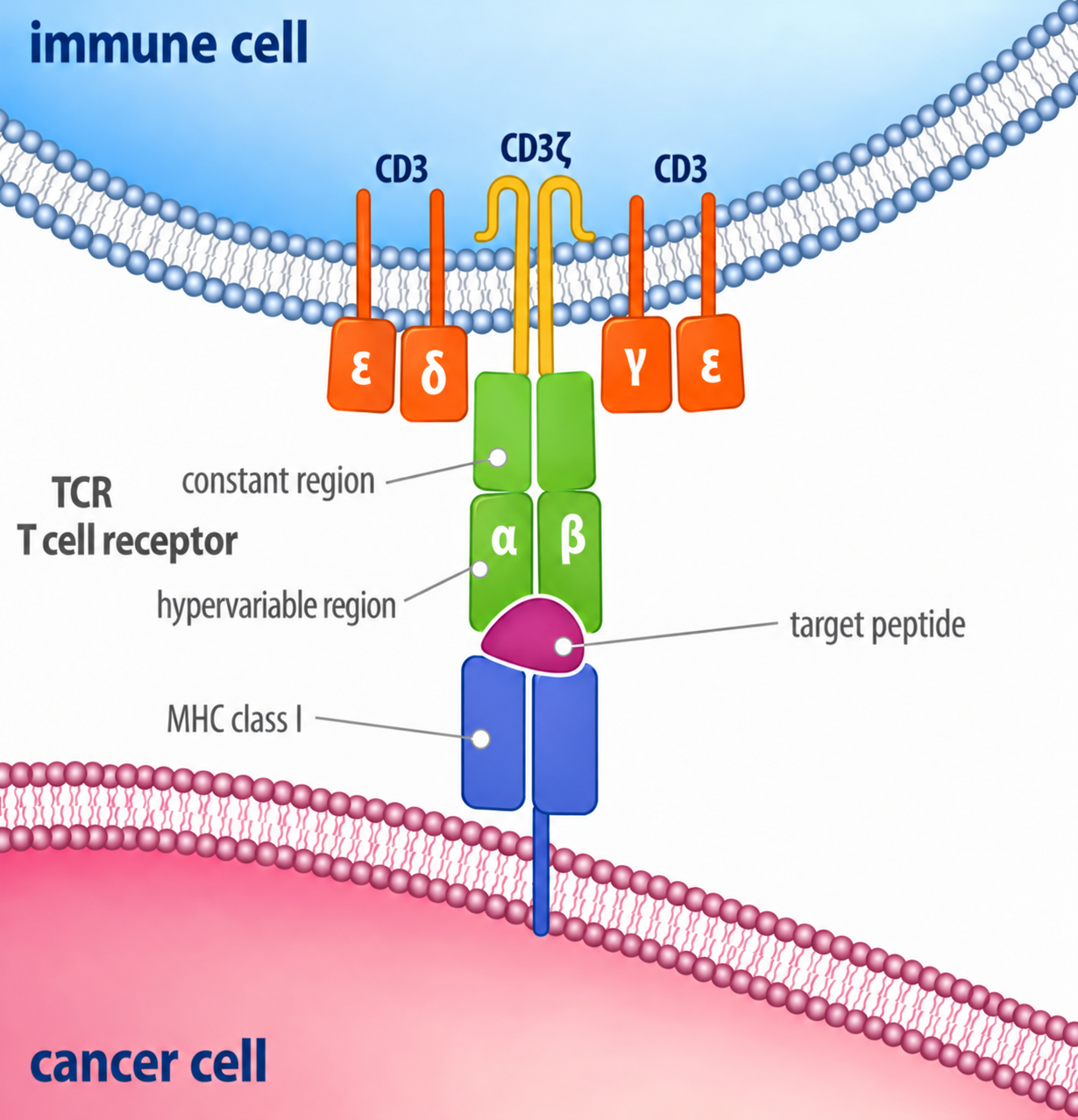}
\caption{Cancer cell interacting with a T cell. 
T cell activation is triggered by the binding of T cell receptors (TCRs)
to peptide major-histocompatibility complex (pMHC) antigens.}
\label{fig:Tcell}
\efig

The experiments reported in~\cite{dushek2019} investigated how primary
human CD8$^+$ T cells respond to sustained antigen stimulation.
Specifically, CD8$^+$ T cells expressing the c58c61 TCR were stimulated
with recombinant pMHC ligand (the cancer-associated peptide 9V)
immobilized on plates, providing a controlled and constant antigen
input $u$ over a wide range of concentrations. This antigen is commonly
used in studies of T cell binding and antigen discrimination. The
measured output was the cumulative amount of secreted TNF-$\alpha$,
denoted by
\[
z(t)=\int_0^t y(s)\,ds,
\]
where $y(t)$ represents the instantaneous cytokine secretion rate.
Figure~\ref{fig:individual_cDR} displays the experimentally measured
cumulative TNF-$\alpha$ levels as functions of antigen concentration,
evaluated at multiple time points ranging from one to eight hours.
These cumulative dose responses exhibit strikingly non-monotonic and,
in some cases, oscillatory behavior.
\bfigstar
\picc{0.18}{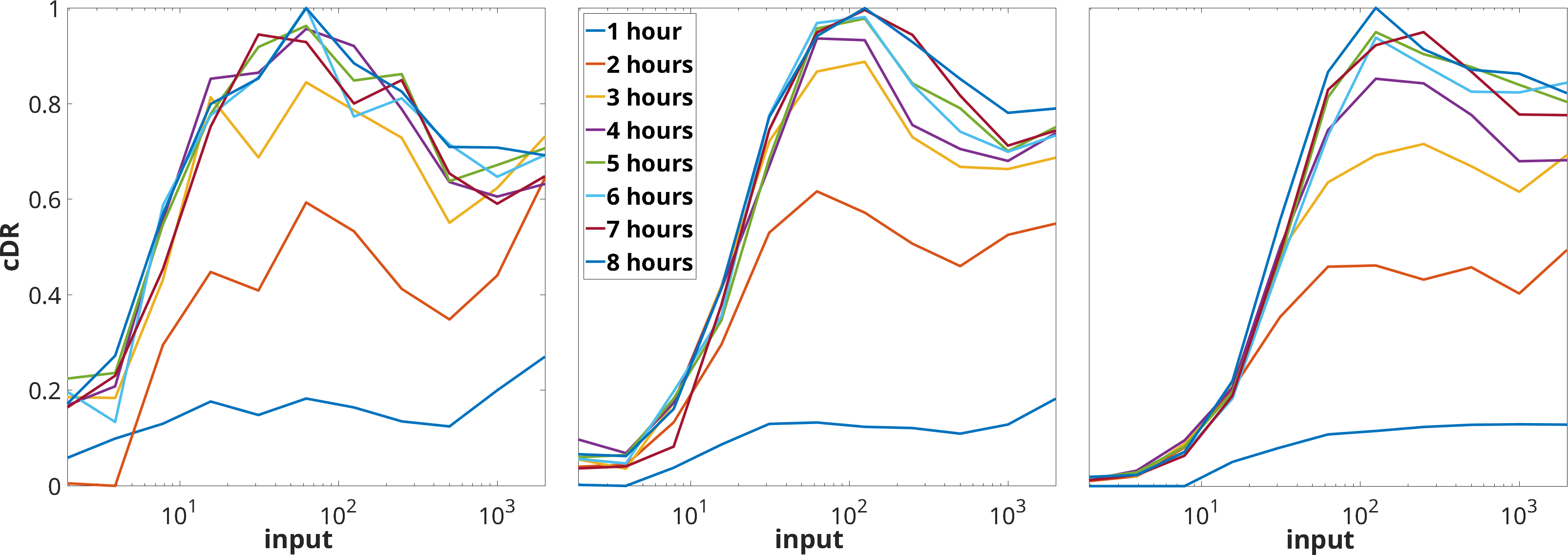}
\caption{Experimental cumulative dose responses in T cell activation,
  plotted for various times ($t =1$ to $8$ hours), with
separate plots from each experiment. Plots done from data used to
generate Figure 1 in the SI of~\cite{dushek2019}.
Horizontal axis denotes concentrations of the input (in units of ligand in ng/well). }
\label{fig:individual_cDR}
\efigstar

Despite the fact that only the integrated output $z(t)$ was directly
measured, the data also provide strong evidence for adaptation. In
particular, for sufficiently large times, the secretion rate $y_u(t)$
appears to become independent of the antigen concentration $u$. To
make this point more explicit, one may form rough estimates of $y(t)$
by numerically differentiating the measured $z(t)$ data using finite
differences and imposing the initial condition $y(0)=0$. Applying this
procedure to the experimental replicates shown in
Figure~\ref{fig:individual_cDR} yields the approximate time courses
displayed in Figure~\ref{fig:trendel_adaptation}. While these estimates
are necessarily coarse, due to the one-hour sampling interval and
experimental noise, they are consistent with the conclusion that the
system adapts in the sense that $y_u(t)$ approaches a common value
(here effectively zero) at long times, regardless of the antigen dose.
Apparent negative values of $y(t)$ at isolated time points are likely
artifacts of numerical differentiation or reflect baseline cytokine
levels unrelated to antigen stimulation.

Beyond adaptation, the experimental data suggest an additional and
stronger property. Over a substantial range of antigen concentrations
(pMHC ligand levels from approximately $1.95$ to $2000$~ng/well; see
Figure~2 of~\cite{dushek2019}), the cumulative responses exhibit a form
of scale invariance: for sufficiently large inputs, the transient
responses collapse onto similar trajectories. This behavior is
consistent with fold-change detection (FCD) at the level of the
output and resembles the dynamics produced by certain
incoherent feedforward architectures (specifically IFF2) or nonlinear
integral feedback mechanisms discussed earlier.

These observations motivate a natural theoretical question: which
network motifs are capable, for appropriate parameter regimes, of
simultaneously exhibiting perfect adaptation and producing the
non-monotonic cumulative dose responses observed in T cell activation?
A non-monotonic cDR immediately rules out linear systems, which
necessarily yield linear and hence monotonic dose responses, as well
as monotone nonlinear systems, whose dose responses are likewise
monotonic. Based on numerical explorations, it was suggested
in~\cite{dushek2019} that incoherent feedforward loops alone are
insufficient to generate non-monotonic cDR behavior without the
addition of explicit thresholding mechanisms, leaving open the search
for alternative or augmented network architectures capable of
explaining these experimental findings.
The work~\cite{gupta_sontag_cdr} set out to investigate this question
for the three prototypical motifs used in adaptation and FCD theory,
and established the following result:

\medskip
\noindent\textbf{Theorem.}
\emph{IFF1 and IFF2 motifs in Figure~\ref{fig:three_motifs} always
have (for all parameters and all times) a monotonic cDR. On the other
hand, the IFB motif in Figure~\ref{fig:three_motifs} may have a
non-monotonic cDR.}

Thus, the two common types of feedforward motifs can never exhibit such
behaviors, because their cDRs are always monotonic. This is
especially surprising for one of them (IFF1) because for such systems
the DR itself can be non-monotonic, yet the cDR is monotonic, in
behavior reminiscent of the cartoon illustrations in
Figure~\ref{fig:example_time}.  This result was complemented by the
new finding that, on the other hand, the standard nonlinear integral
feedback for adaptation is indeed capable of showing non-monotonic
cDR, and thus is potentially a mechanism that is consistent with the
experimentally observed non-monotonic cDR.  In summary:

As an example, consider the following system of two differential
equations, which is a particular example of the IFF1 model.
\beqn
\LDdiffdotI&=& -  x +u \\
\LDdiffdotH & =& -10  x y + u \,.
\eeqn
This is an adapting system: for any given constant input $u>0$, the
steady states are $\barx=u$ and $\bary = 1/10$, which is independent
of $u$, so that the steady
state output $\hat y = 1/10$ is independent of $u$.
Figure~\ref{fig:example_iffl_nonmonotone_DR} shows plots of the DR
(non-monotonic) and the cDR (monotonic).
For an example of IFB, consider the following system of two
differential equations:
\beqn
\LDdiffdotI&=& x (y - 6) \\
\LDdiffdotH & =& \frac{u}{x} - y.
\eeqn
This is also an adapting system: for any given constant input $u>0$, the
steady states are $\barx=u/6$ and $\bary = 6$, so that the steady
state output $\hat y = 6$ is independent of $u$.
The right panel of Figure~\ref{fig:example_iffl_nonmonotone_DR} shows
plots of the (non-monotonic) cDR.
\bfigstar
\minp{0.65}{\picc{0.17}{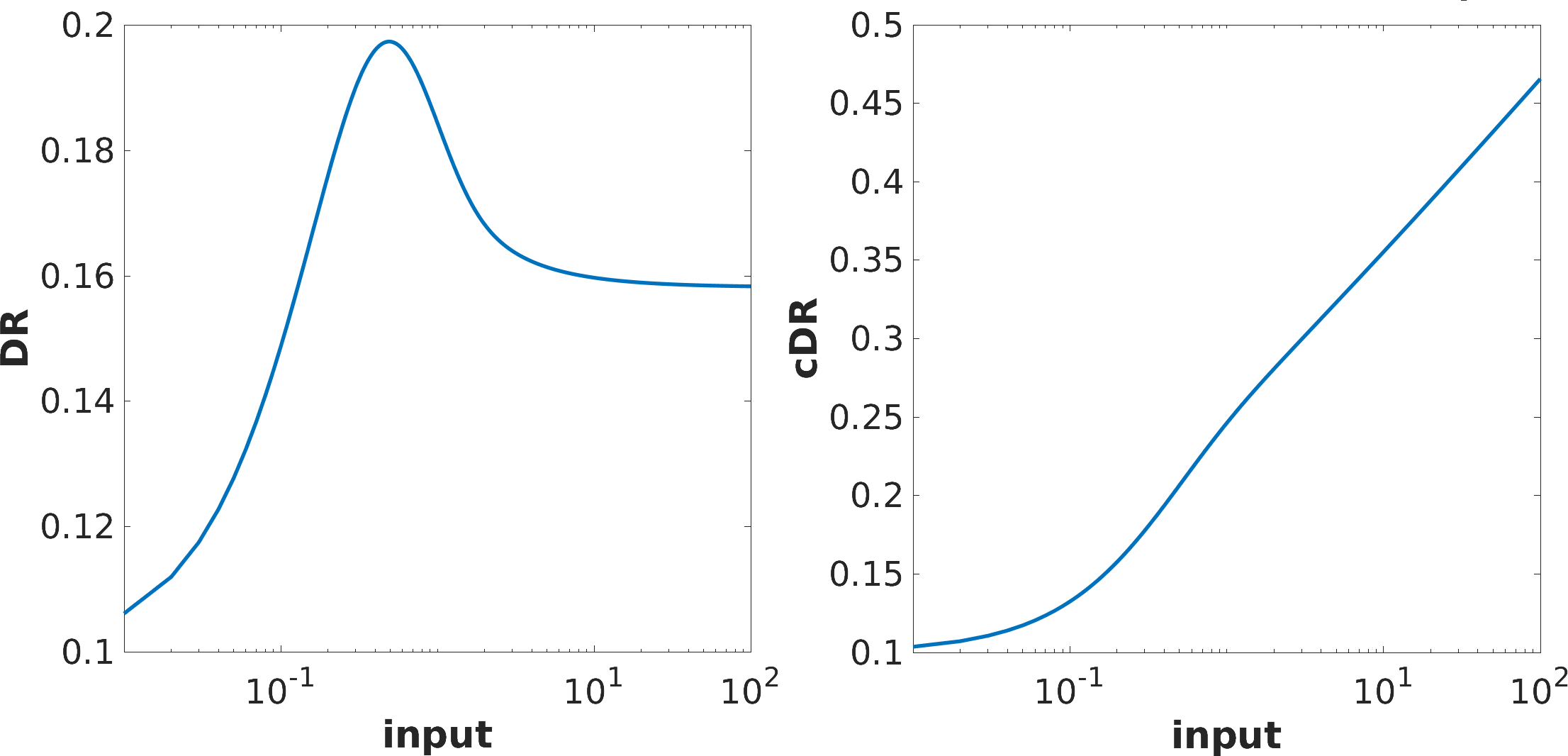}}%
\minp{0,35}{\picc{0.15}{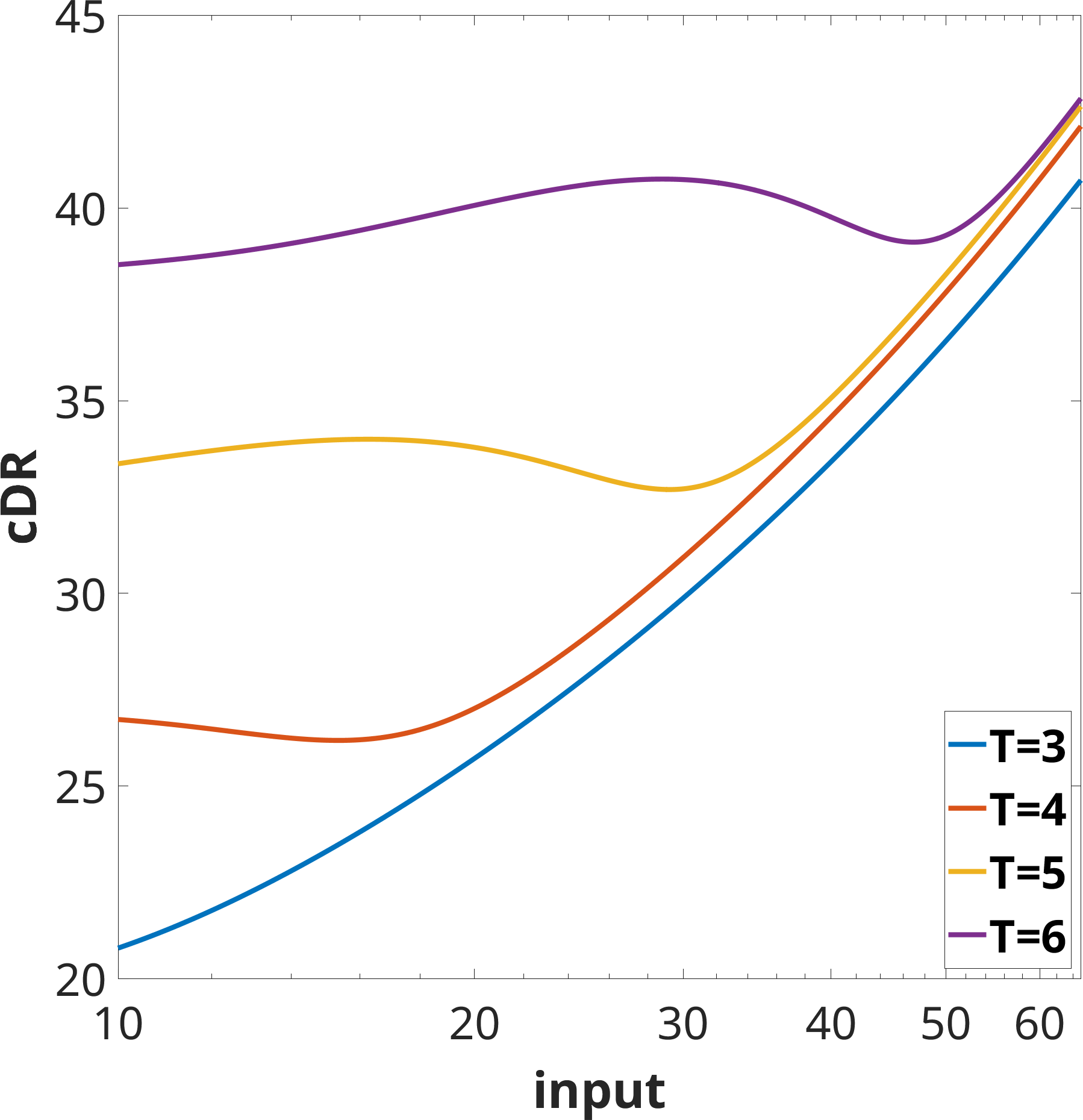}}
\caption{Left and center: Plots of DR ($y(t)$) and cDR ($\int_0^T
  y(t)dt$) for the example $\LDdiffdotI= -  x +u$, $\LDdiffdotH  = -10  x y + u$.
The initial conditions are $x(0)=0$, $y(0)=1/10$, and the time horizon
is $T=1.5$.
Using logarithmic scale on inputs, for comparison with experimental plots.
Observe that the DR is non-monotonic, yet, surprisingly, the cDR is
monotonic.
Right: Plot of cDR ($\int_0^T y(t)dt$) for the integral feedback
example $\LDdiffdotI= x (y - 6)$, $\LDdiffdotH = \frac{u}{x} - y$.
The initial conditions are $x(0)=0.1$, $y(0)=6$, and time horizons
shown are $T=3,4,5,6$. Observe that, just as with the experimental
data plotted in Figure \ref{fig:individual_cDR}, the cDR is more
monotonic (on the shown ranges, at least) for smaller time horizons $T$.
Using logarithmic scale on inputs, for comparison with experimental plots.
}
\label{fig:example_iffl_nonmonotone_DR} 
\efigstar

To show the IFF1 statement in the theorem, one must show that the
map
\[
u \mapsto \int_{0}^T y_{u}(t)\,dt
\]
is nondecreasing (for any fixed $T>0$), and in turn, analyzing the
variational system along any given trajectory, one must show that
\[
\int_{0}^T \partial_u y_u(t)\,dt \,\geq \,0
\]
for all $T,u$, which is done by a careful balancing of $+/-$
contributions to the integral.

The proof for IFF2, on the other hand, can be understood conceptually
as follows.
Let us consider a larger class of systems:
\beqn
\LDdiffdotI &=& \alpha u - \delta x\\
\LDdiffdotH &=& \beta \frac{u}{K+x} - \gamma y
\eeqn
(so that IFF2 is the special case $K=0$).

Now, when $K=0$, one has the equivariance (one-parameter Lie group of
symmetries) $(u,x,y)\mapsto (pu,px,y)$ and this suggests using a new
variable $p:=u/(K+x)$ to map to what turns out to be a monotone system:
\beqn
\LDdiffdotI &=& \alpha u - \delta x\\
\LDdiffdotT &=& p\left(\frac{\delta x}{K+x} - \alpha p\right)\\
\LDdiffdotH &=& \beta p - \gamma y\,.
\eeqn
In this system, solutions depend monotonically on initial states, and
in particular on $p(0)$, and thus also (for a fixed initial state
$x(0)$) monotonically on the (constant) input $u$.
Note that the original system is not monotone, but the transformed
system is.

We remark that the recent work~\cite{26cdc_moh_arthur_eduardo_cdr} builds on~\cite{gupta_sontag_cdr} to provide a more complete classification of IFF motifs in terms of their cDR monotonicity.

\section{Concluding remarks and outlook}

This article has argued that many biologically relevant questions
about network function are answered not by steady states alone, but by
the qualitative structure of \emph{transient} and \emph{input-driven}
behaviors.  We used the term \emph{dynamic phenotypes} to emphasize
experimentally accessible response features, such as overshoots,
biphasic transients, pulse trains, subharmonics, fold-change
invariance, and cumulative-exposure effects, that emerge under rich
classes of probing inputs beyond step stimuli. In this perspective,
dynamic responses become a source of \emph{structural information}: by
observing which qualitative phenotypes are possible (or impossible),
one can rule out entire classes of network interconnection patterns
even when parameters are uncertain and mechanistic detail is
incomplete.

A recurring theme throughout the vignettes has been that \emph{sign
structure} and \emph{feedback organization} place sharp constraints on
what transients can look like.  In particular, monotone (balanced)
architectures support strong qualitative predictions: when initialized
at steady state, monotone input/output systems cannot produce certain
transient signatures (for instance, a nondecreasing input cannot yield
a biphasic output), and the influence of perturbations propagates in
an unambiguous way through the network. When such signatures
\emph{are} observed, the explanation often points to specific
violations of monotonicity, most prominently incoherent feedforward
(IFF) motifs and antagonistic multi-path influences operating on
distinct time scales.  The IFF viewpoint provides a unifying
systems-level explanation for why non-monotonic and delayed responses
are so common across biological scales, from intracellular regulation
to population dynamics, and why modest interventions can have
disproportionately large effects on peak outcomes.

Beyond transients induced by steps, the paper emphasized that
\emph{probing inputs} can be used as a discriminatory lens.  Periodic
and pulsed stimuli can expose dynamical regimes that are invisible in
static dose--response curves, and can separate mechanistic
alternatives that share identical steady-state input/output maps.
Likewise, fold-change detection (scale invariance) and Weber-like
behavior highlight the role of symmetries and invariances as
organizing principles: when a network responds to \emph{relative}
changes rather than absolute levels, that property strongly constrains
admissible model classes.  Finally, cumulative dose and related
integral-type readouts provide an additional experimentally tractable
phenotype that can distinguish between fast adaptive suppression and
true integration of exposure. 

Several broader methodological messages emerge.

\paragraph{Dynamic phenotypes as ``model class tests.''}
In settings where parameter identification is ill-posed, it is often
more productive to ask which \emph{qualitative} behaviors a model
class can or cannot generate.  Dynamic phenotypes can therefore be
used as falsification criteria: a single carefully chosen input family
(e.g., ramps of varying slope, periodic forcing across frequencies,
paired pulses with variable spacing) may exclude large regions of
model space even when data are limited.

\paragraph{Interconnection logic before detailed kinetics.}
Many conclusions can be drawn from the directed signed graph,
time-scale separation, and the presence or absence of particular
motifs (IFF, negative feedback, positive feedback,
sequestration/competition).  This supports a workflow in which
coarse-grained interconnection hypotheses are tested and refined prior
to committing to detailed biochemical kinetics.

\paragraph{Limitations from the interconnection viewpoint.}
On the other hand, a limitation of the viewpoint developed in this
article is that treating models as input-output systems whose interconnection with
upstream and downstream modules is assumed not to qualitatively alter 
the behavior being analyzed is, in biological systems, often not a
valid assumption. This was emphasized in the work on retroactivity and
modularity~\cite{ddv07,mra_book10,delvecchio_qian_murray_sontag_annual_reviews2018}: connecting a biological module to
another module can change its dynamics, because downstream components
may load, sequester, or otherwise perturb the molecular species that
transmit the signal. Thus the input-output behavior of a module
measured in isolation may differ from its behavior once embedded in a
larger network. 
This issue is directly relevant to the themes of the present
paper. Many of the arguments here concern qualitative transient
phenotypes, such as overshoots, biphasic responses, adaptation
kinetics, and peak responses. Retroactivity can modify precisely these
transient features, even if the eventual steady state is less
affected. For example, loading effects may slow down a signaling
response, attenuate an overshoot, or change the apparent time scale
separation between activating and inhibitory pathways. Thus, in
applications of the ideas discussed here, one should keep in mind that
the ``same'' motif may display different dynamic phenotypes depending on
the context in which it is embedded. 

\paragraph{Limitations from deterministic modeling}
A second important limitation is that the discussion in this article is
deterministic. Biological systems are intrinsically stochastic,
especially at the intracellular scale, where molecular copy numbers
may be low, reaction events are discrete, and cell-to-cell variability
can be substantial. Noise can affect not only quantitative features
such as amplitudes and timing, but also qualitative response
classifications. For instance, a response that is biphasic in an
averaged deterministic model may appear differently at the single-cell
level, and stochastic fluctuations may obscure, enhance, or even
generate apparent transient features.
This point is particularly important for feedback mechanisms such as
antithetic integral feedback. The deterministic and stochastic
descriptions of antithetic feedback are not merely two versions of the
same story. In stochastic settings, antithetic integral feedback has
distinctive properties connected with robust perfect adaptation at the
level of expectations, noise propagation, and molecular sequestration
dynamics; see, e.g.~\cite{BriatGuptaKhammash2016}. These questions
are central to the theory of biomolecular control, but they lead to a
different set of mathematical problems than those emphasized here.
For that reason, stochasticity was deliberately not treated in the
present paper. The objective was to isolate deterministic,
input-driven dynamic phenotypes and to show how they can constrain or
discriminate among network architectures. Extending this viewpoint to
stochastic systems would require deciding whether the relevant
phenotype is a sample-path property, a population average, a
distributional feature, or a probability of crossing a threshold. Each
of these choices leads to different notions of transient behavior and
different model-discrimination questions. This is an important
direction, but it is beyond the scope of the present article.

\paragraph{Control-theoretic tools as a language.}
Casting biological circuits in the language of input/output systems
allows one to bring into play concepts such as monotonicity, incremental order
preservation, invariance, and frequency response, not as literal
engineering analogies, but as precise mathematical structures that cut
across domains.

\paragraph{Some possible directions for future work.}
In addition to the many concrete technical questions that one might ask about each of
the topics discussed above, several broader, less sharply defined directions also suggest themselves:

\begin{itemize}
\item \textbf{Design of discriminating experiments.}  A natural next
  step is to formalize, for broad model classes, which families of
  inputs are \emph{maximally informative} for separating motifs that
  are indistinguishable under steps.  This includes principled choices
  of frequencies, pulse patterns, and ramp rates, and quantitative
  notions of identifiability at the level of \emph{network structure}
  rather than parameters.

\item \textbf{Robustness of qualitative inference.}
  While sign-structure arguments are inherently robust to parametric
  uncertainty, measurements are noisy and often indirect.  Developing
  theory that connects qualitative constraints (e.g., impossibility of
  biphasic responses under monotonicity assumptions) to finite-data
  statistical tests is an important open problem.

\item \textbf{Near-monotone networks and ``distance to
  monotonicity.''}  Biological networks often appear close to being
  monotone/balanced, yet small violations can create qualitatively new
  behaviors.  Understanding how dynamic phenotypes degrade as one
  moves away from monotonicity, and how to quantify this ``distance''
  in ways that predict dynamical complexity, remains a rich area for
  both theory and computation.

\item \textbf{Bridging levels of description.}  Many of the most
  compelling biological examples involve multi-scale effects
  (molecular regulation $\rightarrow$ cellular decisions $\rightarrow$
  population dynamics).  A systematic theory of how dynamic phenotypes
  compose across scales, and how coarse-grained motifs emerge from
  fine-grained interactions, would significantly strengthen the use of
  these ideas in practice.
\end{itemize}

In summary, transient behaviors and input-driven response patterns
should be viewed as objects of study in systems biology: they encode
function, they reveal mechanism, and they provide a route to model
discrimination that is often more realistic than full parameter
identification.  By emphasizing dynamic phenotypes and the structural
constraints imposed by interconnection logic, the paper aims to
contribute to a toolkit for reasoning about complex biological
networks in a manner that is both mathematically principled and
experimentally actionable.

More broadly, the examples considered here illustrate a role for control theory and biology that extends beyond the design of controllers.  Control-theoretic concepts can also provide tools for experimental design, model discrimination, and the identification of structural constraints from dynamic data. Developing such tools for the distinctive uncertainty, multiscale organization, and transient behavior of biological systems remains an important opportunity at the intersection of control theory and engineering biology. 

\section{Acknowledgments}

This work was supported in part by grants AFOSR FA9550-21-1-0289 and FA9550-22-1-0316.

The author thanks Nicholas Nolan for his help in verifying the
CC-BY-NC rights to the figures taken from~\cite{bleris10} and~\cite{ShimizuStocker2011}. The author also acknowledges working closely with several AI assistants to brainstorm ideas, search the literature, develop proofs, check arguments, and write
and edit the paper.  The author verified all mathematical statements and proofs and takes full responsibility for the content.

\bibliographystyle{plain}
\bibliography{%
  2026_transient_paper
}
\section*{Appendix}

\subsection*{Sketch of proof of monotone-response theorem}
\label{appendix:proof-monotone}

We start by ``pruning'' those state variables $\xx_j$ which do not lie in any
path from the input node to the output node $\xx_n$.
We now formalize this construction, which is analogous to the ``Kalman
decomposition'' reduction to minimal systems in linear control theory
\cite{mct}.
We start by splitting the set of variables $X$ into four disjoint subsets of
variables $\xx=(x,y,z,w)$, as follows:
\ben
\item
the output node $\xx_n$ is a component of the vector $x$, 
\item
the components of $x$ are reachable and observable,
\item
the components of $y$ are observable but not reachable,
\item
the components of $z$ are reachable but not observable,
and
\item
the components of $w$ are neither reachable nor observable.
\een
We assume without loss of generality that the output node $\xx_n$ is in the
first set of variables, $x$, since otherwise there would be no path from the
input to the output, and the output is then constant when starting from a steady
state.
It is clear that, with this partition, the equations look as follows:
\beqn
\LDdiffdotI &=& f(x,y,u)\\
\LDdiffdotH &=& g(y)\\
\LDdiffdotU &=& h(x,y,z,w,u)\\
\LDdiffdotV &=& k(y,w)
\eeqn
(for example, there can be no dependence on $z$ or $w$ in $f$ and in $g$,
since otherwise the $z$ and/or $w$ variables would be observable).

To prove the Theorem, we need to show, for the original system
$\LDdiffdotW=\ff(\xx,u)$, that if we start from a steady state $\ff(\xx_0,u_0)=0$
and if $u(t)$ is monotonic in time, with $u(0)=u_0$, then $\xx_n$ will be also
monotonic in time (with the same, or opposite, monotonic behavior depending on
parity). 
Write $\xx_0=(x_0,y_0,z_0,w_0)$, so $\ff(\xx_0,u_0)=0$ means that
$
f(x_0,y_0,u_0) = g(y_0) = h(x_0,y_0,z_0,w_0,u_0) = k(y_0,w_0) = 0
$.

The assumption that all directed paths from the input node $u$ to the output
node $\xx_n$ have the same parity applies also to the subsystem given by the
variables in $x$ in which the $y$ variables are set to $y_0$:
\be{eq:reduced}
\LDdiffdotI = {\hat f}(x,u) = f(x,y_0,u)
\ee
with initial state $x(0)=x_0$,
because partial derivatives of $\hat f$ with respect to $x$ and $u$ are also
partial derivatives of the original $\ff$.

\emph{Suppose that we have already proved the theorem for this subsystem in
  which all variables are reachable and observable.}
We claim next that the same is then true for the original system.
Consider the solution $x(t)$ of \eqref{eq:reduced} with input $u=u(t)$
and $x(0)=x_0$.
Consider also the solution of the full system 
$\LDdiffdotW = \ff(\xx,u)$ with $\xx(0)=\xx_0$ and the same input $u$, and write it
in the corresponding block form 
\[
\xx(t) = (\xi (t),\psi (t),\zeta (t),\omega (t)).
\]
We want to prove that $\xi (t)=x(t)$ for all $t\geq 0$, from which the claim
will follow.
This follows because $g(y_0)=0$ implies that $y(t)\equiv y_0$.
(Note that the variables $\zeta (t)$ and $\omega (t)$ do not affect the output variable,
which is a component of $\xi (t)$.)

We now prove the theorem for the $x$-subsystem, for which all variables are
reachable and observable.
For ease of notation, we will write ${\hat f}$ simply as $f$, use $n$ for the
size of $x$, and assume that the output node is $x_n$.
Pick any index $i\in \{1,\ldots ,n\}$.
By reachability, there is at least one path $\pi $ from the input to $x_i$
and, if $i<n$, then by observability there is at least one path $\theta $ from
$x_i$ to the output node $x_n$.
We claim that every other path $\pi '$ from the input to $x_i$ has the same
parity as $\pi $.
Suppose without loss of generality that the parity of $\pi $ is $+1$.
We need to see that every other path $\pi '$ from the input to $x_i$ also has
parity $+1$.
If $i=n$, this is true by assumption (all paths from input to output have the
same parity).  So assume $i<n$.
Suppose that $\pi '$ has parity $-1$.
Then, the path $\pi \theta $ obtained by first following $\pi $ and then following $\theta $
has parity $(+1)*\rho =\rho $, where $\rho $ is the parity of $\theta $, 
and the path $\pi '\theta $ obtained by first following $\pi '$ and then following $\theta $
has parity $(-1)*\rho =-\rho $.
So we have two paths from input to output with different parity, which
contradicts the assumption of the Theorem.
In conclusion, every two paths from the input to any given node have the same
parity.

We assign labels taking the values ``$+1$ or $-1$'', denoted $\sigma _u$ and $\sigma _i$,
$i=1,\ldots ,n$, to the nodes $u$ and each node $x_1,\ldots ,x_n$ respectively, as
follows: $\sigma _u:=+1$, $\sigma _i:=$ sign of any path from $u$ to $x_i$.
A key observation is that, if $\varphi_{ij}=+1$ then $\sigma _i=\sigma _j$, and 
if $\gamma _{i}=+1$ then $\sigma _u=\sigma _i$.
Indeed, if we have a path $\pi $ from the input to $x_i$, then a path $\pi '$ can be
obtained, from the input to $x_j$, by simply adjoining the edge from $i$ to $j$,
which has parity equal to the parity of $\pi $.  Since $\sigma _j$ is the sign of any
path from the input to $x_j$, it follows that $\sigma _i=\sigma _j$, as claimed.
The statement for $\gamma _{i}=+1$ is simply (since we defined $\sigma _u:=+1$) that
$\sigma _i=+1$ if the one-step path from the input to node $x_i$ has parity 1,
which means that all paths have this parity.
Similarly, if  $\varphi_{ij}=-1$ then $\sigma _i=-\sigma _j$ , and 
if $\gamma _{i}=-1$ then $\sigma _u=-\sigma _i$.

Now make the change of variables $x_i \mapsto  \sigma _ix_i$ (i.e., reverse the sign of
variables with a ``$-1$'' label).
Writing the system in the new variables, we have now that
\[
\frac{\partial f_i}{\partial u}(x,u) \geq  0
\quad\mbox{and}\quad
\frac{\partial f_j}{\partial x_i}(x,u) \geq  0
\]
for all $i=1,\ldots,n$ and all $i,j=1,\ldots,n$ respectively.
Thus in the new variables we have what is called a \emph{cooperative system}
\cite{smith}.

We must prove that, if $u=u(t)$ is a monotonically increasing input for a
cooperative system, and if $x(0)=x_0$ is a steady state $f(x_0,u_0)=0$,
then every coordinate $x_i(t)$ of $x(t)$ (and, in particular, the output
node) is monotonically increasing as well.
(In the original coordinates, before sign reversals, $x_i(t)$ will decrease
if $\sigma _i=-1$.)
Similarly if  $u=u(t)$ is a monotonically decreasing input for a
cooperative system, and if $x(0)=x_0$ is a steady state $f(x_0,u_0)=0$,
then every coordinate $x_i(t)$ of $x(t)$ (and, in particular, the output node)
is monotonically decreasing as well.
We prove the increasing statement, since the second statement is proved
analogously.
From now on, for any two vectors $a,b\in \R^n$, we write simply $a\leq b$ to mean
that $a_i\leq b_i$ for each $i=1,\ldots ,n$.

We let $\varphi(t,x_0,v)$ denote the solution of $\LDdiffdotI=f(x,u)$ at time $t>0$ with
initial condition $x(0)=x_0$ and input signal $v=v(t)$.
\emph{Kamke's Comparison Theorem} (see \cite{smith} for
systems without inputs, and \cite{monotoneTAC} for an extension to systems
with inputs), asserts as follows:
Let $y(t)$ and $z(t)$ be two solutions of the system $\LDdiffdotI=f(x,u)$
corresponding, respectively, to an input $v(t)$ and an input $w(t)$.
Suppose that $y(0)\leq z(0)$ and that $v(t)\leq w(t)$ for all $t\geq 0$.
Then, $y(t)\leq z(t)$ for all $t\geq 0$.

Now pick an input $v$ that is non-decreasing in time and an initial state $x_0$
that is a steady state with respect to $v_0=v(0)$, that is, $f(x_0,v_0)=0$. 
Since $v(t)$ is non-decreasing, we have that $v(t)\geq  v(0)$
so that, by comparison with the input that is identically
equal to $v(0)$, we know that
\[
\varphi(h,x_0,v) \geq  \varphi(h,x_0,v_0)
\]
for all $h \geq 0$, where, by a slight abuse of notation, ``$v_0$'' is the
function that has the constant value $v_0$. 
We used the comparison theorem with respect to inputs and with the same initial
state. 
The assumption that the system starts at a steady state gives that
$\varphi(h,x_0,v_0) = x_0$ for all $h \geq 0$.  Therefore:
\be{eq:1}
x(h) \,\geq  \, x(0) \quad \quad \mbox{for all} \; h \geq 0 \,.
\ee
Next, we consider any two times $t \leq t+h$.  We wish to show that
$x(t) \leq  x(t+h)$.
Using~(\ref{eq:1}) and the comparison theorem now applied with respect to
initial states and the same input, we have that:
\[
x(t+h) \,=\, \varphi(t,x(h),v_h) \,\geq  \, \varphi(t,x(0),v_h) \,,
\]
where $v_h$ is the ``tail'' of $v$, defined by: $v_h(s)=v(s+h)$.
On the other hand, since the function $v$ is non-decreasing, it holds that
$v_h\geq  v$, in the sense that the inputs are ordered:
$v_h(t)\geq  v(t)$ for all $t \geq 0$.
Therefore, using once again the comparison theorem with respect to inputs and
with the same initial state, we have that
\[
\varphi(t,x(0),v_h) \,\geq  \, \varphi(t,x(0),v)= x(t)
\]
and thus we proved that $x(t+h)\geq  x(t)$.
So $x$ is a non-decreasing function.
This concludes the proof.

Sometimes we only care about conditional monotonicity, depending on monotonic
behavior of a particular node, even if the input is not monotonic.  The
following theorem from \cite{igoshin_et_al2015_monotone} is useful in that context.

{\bf Theorem.}
If the system is initially in steady state, the output
$x_n(t)$ will increase or decrease monotonically in
response to changes
in the input $u(t)$ if all directed paths from the input node to the
output node pass through an internal node $x_i(t)$ whose dynamics are themselves monotonic,
 and all the directed paths from node
$x_i(t)$ to the output node $x_n(t)$ have the same parity. Furthermore,
monotonically increasing (decreasing) $x_i(t)$ will trigger the monotonic
increase (respectively, decrease) in $x_n(t)$ if the parity is positive or will
trigger monotonic decrease (respectively, increase) if parity is negative.

A proof is as follows.
The assumption that all directed paths from the input node $u$ to the output
node $\xx_n$ must pass through the internal node $\xx_i$ can be formalized by
splitting the set of nodes $\xx$ into three subsets,
$\xx=(x,y,z)$, 
where
the components of $x$ are those nodes $\xx_j$, $j\not= i$, for which there is at
least one path from the input node $u$ to $\xx_j$ which does not pass through
node $\xx_i$,
$y=\xx_i$, and
the components of $z$ are all remaining nodes, including $\xx_n$.
For this partition, the equations look as follows:
\beqn
\LDdiffdotI &=& f(x,y,z,u)\\
\LDdiffdotH &=& g(x,y,z,u)\\
\LDdiffdotU &=& h(z,y)
\eeqn
because, if there were any dependence of $h$ on some coordinate $x_j$, then
there would be a path from the input to some component of $z$ (follow a path to
$x_j$ and concatenate it with an edge from $x_j$ to this component).

The condition that all the directed paths from $y=\xx_i$ to the
output node $\xx_n$ have the same parity means that in the system
$\LDdiffdotU = h(z,v)$
(where we now view $y(t)$ as an input, which we write as ``$v(t)$''
to avoid confusion) all paths from the input to the output have the same
parity, as in the hypothesis of the Theorem.
Suppose that we consider an input $u$, starting from a steady state
$(x_0,y_0,z_0)$.  Think of $v(t)=y(t)$ as an input.
Since we started from a steady state, we know that $h(v(0),z_0)=0$.
Thus, if $v(t)$ is monotonic, the previous theorem gives us that the output
is monotonic, increasing or decreasing depending on parity and on the increasing
or decreasing character of the input. 
\begin{figure}[ht]
\picc{0.3}{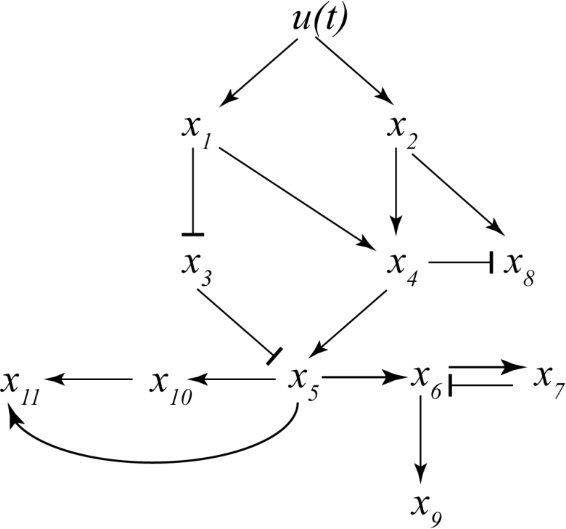}
\caption{Example to illustrate monotone dependence results.}
\label{fig:monotone_graph_example}
\end{figure}%

For example, on the system shown in Fig.~\ref{fig:monotone_graph_example}, the first
theorem allows us to conclude that monotonically increasing input $u(t)$ will
ensure monotonic increase of $\xx_1, \xx_2, \xx_4, \xx_5, \xx_{10}, \xx_{11}$
(since all directed paths from $u$ to the respective node have positive parity),
and monotonic decrease is ensured for $\xx_3$, but monotonicity cannot be
guaranteed for $\xx_6, \xx_7,\xx_8, \xx_9$.  On the other hand, if we do not
know whether the input signal $u(t)$ is monotonic, or if in case an additional
negative path in the network from $u(t)$ to $\xx_5$ is added, we may still use
the second formulation to conclude that if $\xx_5(t)$ is monotonic, then
$\xx_{10}$ and $\xx_{11}$ will also be monotonic. Indeed, all the paths to $\xx_{10}$ and $\xx_{11}$ from
input $u(t)$ pass through $\xx_5$ and all the paths from $\xx_5$ to $\xx_{10}$
and $\xx_{11}$ have positive parity. The argument does not work for $\xx_9$ due
to a negative feedback loop between $\xx_6$ and $\xx_7$ (a directed path
that goes around this loop will have the opposite parity from the path that
does not).

\endarticle

\end{document}